\documentclass[12pt,letterpaper]{article}
 
\usepackage{epsfig}
\usepackage{axodraw}
 
\newcommand{\mrm}[1]{\mathrm{#1}}
\newcommand{\mbf}[1]{\mathbf{#1}}

\newcommand{\ttt}[1]{\texttt{#1}}

\newlength{\tmplen}

\def\lsim{\mathrel{\rlap{\lower4pt\hbox{\hskip1pt$\sim$}}
    \raise1pt\hbox{$<$}}}           
\def\gsim{\mathrel{\rlap{\lower4pt\hbox{\hskip1pt$\sim$}}
    \raise1pt\hbox{$>$}}}           
\newcommand{\avg}[1]{\ensuremath{\langle #1 \rangle}}
 
\newcommand{\alphas}{\alpha_{\mathrm{s}}}

\newcommand{\as}{\alpha_{\mathrm{s}}}

\newcommand{\kT}{k_{\perp}}
\newcommand{\pT}{p_{\perp}}
\newcommand{\pTs}{p^2_{\perp}}
\newcommand{\pTo}{p_{\perp 0}}
\newcommand{\pTmin}{p_{\perp\mathrm{min}}}
\newcommand{\pTsmin}{p^2_{\perp\mathrm{min}}}
\newcommand{\pTmax}{p_{\perp\mathrm{max}}}
\newcommand{\pTsmax}{p^2_{\perp\mathrm{max}}}
\newcommand{\pThard}{p_{\perp\mathrm{hard}}}
\newcommand{\pTZ}{p_{\perp\mathrm{Z}}}

\newcommand{\pTslc}{p^2_{\perp\mathrm{LC}}}
\newcommand{\pTnum}{p_{\perp 1,2}}
\newcommand{\pTsnum}{p^2_{\perp 1,2}}

\newcommand{\pTsbc}{p^2_{\perp b,c}}
\newcommand{\pTe}{p_{\perp\mrm{evol}}}
\newcommand{\pTse}{p^2_{\perp\mrm{evol}}}
\newcommand{\pTL}{p_{\perp\mathrm{L}}}
\newcommand{\pTsL}{p^2_{\perp\mathrm{L}}}
\newcommand{\pTD}{p_{\perp\mathrm{D}}}

\newcommand{\pTsA}{p^2_{\perp\mathrm{A}}}
\newcommand{\ET}{E_{\perp}}
\newcommand{\mQ}{m_{\mathrm{Q}}}
 
\renewcommand{\b}{\mathrm{b}}
\renewcommand{\c}{\mathrm{c}}
\renewcommand{\d}{\mathrm{d}}
\newcommand{\e}{\mathrm{e}}

\newcommand{\g}{\mathrm{g}}

\newcommand{\p}{\mathrm{p}}
\newcommand{\q}{\mathrm{q}}

\renewcommand{\H}{\mathrm{H}}

\newcommand{\Q}{\mathrm{Q}}
\newcommand{\W}{\mathrm{W}}
\newcommand{\Z}{\mathrm{Z}}

\newcommand{\pbar}{\overline{\mathrm{p}}}
\newcommand{\qbar}{\overline{\mathrm{q}}}

\newcommand{\Qbar}{\overline{\mathrm{Q}}}

\newcommand{\qsea}{\ensuremath{\q_{\mrm{s}}}}
\newcommand{\qcmp}{\ensuremath{\q_{\mrm{c}}}}

\newcommand{\nint}{\ensuremath{n_{\mathrm{INT}}}}
\newcommand{\nmi}{\ensuremath{n_{\mathrm{MI}}}}
\newcommand{\nisr}{\ensuremath{n_{\mathrm{ISR}}}}
\newcommand{\nfsr}{\ensuremath{n_{\mathrm{FSR}}}}

\newenvironment{Itemize}{\begin{list}{$\bullet$}%
{\setlength{\topsep}{0.2mm}\setlength{\partopsep}{0.2mm}%
\setlength{\itemsep}{0.2mm}\setlength{\parsep}{0.2mm}}}%
{\end{list}}
\newcounter{enumct}
\newenvironment{Enumerate}{\begin{list}{\arabic{enumct}.}%
{\usecounter{enumct}\setlength{\topsep}{0.2mm}%
\setlength{\partopsep}{0.2mm}\setlength{\itemsep}{0.2mm}%
\setlength{\parsep}{0.2mm}}}{\end{list}}
 
\newlength{\abstwidth}
\begin{document}
\sloppy
 
\pagestyle{empty}
 
\begin{flushright}
LU TP 04--29\\
hep-ph/0408302\\
August 2004
\end{flushright}
 
\vspace{\fill}
 
\begin{center}
{\LARGE\bf Transverse-Momentum-Ordered Showers}\\[4mm]
{\LARGE\bf and Interleaved Multiple Interactions}\\[10mm]
{\Large T. Sj\"ostrand\footnote{torbjorn@thep.lu.se} and %
P. Z. Skands\footnote{peter.skands@thep.lu.se}} \\[3mm]
{\it Department of Theoretical Physics,}\\[1mm]
{\it Lund University,}\\[1mm]
{\it S\"olvegatan 14A,}\\[1mm]
{\it SE-223 62 Lund, Sweden}
\end{center}
 
\vspace{\fill}
 
\begin{center}
{\bf Abstract}\\[2ex]
\begin{minipage}{\abstwidth}
We propose a sophisticated framework for high-energy hadronic collisions,
wherein different QCD physics processes are interleaved in a common
sequence of falling transverse-momentum values. Thereby phase-space
competition is introduced between multiple parton--parton interactions
and initial-state radiation. As a first step we develop new
transverse-momentum-ordered showers for initial- and final-state
radiation, which should be of use also beyond the scope of the
current article. These showers are then applied in the context of
multiple interactions, and a few tests of the new model are presented.
The article concludes with an outlook on further aspects, such as
the possibility of a shower branching giving partons participating in
two different interactions. 
\end{minipage}
\end{center}
 
\vspace{\fill}
 
\clearpage
\pagestyle{plain}
\setcounter{page}{1}
 
\section{Introduction}
 
High-energy hadronic collisions offer a busy environment.
The incoming hadrons seethe with activity as partons
continuously branch and recombine. At the moment of
collision, several partons from the two incoming hadrons
may undergo interactions, that scatter the partons in different
directions. The scattered partons may radiate, and all
outgoing partons, including the beam remnants, 
hadronize in a correlated fashion to produce the observable
high-multiplicity events. The physics involves a subtle blend
of many perturbative and nonperturbative phenomena. No wonder
that there is no simple, standard description to be offered!
 
What often saves the day is that most of the above activity
is soft, i.e.\ confined to small transverse momenta $\pT$.
When the processes of interest occur at large momentum
transfers they therefore stand out, by producing jets,
leptons or photons at large $\pT$. To first approximation, 
the rest of the activity, which we refer to as the 
\emph{underlying event}, may then be disregarded. For precision
studies, however, the problem remains: minijets from the
underlying event may e.g.\ affect the jet energy calibration and
the lepton and photon isolation criteria. Quite apart from the
interesting challenge of better understanding the complex
(semi-)soft processes for their own sake, this motivates an effort
to investigate and model as well as possible the underlying event
physics (when a selective trigger is used) and minimum-bias physics
(for the inclusive sample of multihadronic events).
 
The basic building blocks needed to describe
hadron--hadron collisions include hard-scattering matrix
elements, parton density functions, initial- and final-state parton
showers, and a hadronization scheme. Each of these deserve study in its
own right, but additionally there is the question of how they should be
combined. It is this latter aspect that we take aim at here. More 
specifically, we concentrate on the non-trivial interplay between multiple
parton--parton interactions and initial-state parton showers, 
extending previous models for multiple interactions and developing new
models for $\pT$-ordered initial- and final-state parton showers in the
process.
 
\begin{figure}[t]
\begin{picture}(420,400)(-40,-20)
\SetWidth{0}
\GBox(0,0)(320,20){0.8}
\SetWidth{1.5}
\LongArrow(0,0)(330,0)\Text(340,2)[l]{interaction}
\Text(340,-12)[l]{number}
\LongArrow(0,0)(0,350)\Text(0,365)[]{$\pT$}
\Vertex(40,300){5}\Text(45,310)[l]{hard int.}
\Line(40,20)(40,300)\Text(40,-12)[]{1}
\Vertex(120,210){4}\Text(125,220)[l]{mult. int.}
\Line(120,90)(120,210)\Text(120,-12)[]{2}
\Vertex(210,140){4}\Text(215,150)[l]{mult. int.}
\Gluon(210,90)(210,140){5}{5}\Text(210,-12)[]{3}
\Line(120,90)(165,90)\Line(165,90)(165,20)
\Vertex(165,90){3}\Gluon(165,90)(210,90){5}{4}
\Vertex(290,55){4}\Text(295,65)[l]{mult int.}
\Gluon(290,20)(290,55){5}{3}\Text(290,-12)[]{4}
\DashLine(0,330)(320,330){5}\Text(-5,330)[r]{$\pTmax$}
\DashLine(0,20)(320,20){5}\Text(-5,20)[r]{$\pTmin$}
\SetWidth{1.0}
\DashLine(0,300)(320,300){5}\Text(-10,300)[r]{$p_{\perp 1}$}
\DashLine(0,210)(320,210){5}\Text(-10,210)[r]{$p_{\perp 2}$}
\DashLine(0,140)(320,140){5}\Text(-10,140)[r]{$p_{\perp 3}$}
\DashLine(0,90)(320,90){5}\Text(-8,90)[r]{$p_{\perp 23}$}
\DashLine(0,55)(320,55){5}\Text(-10,55)[r]{$p_{\perp 4}$}
\Vertex(40,250){3}\Gluon(40,250)(85,250){4}{5}\Text(65,262)[]{ISR}
\Vertex(40,160){3}\Gluon(40,160)(85,160){4}{5}
\Vertex(40,105){3}\Gluon(40,105)(85,105){4}{5}
\Vertex(40,70){3}\Gluon(40,70)(85,70){4}{5}
\Vertex(120,180){3}\Gluon(120,180)(165,180){4}{5}\Text(145,192)[]{ISR}
\Vertex(120,125){3}\Gluon(120,125)(165,125){4}{5}
\Vertex(210,115){3}\Gluon(210,115)(255,115){4}{5}\Text(235,127)[]{ISR}
\Vertex(165,40){3}\Gluon(165,40)(210,40){4}{5}\Text(190,52)[]{ISR}
\SetWidth{0.5}
\DashLine(0,250)(320,250){5}\Text(-9,250)[r]{$p_{\perp 1}'$}
\DashLine(0,180)(320,180){5}
\DashLine(0,160)(320,160){5}
\DashLine(0,125)(320,125){5}
\DashLine(0,115)(320,115){5}
\DashLine(0,105)(320,105){5}
\DashLine(0,70)(320,70){5}
\DashLine(0,40)(320,40){5}
\end{picture}
\caption{Schematic figure illustrating one incoming hadron in
an event with a hard interaction occurring at $p_{\perp 1}$ and
three further interactions at successively lower $\pT$ scales, each
associated with (the potentiality of) initial-state radiation,
and further with the possibility of two interacting partons (2 and 3
here) having a common ancestor in the parton showers. Full lines
represent quarks and spirals gluons. The vertical $\pT$ scale is
chosen for clarity rather than realism; most of the activity is
concentrated to small $\pT$ values.
\label{fig:intertwined}}
\end{figure}
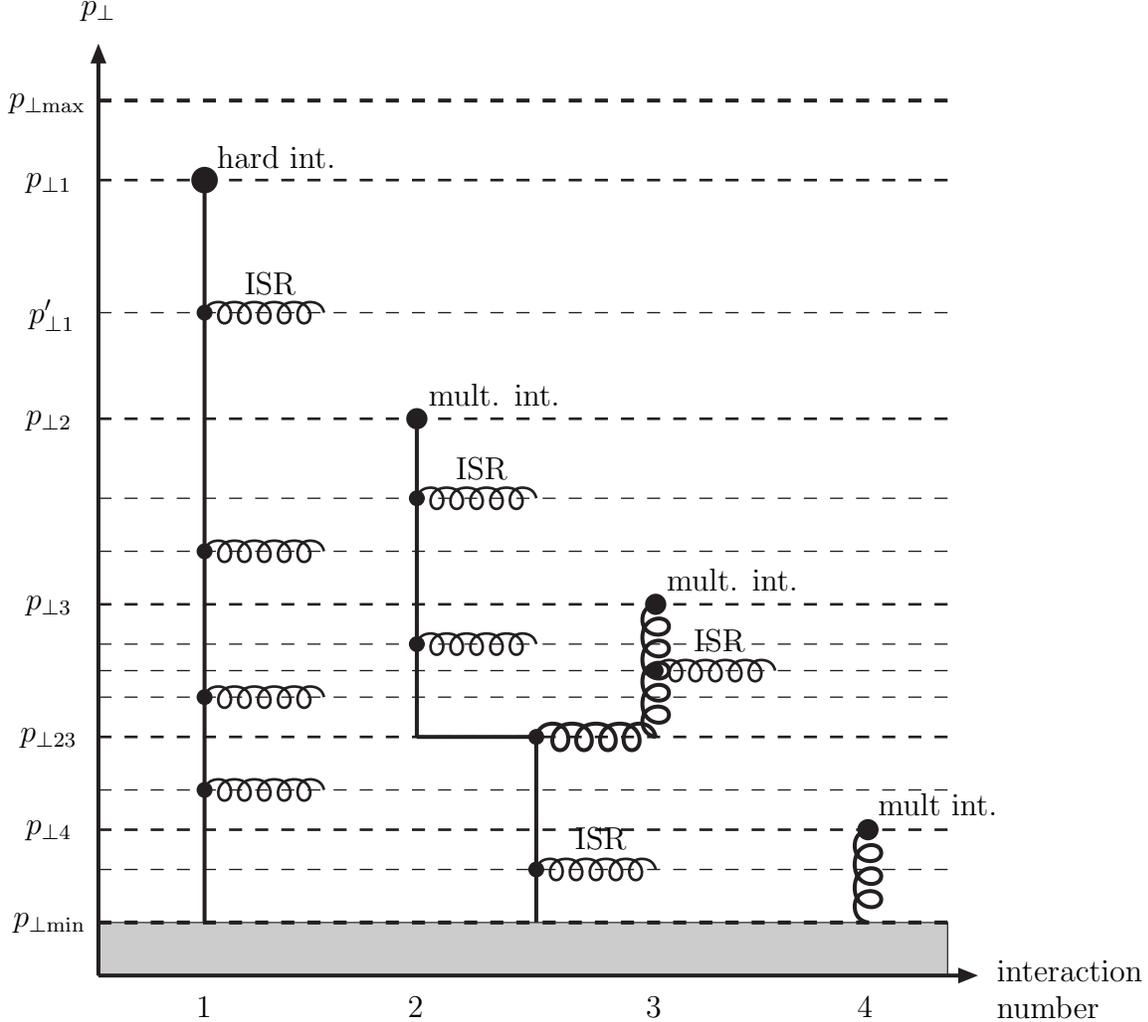
 
A good starting point for the discussion is offered by
Fig.~\ref{fig:intertwined}. Based on the composite nature of hadrons
we have here depicted \textit{multiple interactions} (MI) between
several pairs of incoming partons, see
ref.~\cite{multint} for a minireview. The structure of an incoming
hadron is illustrated, with the $\pT$ evolution of some partons
from a nonperturbative border at $\pTmin$ up to the different
perturbative interactions. The $\pTs = \hat{t}\hat{u}/\hat{s}$ scale
is a convenient measure of hardness, since the $t$ (and $u$) channel
gluon  exchange processes $\q \q' \to \q \q'$, $\q \g \to \q \g$ and
$\g \g \to \g \g$ dominate the cross section. One has to imagine a
corresponding picture for the other hadron --- omitted for clarity ---
with the two incoming sides joined at the interactions.
 
The next immediate issue that arises is how to describe hadronic objects
under such conditions. In general,
cross section calculations rely on parton density
functions to describe the initial state. For the 
joint cross section of several simultaneous interactions one thus needs
multi-parton densities, categorized by flavour content and fully
differential in all $x$ and $Q^2 \approx \pTs$ values. Obviously such
densities are almost entirely unconstrained, with neither data nor
first-principles theory giving more than the roughest guidelines.
To develop a realistic approximate framework,
it is natural to consider first the hardest interaction, which after all
should be the most important one in terms of experimental
consequences. Moreover, self-consistency ensures that
this is also the interaction for which the standard `one-parton-inclusive'
pdf's should be applicable; when averaging over all configurations of
softer partons, the standard QCD phenomenology should be obtained for the
ones participating in the hardest interaction, this being the way the 
standard parton densities have been measured. Thus it makes sense to
order and study the interactions in a sequence of falling `hardness', for
which we shall here take $\pT$ as our measure, i.e.\ we consider the
interactions in a sequence $p_{\perp 1} > p_{\perp 2} > p_{\perp 3} >
p_{\perp 4}$.  The normal parton densities can
then be used for the scattering at $p_{\perp 1}$, and
correlation effects, known or estimated, can be introduced in the
choice of `subsequent' lower-$\pT$ scatterings.
 
In ref.~\cite{multint} we developed a new and sophisticated model to
take into account such correlations in momentum and flavour.
In particular, contrary to the earlier model described in ref.~\cite{Zijl},
the new model allows for more than one valence quark to be kicked out, and
also takes into account the fact that sea quarks come in pairs.
The beam remnant structure and colour flow topologies can become quite
complicated, and so-called string junctions have to be handled, 
see \cite{BNV}.
 
In addition, the more sophisticated machinery allowed a more complete
treatment of \textit{initial-state radiation} (ISR) and
\textit{final-state radiation} (FSR). That is, each simple $2 \to 2$ 
interaction could be embedded in the 
center of a more complicated $2 \to n$ process, $n \geq 2$, where additional
partons are produced by ISR or FSR. In order to avoid
doublecounting, this additional radiation should be softer than
the core $2 \to 2$ interaction. Here $\pTs$ is again a convenient
measure for hardness ordering, but not a unique one.
 
In this article, we introduce an additional interplay,
between multiple parton interactions and ISR. ISR is the mechanism whereby
parton densities evolve and become 
scale-dependent. The paradigm is that parton densities at a scale
$Q^2$, in our case identified with $\pTs$, probe the resolved
partonic content at that scale. Therefore the issue of multi-parton
densities is mixed in with the handling of ISR. For instance,
if an ISR branching related to the first interaction occurs
at a $p_{\perp 1}' < p_{\perp 1}$ then that reduces the available phase space
for a second interaction at $p_{\perp 2} < p_{\perp 1}'$.
In the complementary region $p_{\perp 2} > p_{\perp 1}'$, it is
instead the momentum carried away by the second interaction that reduces
the phase space for the ISR branching of the first. Thus, a consistent
choice is to consider ISR (on both of the two incoming hadron sides)
and MI in parallel, in one common sequence of decreasing $\pT$ values,
where the partonic structure at one $\pT$ scale defines what is allowed
at lower scales. Again this approach of \textit{interleaved evolution} 
is intended to accurately reproduce
measurements at $\pT$ values corresponding to the hardest scales in the 
event, and fits well with the backwards
evolution approach to ISR \cite{backwards}. (One could have devised
alternative procedures with forward evolution from lower to higher $\pT$
values, which would have offered a more intuitive physics picture,
but with problems of its own.)
 
To the best of our knowledge, a scenario of this kind has never
before been studied. In the early multiple interactions modelling
\cite{Zijl} ISR and FSR was only included for the hardest interaction,
and this before additional interactions were at all considered. In our
more recent study \cite{multint} all interactions included ISR and FSR,
but again separately for each interaction.
 
An additional difference is that, in our previous studies, spacelike
(for ISR) or timelike (for FSR) virtuality was used as evolution and
ordering variable in the showers. In the framework we shall present here, 
an essential
ingredient is the use of $\pT$-ordered showers, such that the proposed
competition between MI and ISR can be introduced in terms of a common
ordering variable. We have therefore
completed the rewriting begun in \cite{pTshower} of the existing
\textsc{Pythia} showering algorithms \cite{Pythia} to $\pT$-ordering.
These new models have interesting features in their own right,
quite apart from the application to interleaved multiple
interactions.
 
This article should be viewed as one step on the way towards a better
understanding of hadronic physics, but not as the final word. Further 
issues abound. The downwards evolution in $\pT$ may also reveal
that two seemingly separately interacting partons actually have a
common origin in the branching of a single parton at a lower $\pT$
scale ($p_{\perp 23}$ in Fig.~\ref{fig:intertwined}), and a single 
parton may scatter twice against partons in the other hadron. We shall refer
to such possibilities specifically as \textit{intertwined multiple
  interactions}, to distinguish them somewhat from the \textit{interleaved} 
evolution that will be our main focus here.
 
In this article we begin, in Section~\ref{sec:showers},
with a description of the new showering framework. This is followed,
in Section~\ref{sec:imi}, by a discussion on the model for interleaving
MI and ISR, and a few results are presented in Section~\ref{sec:results}. 
The outlook in Section~\ref{sec:outlook} contains a first estimate
of the significance of the backward evolution joining several 
interactions. Finally Section~\ref{sec:end} gives our conclusions.
 
\section{New Transverse-Momentum-Ordered Showers}
\label{sec:showers}
 
In this section we describe the new framework for timelike FSR and
spacelike ISR in the context of a single hard-scattering process. We
start by a brief review of the main existing showering algorithms, to
introduce the basic terminology and ideas we will make use of.
Thereafter the philosophy underlying the new algorithms is outlined.
The more technical details are then described separately, first for
timelike showers and then for spacelike ones, the latter as a rule
being the more complicated.
 
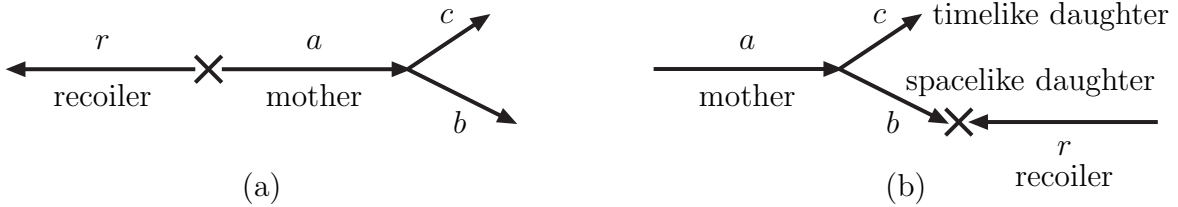
\begin{figure}[t]
\begin{picture}(220,80)(-90,-50)
\SetWidth{1.5}
\Line(-5,-5)(5,5)\Line(-5,5)(5,-5)
\LongArrow(5,0)(75,0)\Text(40,10)[]{$a$}\Text(40,-10)[]{mother}
\LongArrow(75,0)(115,-20)\Text(95,-20)[]{$b$}
\LongArrow(75,0)(105,20)\Text(90,20)[]{$c$}
\LongArrow(-5,0)(-75,0)\Text(-40,10)[]{$r$}\Text(-40,-10)[]{recoiler}
\Text(20,-45)[]{(a)}
\end{picture}\hfill
\begin{picture}(220,80)(-130,-30)
\SetWidth{1.5}
\Line(-5,-5)(5,5)\Line(-5,5)(5,-5)
\LongArrow(-115,20)(-45,20)\Text(-80,30)[]{$a$}\Text(-80,10)[]{mother}
\LongArrow(-45,20)(-5,0)\Text(-25,0)[]{$b$}
\Text(-20,15)[l]{spacelike daughter}
\LongArrow(-45,20)(-15,40)\Text(-30,40)[]{$c$}
\Text(-10,40)[l]{timelike daughter}
\LongArrow(75,0)(5,0)\Text(40,-10)[]{$r$}\Text(40,-20)[]{recoiler}
\Text(-20,-25)[]{(b)}
\end{picture}
\caption{Schematic figure with our standard terminology for
(a) a final-state and (b) an initial-state branching $a \to b c$,
with a cross marking the central hard process and a recoiling
parton $r$ moving out to or coming in from the other side.
\label{fig:terminology}}
\end{figure}
 
\subsection{Shower minireview}
 
In the shower approach, the evolution of a complex multi-parton final
state is viewed as a succession of simple parton branchings. Thus a
$2 \to n$ process can be viewed as consisting of a simple high-virtuality
process,
often $2 \to 2$, that approximately defines the directions and
energies of the hardest jets of the process, combined with shower
branchings at lower virtuality scales. The shower branchings thus
add details to the
simple answer, both by the production of additional jets and by a
broadening of the existing ones. We distinguish between initial-state
showers, whereby the incoming partons to the hard process build up
increasingly spacelike virtualities $Q^2$, and final-state showers,
where outgoing partons, including the non-colliding partons emitted
from the initial state, may have timelike virtualities $Q^2$ that
decrease in the cascade down to on-shell partons.
 
To first order, both cascade types are governed by the same DGLAP
evolution equations \cite{DGLAP}
\begin{equation}
\d {\cal P}_a(z, Q^2) =  \frac{\d Q^2}{Q^2} \, \frac{\alphas}{2 \pi} \,
P_{a \to bc}(z) \, \d z ~,
\label{eq:DGLAP}
\end{equation}
expressing the differential probability that a `mother' parton $a$
will branch to two `daughter' partons $b$ and $c$, at a virtuality
scale $Q^2$, and with parton $b$ taking a fraction $z$ of the $a$
energy, and $c$ a fraction $1-z$, cf.~Fig.\ref{fig:terminology}. The
splitting kernels $P_{a \to bc}(z)$ are (for massless quarks)
\begin{eqnarray}
 P_{\q \to \q\g}(z)&=&\frac{4}{3} \, \frac{1+z^2}{1-z} ~, \\
 P_{\g \to \g\g}(z)&=&3 \, \frac{(1-z(1-z))^2}{z(1-z)}  ~, \\
 P_{\g \to \q\qbar}(z)&=& \frac{n_f}{2 } \, (z^2 + (1-z)^2) ~,
\label{eq:APker}
\end{eqnarray}
where $n_f$ is the number of quark flavours kinematically allowed.
The kernels can be viewed as the universal collinear limit of the
behaviour of relevant matrix-element expressions. In such a context
it is natural to associate $Q^2$ with $|m^2|$, the virtuality of
an intermediate off-shell parton, since a $1/m^2$ comes from the
propagator of the virtual particle.  This is a free choice, however:
if $Q^2 = f(z) \, m^2$, then for any (nice) function $f(z)$ it holds
that $\d Q^2/Q^2 \, \d z = \d m^2/m^2 \, \d z$. At this stage,
several equivalent choices are therefore possible.
 
Note that eq.~(\ref{eq:DGLAP}) formally corresponds to the emission of an 
infinite number of partons. However, very soft and collinear gluons will
not be resolved in an infrared safe fragmentation framework such as the
string one \cite{string}, so we are free to introduce some effective $Q_0$
cut-off scale, of the order of 1~GeV or $\Lambda_{\mrm{QCD}}$, below which
perturbative
emissions need not be considered (to first approximation).
 
The remaining
total emission probability is still normally above unity, which
is allowed for an inclusive rate since several emissions can occur.
For an exclusive parton shower it is then convenient to introduce
 a `time' ordering, i.e.\ to decide which of the allowed emissions
occur `first'. This is encompassed in the Sudakov form factor
\cite{Sudakov}, expressing the probability that no emissions occur between
the initial
maximum scale $Q^2_{\mrm{max}}$ and a given $Q^2$, and within
limits $z_{\mrm{min}} < z < z_{\mrm{max}}$ that depend on the
kinematics and the $Q_0$ cutoff,
\begin{equation}
{\cal P}^{\mrm{no}}_a(Q^2_{\mrm{max}},Q^2) = \exp \left( -
\int_{Q^2}^{Q^2_{\mrm{max}}} \int_{z_{\mrm{min}}}^{z_{\mrm{max}}}
\d {\cal P}_a(z', {Q'}^2)    \right) ~,
\label{eq:Sudakov}
\end{equation}
so that the differential probability for the first branching to
occur at a $Q^2 = Q_a^2$ is given by $\d {\cal P}_a(z, Q_a^2) \, %
{\cal P}^{\mrm{no}}_a(Q^2_{\mrm{max}},Q_a^2)$.
Once the parton $a$ has branched, it is now the daughters $b$ and $c$
that can branch in their turn, with their $Q^2_{\mrm{max}}$ given by
$Q_a^2$, and so on until the cutoff scale is reached. Thus the shower
builds up.
 
Obviously, at this stage different $Q^2$ choices are no longer
equivalent: since $a$ will only branch once, those regions of
phase space considered at a later stage will be suppressed by
a Sudakov factor relative to those considered earlier.
 
For ISR, the most commonly adopted approach is that of
backwards evolution \cite{backwards}, wherein branchings are reconstructed
backwards in time/virtuality from the hard interaction to the shower
initiators. The starting point is the DGLAP equation for the
$b$ density
\begin{equation}
\d f_b(x,Q^2) =  \frac{\d Q^2}{Q^2}  \, \frac{\alphas}{2 \pi}
\, \int \frac{\d x'}{x'} \, f_a(x',Q^2)\,
P_{a \to bc} \left( \frac{x}{x'} \right) ~.
\label{dglap}
\end{equation}
This expresses that, during a small increase $\d Q^2$ there is a
probability for parton $a$ with momentum fraction $x'$ to become
resolved into parton $b$ at $x = z x'$ and another parton $c$ at
$x' - x = (1-z) x'$. Correspondingly, in backwards evolution, during
a decrease $\d Q^2$ a parton $b$ may become `unresolved' into parton
$a$. The relative probability $\d {\cal P}_b$ for this to happen is
given by the ratio $\d f_b / f_b$, which translates into
\begin{equation}
\d {\cal P}_b(x,Q^2)  = \left| \frac{\d Q^2}{Q^2} \right| \,
\frac{\alphas}{2 \pi} \, \int \d z \,
\frac{x' f_a(x',Q^2)}{x f_b(x,Q^2)} \, P_{a \to bc}(z) ~.
\label{dglapback}
\end{equation}
Again, ordering the evolution in $Q^2$ implies that
this `naive probability' should be multiplied by the probability
${\cal P}^{\mrm{no}}_b(x,Q^2_{\mrm{max}},Q^2)$ for no emissions to occur
at scales higher than $Q^2$, obtained from $\d {\cal P}_b$ by exponentiation
like  in eq.~(\ref{eq:Sudakov}). As for the timelike showers, additional
sophistication can be added by coherence constraints and
matrix-element merging, but ISR remains less well understood
than FSR \cite{smallx}.
 
\subsection{Existing approaches}
 
Of the three most commonly used final-state shower algorithms,
\textsc{Pythia} uses $m^2$ as evolution variable
\cite{PythiaFSR,PythiaMEmatch}, while \textsc{Herwig} uses an
energy-weighted emission angle,
$E^2(1-\cos\theta) \sim m^2/(z(1-z))$ \cite{Herwig}, and
\textsc{Ariadne} a squared transverse momentum, $\sim z(1-z) m^2$
\cite{pTcoherence,GGUP,Ariadne}. Thus the three programs give priority
to emissions with large invariant mass, large emission angle and large
transverse momentum, respectively.
 
The \textsc{Herwig} algorithm makes angular ordering a direct
part of the evolution process, and thereby correctly (in an
azimuthal-angle-averaged sense) takes into account coherence
effects in the emission of soft gluons \cite{coherence}. Branchings
are not ordered in hardness: often the first emission is that of
a soft gluon at wide angles. The algorithm does not populate the
full phase space but leaves a `dead zone' in the hard three-jet
region, that has to be filled up separately \cite{HerwigMEmatch}.
The kinematics of a shower is only constructed at the very
end, after all emissions have been considered.
 
The \textsc{Pythia} algorithm is chosen such that the shower
variables closely match the standard three-jet phase space in
$\e^+\e^- \to \q\qbar\g$, and such that the shower slightly
overpopulates the hard three-jet region, so that a simple
rejection step can be used to obtain a smooth merging of all
relevant first-order gluon-emission matrix elements with the
shower description \cite{PythiaMEmatch}. The mass-ordering of
emissions is one possible definition of hardness-ordering.
The main limitation of the algorithm is that it does not
automatically include coherence effects. Therefore angular
ordering is imposed by an additional veto, but then cuts away
a bit too much of the soft-gluon phase space \cite{GGonPythia}.
The kinematics of a branching is not constructed until the
daughters have been evolved in their turn, so that their
virtualities are also known.
 
The \textsc{Ariadne} algorithm differs from the above two
in that it is formulated in terms of dipoles, consisting of parton pairs,
rather than in terms of individual partons. The two partons that make up a
dipole may then collectively emit a gluon, causing the dipole to split in
two. Thus the basic process
is that of one dipole branching into two dipoles, rather than
of one parton branching into two partons. Emissions are ordered in
terms of a decreasing transverse momentum, which automatically
includes coherence effects \cite{pTcoherence}, and also is a good
measure of hardness. Kinematics can be constructed, in a Lorentz
invariant fashion, immediately after each branching, with individual
partons kept on mass shell at each stage. This makes it easy to stop
and restart the shower at some intermediate $\pT$  scale. The
implementation of an (L)CKKW-style matching of  matrix elements with
parton showers \cite{CKKWL} is therefore simplified, and in particular 
Sudakov factors can be generated dynamically to take into account the 
full kinematics of the branching history. A disadvantage is that
$\g \to \q\qbar$ branchings do not fit naturally into a dipole
framework, since they cannot be viewed as one dipole branching
into two.
 
In experimental tests, e.g.\ compared with LEP data \cite{LEPstudy},
the three final-state algorithms all offer acceptable
descriptions. If \textsc{Herwig} tends to fare the worst, it
could partly reflect differences in the hadronization
descriptions, where the \textsc{Herwig} cluster approach is
more simplistic than the \textsc{Pythia} string one, also used
by \textsc{Ariadne}. Among the latter two, \textsc{Ariadne}
tends to do somewhat better.
 
The above three programs also can be used for initial-state showers.
For \textsc{Herwig} the evolution variable is again angular-defined,
and for \textsc{Pythia} now $Q^2 = - m^2$. Both programs make use of
backwards evolution, as described above.
 
By contrast, the \textsc{Ariadne} approach defines radiating
dipoles spanned between the remnants and the hard scattering
\cite{AriadneIS}, and thereby cannot easily be related to the
standard DGLAP formalism. \textsc{Ldcmc} is a more sophisticated
approach \cite{LDCMC}, based on forward evolution and unintegrated
parton densities, and equivalent to the CCFM equations
\cite{CCFM}.
 
\subsection{The new approach}
 
In this article we wish to modify/replace the existing \textsc{Pythia}
shower routines so that emissions are ordered in $\pTs$ rather
than in $Q^2 = \pm m^2$, and also include some of the good points
of the dipole approach within the shower formalism. Specifically
we
\begin{Itemize}
\item retain the shower language of one parton branching into two,
such that $\g \to \q\qbar$ appears on equal footing with other
branchings,
\item make use of a simplified $\pTs$ as evolution variable,
picked such that the translation $\pTs \leftrightarrow \pm m^2$
is trivial, thereby preserving all the sophistication of the existing
matrix-element-merging,
\item construct a preliminary kinematics directly after each
branching, with currently unevolved partons explicitly on mass
shell,
\item define a recoil partner, `recoiler', for each branching
parton, `radiator', to keep the total energy and momentum
of the radiator+recoiler `dipole' preserved whenever a parton
previously put on mass shell is assigned a virtuality, and
\item ensure that the algorithms can be stopped and restarted at
any given intermediate $\pT$ scale without any change of the final
result, so that they can be used for interleaving showers and
multiple interactions (and also for (L)CKKW-style matching,
although this will not be made use of here).
\end{Itemize}
 
\subsubsection{Transverse momentum definitions}
 
So far, we have used $\pT$ to denote a general kind of `transverse
momentum', without specifying further the details of which momentum
we are talking about and which direction it is transverse to.
It is now our purpose to specify more closely which precise
definition(s) we have in mind, and to give a comparison to some
other commonly encountered $\pT$ definitions.
 
To specify a $\pT$ suitable for a branching $a \to b c$, consider
lightcone kinematics, $p^{\pm} = E \pm p_z$, for which $p^+ p^- =
m_{\perp}^2 = m^2 + \pTs$. For $a$ moving along the $+z$ axis, with
$p_b^+ = z p_a^+$ and $p_c^+ = (1-z) p_a^+$, $p^-$ conservation then
gives
\begin{equation}
m_a^2 = \frac{m_b^2 + \pTs}{z} + \frac{m_c^2 + \pTs}{1-z}
\end{equation}
or equivalently
\begin{equation}
\pTs = z (1-z) m_a^2 - (1-z) m_b^2 - z m_c^2 = \pTslc~.
\label{eq:pTlightcone}
\end{equation}
For a timelike branching $Q^2 = m_a^2$ and $m_b = m_c = 0$,
so then $\pTslc = z(1-z) Q^2$. For a spacelike branching
$Q^2 = - m_b^2$ and $m_a = m_c = 0$, so instead
$\pTslc = (1-z) Q^2$. We use these relations to
define abstract evolution variables $\pTse = z(1-z)Q^2$ or $= (1-z)Q^2$,
in which to order the sequence of shower emissions.
 
However, this is \textit{not} the $z$ definition we will use
to construct the kinematics of the branchings. For this, we
interpret $z$ to give the energy sharing between the daughters,
in the rest frame of the radiator+recoiler system, $E_b = z E_a$
and $E_c = (1-z) E_a$. The latter $z$ interpretation gives nice Lorentz
invariance properties --- energies in this frame are easily related
to invariant masses, $2E_i/m_{ijk} = 1 - m_{jk}^2/m_{ijk}^2$
for the $ijk$ three-parton configuration after the radiation
--- but gives more cumbersome kinematics relations, specifically for
$\pT$. This is the reason we use the lightcone relations to define
the evolution variable while we use the energy
definition of $z$ to construct the actual kinematics of the branchings.
 
\begin{figure}[t]
\begin{picture}(260,80)(-0,-40)
\SetWidth{1.5}
\LongArrow(0,0)(140,0)\Text(70,10)[]{$\mbf{p}_1 + \mbf{p}_2$}
\LongArrow(140,0)(220,30)\Text(185,25)[]{$\mbf{p}_1$}
\LongArrow(140,0)(200,-30)\Text(175,-25)[]{$\mbf{p}_2$}
\SetWidth{0.75}
\DashLine(140,0)(240,0){5}
\LongArrow(220,0)(220,30)\Text(225,15)[l]{$\pTnum$}
\LongArrow(200,0)(200,-30)\Text(205,-15)[l]{$\pTnum$}
\CArc(140,0)(30,-27,22)\Text(175,-10)[l]{$\theta_{12}$}
\Text(130,-35)[]{(a)}
\end{picture}\hfill
\begin{picture}(200,80)(-40,-40)
\SetWidth{1.5}
\LongArrow(0,0)(40,0)\Text(20,10)[]{$\mbf{p}_1 + \mbf{p}_2$}
\LongArrow(40,0)(130,20)\Text(70,20)[]{$\mbf{p}_1$}
\LongArrow(40,0)(-10,-20)\Text(10,-20)[]{$\mbf{p}_2$}
\SetWidth{0.75}
\DashLine(40,0)(150,0){5}\DashLine(0,0)(-30,0){5}
\LongArrow(130,0)(130,20)\Text(135,10)[l]{$\pTnum$}
\LongArrow(-10,0)(-10,-20)\Text(-15,-10)[r]{$\pTnum$}
\CArc(40,0)(20,-155,15)\Text(64,-10)[l]{$\theta_{12}$}
\Text(60,-35)[]{(b)}
\end{picture}
\caption{(a) Schematic figure of the clustering of two particles.
(b) A topology with a large $\theta_{12}$ but a small $\pTnum$.
\label{fig:clusterterm}}
\end{figure}
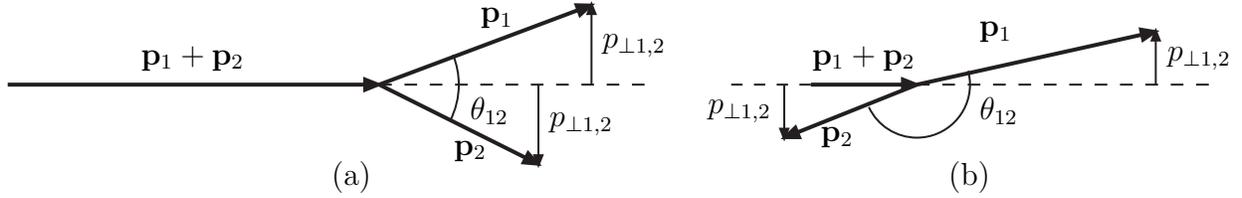
 
The deliberate choice of maintaining this dichotomy can be better understood
by examining a few different $\pT$ definitions in common use, in
particular those in clustering algorithms. To this end consider
first the situation depicted in Fig.~\ref{fig:clusterterm}a:
With the two particles massless, so that $E_1 = |\mbf{p}_1|$ and
$E_2 = |\mbf{p}_2|$,  the
momentum transverse to the vector sum $\mbf{p}_1 + \mbf{p}_2$,
which would correspond to the momentum of an imagined mother, is
\begin{equation}
\pT  =
\frac{| \mbf{p}_1 \times \mbf{p}_2 |}{| \mbf{p}_1 + \mbf{p}_2 |}
 = \frac{E_1 E_2 \sin\theta_{12}}%
{\sqrt{E_1^2 + E_2^2 + 2E_1 E_2 \cos\theta_{12}}} = \pTnum ~.
\end{equation}
There is one troubling feature of this $\pTnum$: not only does
it vanish when the opening angle $\theta_{12}$ goes to zero, but it also
vanishes for $\theta_{12} \to \pi$ (unless $E_1 \equiv E_2$).
Physically it is clear what is happening in this limit: the parton
with larger energy is going along the $\mbf{p}_1 + \mbf{p}_2$
direction and the one with smaller energy is just opposite to it,
Fig.~\ref{fig:clusterterm}b.
In a clustering algorithm, where the idea is to combine `nearby'
particles, a measure with such a behaviour clearly is undesirable.
Even when the starting point would be to have a $\pT$-related
measure for small $\theta_{12}$, we would prefer to have this measure
increase monotonically for increasing $\theta_{12}$, given fix $E_1$
and $E_2$, and behave a bit more like the invariant mass at large
angles. Therefore, in the \textsc{Luclus} algorithm \cite{Luclus},
the replacements $\sin\theta_{12} \to 2\sin(\theta_{12}/2)$ and
$| \mbf{p}_1 + \mbf{p}_2 | \to E_1 + E_2$ are performed, so that
\begin{equation}
\pT  =
\frac{| \mbf{p}_1 \times \mbf{p}_2 |}{| \mbf{p}_1 + \mbf{p}_2 |}
\to \frac{E_1 E_2 2 \sin(\theta_{12}/2)}{E_1 + E_2} = \pTL  ~.
\end{equation}
But, since $\sin^2(\theta_{12}/2) = (1-\cos\theta_{12})/2$, it also follows
that
\begin{equation}
\pTsL = \frac{E_1}{E_1 + E_2} \, \frac{E_2}{E_1 + E_2}
2 E_1 E_2 (1 - \cos\theta_{12}) \simeq  z (1-z) m^2 = \pTse  ~,
\end{equation}
given our $z$ definition in the shower as being one of energy sharing.
 
The $\pTL$ and $\pTe$ are not completely equivalent: for the
shower algorithm to be Lorentz invariant it is essential that the
energies in the $z$ definition are defined in the radiator+recoiler
rest frame, whereas the \textsc{Luclus} algorithm normally would be
applied in the rest frame of the event as a whole. Nevertheless, we
gain some understanding why the choice of $\pTse$ as evolution
variable actually may be more physically meaningful than $\pTsnum$.
Specifically, for the emission of a gluon off a $\q\qbar$ dipole,
say, we retain the subdivision of radiation from the mass-ordered
algorithm, roughly in proportions $1/m_{\q\g}^2 : 1/m_{\qbar\g}^2$
for $\q \to \q\g : \qbar \to \qbar\g$. With the $\pTsnum$ measure,
$\q$ radiation close to the $\qbar$ would not be disfavoured, since
also $\theta_{12} \to \pi$ would be classified as a collinear emission
region.
 
The Durham clustering algorithm \cite{Durham} is intended to
represent the transverse momentum of the lower-energy parton
relative to the direction of the higher-energy one, but again
modified to give a sensible behaviour at large angles:
\begin{equation}
p_{\perp\mrm{rel}} = \min(E_1, E_2) \, \sin\theta_{12} \to
\min(E_1, E_2) \, 2\sin(\theta_{12}/2) = \pTD
\end{equation}
Thereby it follows that
\begin{equation}
(\pTe \simeq)~ \pTL = \frac{\max(E_1, E_2)}{E_1 + E_2} \, \pTD
\end{equation}
so the two $\pT$ measures never disagree by more than a factor
of two, and coincide in the soft-gluon limit.
 
In the \textsc{Ariadne} dipole emission approach, finally, the
$\pT$ is defined as the momentum of the emitted parton relative
to the axis of the emitting partons \cite{Ariadne}. For the
emission of a soft parton 3 from the 1 and 2 recoiling parton
dipole one can then derive
\begin{equation}
\pTsA = \frac{m_{13}^2 m_{23}^2}{m_{123}^2} ~.
\end{equation}
When the $m^2 = m_{13}^2 \to 0$ limit is considered, this
corresponds to $\pTsA \approx (1-z) m^2$, rather than the
$\pTsL \approx z (1-z) m^2$. That is, for the soft-gluon limit
$z \to 1$ the two measures agree, while they disagree in the
hard-gluon limit $z \to 0$:
$\pTsA \approx m^2 \gg z m^2 \approx \pTsL$.
It is not clear whether this difference by itself would have any
visible consequences, but it illustrates that the meaning of
`$\pT$-ordered emission' is not uniquely defined.
 
\subsubsection{The new algorithms}
 
Taking into account the above considerations, the basic strategy
of the algorithms therefore can be summarized as follows:
\begin{Enumerate}
\item Define the evolution variable $\pTse$,
\begin{eqnarray}
\mrm{FSR:} & \hspace{-4mm} & \pTse = z(1-z)Q^2 ~,\\
\mrm{ISR:} & \hspace{-4mm} & \pTse = (1-z)Q^2 ~.
\end{eqnarray}
\item Evolve all radiators downwards in $\pTse$, from a $\pTsmax$
defined either by the hard process or by the preceding shower
branching, to find trial branchings according to the respective
evolution equation,
\begin{eqnarray}
\hspace{-8mm}\mrm{FSR:} & \hspace{-4mm} &
\d\mathcal{P}_a = \frac{\d\pTse}{\pTse} \, \frac{\as(\pTse)}{2\pi} \,
P_{a \to bc}(z) \, \d z \; \mathcal{P}^{\mrm{no}}_a(\pTsmax,\pTse) ~,
\label{eq:evforwfin}\\
\hspace{-8mm}\mrm{ISR:} & \hspace{-4mm} &
\d\mathcal{P}_b = \frac{\d\pTse}{\pTse} \, \frac{\as(\pTse)}{2\pi} \,
\frac{x' f_a(x',\pTse)}{x f_b(x,\pTse)} \, P_{a \to bc}(z) \, \d z \;
\mathcal{P}^{\mrm{no}}_b(x,\pTsmax,\pTse) ~.
\label{eq:evbackwfin}
\end{eqnarray}
Note that we have chosen $\pTse$ as scale both for parton densities
and $\as$ \cite{asscale}. The Sudakov form factors are, as before,
obtained by exponentiation of the respective real-emission expressions.
\item Select the radiator+recoiler set with the largest trial $\pTse$
to undergo the next actual branching.
\item For this branching, use the picked $\pTse$ and $z$ values to
derive the virtuality $Q^2$,
\begin{eqnarray}
\mrm{FSR:} & \hspace{-4mm} & m_a^2 = Q^2 = \frac{\pTse}{z(1-z)} ~,\\
\mrm{ISR:} & \hspace{-4mm} & - m_b^2 = Q^2 = \frac{\pTse}{1-z} ~.
\end{eqnarray}
\item Construct kinematics based on $Q^2$ and $z$\\
a) in the radiator+recoiler rest frame,\\
b) defining $z$ in terms of energy fractions, or equivalently
mass ratios,\\
c) assuming that yet unbranched partons are on-shell and that
the current two `earliest' ISR partons are massless, and\\
d) shuffling energy--momentum from the recoiler as required.
\item Iterate towards lower $\pTse$ until no further branchings
are found above the lower cutoff scale $\pTsmin$.
\end{Enumerate}
We now proceed to fill in the details for the respective
algorithms.
 
\subsection{Timelike showers}
 
\subsubsection{The basic formalism}
 
At each step of the evolution there is a set of partons that are candidates 
for further branching. Each such radiator defines dipoles together with
one or several recoiler partons. Normally these recoilers are defined
as the parton carrying the anticolour of the radiator, where
colour indices in a cascade are traced in the $N_C \to \infty$
limit. A gluon, with both a colour and an anticolour index, thus
has two partners, and the nominal emission rate is split
evenly between these two. Since the kinematics constraints in the
two radiator+recoiler dipoles normally will be different, the
actual emission probabilities will not agree, however.
 
To illustrate, consider
$\e^+ \e^- \to \gamma^*/\Z^0 \to \q\qbar\g$,
where one gluon has already been radiated. The quark is then
a radiator, with the gluon as recoiler, but also the gluon is a
radiator with the quark as recoiler. Similarly for the
antiquark--gluon pair. There is no colour dipole directly
between the quark and the antiquark. On the other hand, we
may also allow photon emission via the shower branching
$\q \to \q\gamma$, similarly to $\q \to \q\g$, and for such
branchings indeed the quark and the antiquark are each other's
recoilers, while the uncharged gluon is not involved at all.
In total, this configuration thus corresponds to six possible
radiator+recoiler sets. Each of these are to be evolved downwards
from the $\pTse$ scale of the first gluon emission, and the one with
largest new $\pTse$ is chosen as the next evolution step to be realized.
Thereafter the whole procedure is iterated, to produce one common
sequence of branchings
with $\pTmax > p_{\perp 1} > p_{\perp 2} > \ldots > \pTmin$.
 
A special case is where a narrow coloured resonance is concerned,
as for instance in top decay to $\b \W^+$. Here, gluon emissions
with energies above the width of the top should not
change the top mass. They are constrained inside the top
system. (In fact, when $\b \to \b\g$, the other end of the colour
dipole is rather defined by the decaying top itself.) Technically,
the $\W^+$ may then be chosen as the recoiler to the $\b$, to
ensure that the top mass remains unchanged. In this case all
radiation is off the $\b$, i.e.\ the system only contains one
gluon radiator, and this is enough to reproduce the desired
rate \cite{PythiaMEmatch}.
 
Once a recoiler has been assigned, the kinematics of a branching
is suitably defined in the rest frame
of the radiator+recoiler system, with the radiator $a$ (recoiler $r$)
rotated to move out along the $+z$ ($-z$) axis. Then one may define
$m_{ar}^2 = (p_a +p_r)^2$. For massless partons, the introduction of
an off-shell $Q^2 = m_a^2 = \pTse/z(1-z)$ increases $E_a$ from
$m_{ar}/2$ to $( m_{ar}^2 + Q^2)/2 m_{ar}$, with $E_r$ reduced by the
same amount. The two daughters share the energy according to
$E_b = z E_a$ and $E_c = (1-z) E_a$. With the modified $a$ still along
the $+z$ axis, the transverse momentum of the two daughters then becomes
\begin{equation}
\pTsbc = \frac{z(1-z)(m_{ar}^2 + Q^2)^2 - m_{ar}^2 Q^2}%
{(m_{ar}^2 - Q^2)^2} \, Q^2 ~\leq~ z(1-z) Q^2 = \pTse ~.
\end{equation}
The kinematics can now be completed, rotating and boosting
the two daughters and the modified recoiler back to the original frame.
 
Note that $\pTsbc$ and $\pTse$ always coincide for $z = 1/2$, and
agree well over an increasing $z$ range as $Q^2/m_{ar}^2 \to 0$.
We have already explained why $\pTse$ is a better evolution variable
than $\pTsbc$. In addition, there are technical advantages: had
evolution been performed in $\pTsbc$, the extraction of a $Q^2$ from
$\pTsbc$ would require solving a third-degree equation, which would be
messy and possibly give several solutions. The allowed $z$ range would
also be nontrivially defined. As it is now, the requirement
$Q^2 < m_{ar}^2$ easily leads to a
range $z_{\mrm{min}} < z < z_{\mrm{max}}$ for $\pTse$, with
\begin{equation}
z_{\mrm{min},\mrm{max}} = \frac{1}{2} \left(
1 \mp \sqrt{1 - \frac{\pTse}{m_{ar}^2}} \right)  ~.
\end{equation}
Once a trial $\pTse$ and $z$ has been picked, and thereby $Q^2$ is
known, an acceptable solution has to be in the smaller range
\begin{equation}
z_{\mrm{min},\mrm{max}} = \frac{1}{2} \left(
1 \mp \frac{m_{ar}^2 - Q^2}{m_{ar}^2 + Q^2} \right)
\end{equation}
for $\pTsbc > 0$ to be valid.
 
It is the choice of a dipole-style phase space in conjunction with
$\pTs$ as evolution variable that ensures the angular ordering
required for coherence \cite{pTcoherence,GGonPythia}.
 
\subsubsection{Further details}
 
\textit{(i)} The colour topology of an event needs to be updated after
each branching, so as to define possible recoilers for the next step
of the evolution, and also for the subsequent hadronization.
Most of this is trivial, since we work in the  $N_C \to \infty$
limit: for $\q \to \q\g$ the original quark colour is inherited by
the gluon and a new colour dipole is created between the two
daughters, while for $\g \to \q \qbar$ the (anti)quark takes the
gluon (anti)colour. Somewhat more tricky is $\g \to \g\g$, where
two inequivalent possibilities exist. We here use the rewriting
of the splitting kernel \cite{GGUP}, $(1 - z(1-z))^2/z(1-z) =
(1+z^3)/(1-z) + (1+(1-z)^3)/z \simeq 2(1+z^3)/(1-z)$, to associate
a $1-z$ picked according to the right-hand side with the energy
fraction of the `radiated' gluon that carries away the `radiating'
(anti)colour of the original gluon.
 
\textit{(ii)} The above $\pT$ equations have been written for the case
of massless partons. It is straightforward to generalize to massive
partons, however, starting from the formalism presented in
ref.~\cite{PythiaMEmatch}. There it was shown that the natural
variable for mass-ordered evolution of a parton $a$ with on-shell
mass $m_{a,0}$ is $Q^2 = m_a^2 - m_{a,0}^2$, since this reproduces
relevant propagators. Now the generalization is
\begin{equation}
\pTse = z(1-z)(m_a^2 - m_{a,0}^2) ~.
\end{equation}
Furthermore, in the handling of kinematics, the $z$ variable is
reinterpreted to take into account masses \cite{PythiaMEmatch}.
 
\textit{(iii)} Whether radiation off massive or massless partons is
considered, matrix-element expressions are available for the one-gluon
emission corrections in $a \to b c$ decays in the standard model and
its minimal supersymmetric extension, say $\gamma^*/\Z^0 \to \q \qbar$
or $\tilde{\g} \to \tilde{\q}\qbar$ \cite{PythiaMEmatch}. Since the
shower overpopulates phase space relative to these expressions, a
simple veto step can be used to smoothly merge a matrix-element
behaviour for hard non-collinear emissions with the shower picture
for soft and collinear ones. When the $\b + \c$ system radiates
repeatedly, the matrix-element corrections are applied to the
system at the successively reduced energy. This ensures that a good
account is given of the reduced radiation in the collinear region
by mass effects. For $\g \to \q\qbar$ branchings, mass effects and
subsequent gluon emissions off the quarks are given the same corrections
as for $\gamma^* \to \q\qbar$ branchings, i.e.\ disregarding the
difference in colour structure.
 
\textit{(iv)} Azimuthal $\varphi$ angles are selected isotropically in
$\q \to \q\g$ branchings, but nonisotropically for  $\g \to \g\g$ and
 $\g \to \q\qbar$ to take into account gluon polarization effects
\cite{Webberrev}. Anisotropies from coherence conditions are not
included explicitly, since some of that is implicitly generated
by the dipole kinematics.
 
\textit{(v)} We use a first-order $\alphas(\pTs) =
12\pi/ ((33-2n_f) \ln(\pT^2/\Lambda_{(n_f)}^2)$, matched
at the $m_{\c}$ and $m_{\b}$ mass thresholds, where default is
$m_{\c} = 1.5$~GeV and $m_{\b} = 4.8$~GeV.
 
\subsubsection{Algorithm tests}
 
Ultimately, the usefulness of a shower algorithm is gauged by its
ability to describe data. Obviously, we have  checked that the
results of the new routine qualitatively agree with the old program,
which is known to describe data reasonably well. A more detailed study
has been performed by G.\ Rudolph \cite{Rudolph}, who has compared our
algorithm with
ALEPH data at the $\Z^0$ peak \cite{LEPstudy}. A tune to a set of
event shapes and particle spectra gives a total $\chi^2$ that is
roughly $2/3$ of the corresponding value for the old mass-ordered
evolution, i.e.\ a marked improvement. Of the distributions
considered, the only one that does not give a decent description
is the single-particle $p_{\perp\mrm{out}}$ spectrum, i.e. the
transverse momentum out of the event plane, in the region
$p_{\perp\mrm{out}} > 0.7$~GeV. This is a common problem for showering
algorithms, and in fact was even bigger in the mass-ordered one. 
With the exception
of this region, the $\chi^2$ per degree of freedom comes down to
the order of unity, if one to the experimental statistical and
systematical errors in quadrature adds an extra term of 1\% of the
value in each point. That is, it appears plausible that the overall
quality of the algorithm is at the 1\% level for most observables
at the $\Z^0$ peak.
 
Some of the tuned values have changed relative to the old algorithm.
Specifically the first-order five-flavour $\Lambda$ is roughly halved
to 0.140 GeV, and the cutoff parameter is reduced from
$m_{\mrm{min}} \approx 1.6$~GeV to $2 \pTmin \approx 0.6$ GeV.
The former represents a real enough difference in the capability of
the algorithms to populate the hard-emission region, while the latter
is less easily interpreted and less crucial, since it deals with
how best to match perturbative and nonperturbative physics, that is
largely compensated by retuned hadronization parameters.
 
\subsection{Spacelike showers}
 
\subsubsection{The basic formalism}
 
At any resolution scale $\pTse=(1-z)Q^2$ the ISR algorithm will
identify two initial partons, one from each incoming hadron, that are
the mothers of the respective incoming cascade to the hard interaction.
When the resolution scale is reduced, using backwards evolution according
to eq.~(\ref{eq:evbackwfin}), either of these two partons may turn out
to be the daughter $b$ of a previous branching $a \to b c$. The
(currently resolved) parton $r$ on the other side of the event takes
on the role of recoiler, needed for consistent reconstruction of the
kinematics when the parton $b$ previously considered massless now
is assigned a spacelike virtuality $m_b^2 = - Q^2$. This redefinition
should be performed in such a way that the invariant mass of the $b+r$
system is unchanged, since this mass corresponds to the set of outgoing
partons already defined by the hard scattering and by partons emitted
in previously considered branchings. The system will have to be rotated
and boosted as a whole, however, to take into account that $b$ not only
acquires a virtuality but also a transverse momentum; if previously
$b$ was assumed to move along the event axis, now it is $a$ that should
do so.
 
At any step of the cascade, the massless mothers suitably should have
four-momenta given by $p_i = x_i \, (\sqrt{s}/2) \, (1; 0, 0, \pm 1)$
in the rest frame of the two incoming beam particles, so that
$\hat{s} = x_1 x_2 s$. If this relation is to be preserved in the
$a \to b c$ branching, the $z = x_b/x_a$ should fulfil
$z = m_{br}^2/m_{ar}^2 = (p_b + p_r)^2/(p_a + p_r)^2$. As we have
already noted, $z$ definitions in terms of squared mass ratios are
easily related to energy sharing in the  rest frame of the process.
This is illustrated by explicit construction of the kinematics in the
$a + r$ rest frame, assuming $a$ moving along the $+z$ axis and
$c$ massless:
\begin{eqnarray}
p_{a,r} & = & \frac{m_{ar}}{2} \left( 1 ; 0, 0, \pm 1 \right) ~, \\
p_b & = &  \left( \frac{m_{ar}}{2} \, z ;
\sqrt{ (1-z)Q^2 - \frac{Q^4}{m_{ar}^2} }, 0, \frac{m_{ar}}{2}
\left(z + \frac{2Q^2}{m_{ar}^2} \right) \right) ~,
\label{pbspacebranch} \\
p_c & = &  \left( \frac{m_{ar}}{2} \, (1- z) ;
- \sqrt{ (1-z)Q^2 - \frac{Q^4}{m_{ar}^2} }, 0, \frac{m_{ar}}{2}
\left(1 - z - \frac{2Q^2}{m_{ar}^2} \right) \right) ~.
\end{eqnarray}
For simplicity we have here put the azimuthal angle $\varphi = 0$.
 
Note that
\begin{equation}
\pTsbc = (1-z)Q^2 - \frac{Q^4}{m_{ar}^2} < (1-z)Q^2 = \pTse ~.
\label{eq:pTsbcspa}
\end{equation}
For small $Q^2$ values the two measures $\pTsbc$ and $\pTse$ agree well,
but with increasing $Q^2$ the $\pTsbc$ will eventually turn over and
decrease again (for fixed $z$ and $m_{ar}$). Simple inspection shows
that the maximum $\pTsbc$ occurs for $p_{\parallel c} = 0$ and that the
decreasing $\pTsbc$ corresponds to increasingly negative
$p_{\parallel c}$. The drop of $\pTsbc$ thus is deceptive, and does not
correspond to our intuitive picture of time ordering. Like for the FSR
algorithm, $\pTse$ therefore makes more sense than $\pTsbc$ as evolution
variable, in spite of it not always having as simple a kinematics
interpretation. One should note, however, that emissions with negative
$p_{\parallel c}$ are more likely to come from radiation off the other
incoming parton, where it is collinearly enhanced, so in practice the
region of decreasing $\pTsbc$ is not so important.
 
The allowed range $z_{\mrm{min}} < z < z_{\mrm{max}}$ is from below
constrained by $x_a = x_b/z < 1$, i.e.\ $z_{\mrm{min}} = x_b$, and
from above by $\pTsbc > 0$, which gives
\begin{equation}
z_{\mrm{max}} = 1 - \frac{\pTe}{m_{br}} \left(
\sqrt{1 + \frac{\pTse}{4 m_{br}^2}} - \frac{\pTe}{2 m_{br}} \right) ~. 
\label{eq:zmax_massless}
\end{equation}
 
When the $a \to b c$ kinematics is constructed, the above equations
for $p_{a,b,c,r}$ are not sufficient. One also needs to boost and
rotate all the partons produced by the incoming $b$ and $r$ partons.
The full procedure then reads
\begin{Enumerate}
\item Go to the $b + r$ rest frame, with
$p_{b,r} = (m_{br}/2) \, (1; 0, 0, \pm 1)$.
\item Rotate by a randomly selected azimuthal angle $- \varphi$.
\item Put the $b$ off mass shell, $Q^2 = -m_b^2 = \pTse/(1-z)$,
while preserving the total  $b + r$  four-momentum, i.e.
$p_{b,r} = ( (m_{br}^2 \mp Q^2)/2m_{br} ; 0, 0,
\pm (m_{br}^2 + Q^2)/2m_{br} )$.
\item Construct the massless incoming $p_a$ in this frame, and the
outgoing $c$, from the requirements $(p_a + p_r)^2 = m_{br}^2/z$ and
$p_c^2 = (p_a - p_b)^2 = m_c^2 (=0)$, and with transverse momentum
in the $x$ direction \cite{backwards}.
\item Boost everything to the $a + r$ rest frame, and thereafter
rotate in $\theta$ to have $a$ moving along the $+z$ axis.
\item Finally rotate $+ \varphi$ in azimuth, with the same $\varphi$
is in point 2. This gives $c$ a random $\varphi$ distribution, while
preserving the $\varphi$ values of the $b+r$ daughters, up to
recoil effects.
\end{Enumerate}
 
Apart from the change of evolution variable, the major difference
relative to the old algorithm \cite{backwards} is that kinematics is
now constructed with the recoiler assumed massless, rather than only
after it has been assigned a virtuality as well.
 
Currently a smooth merging with first-order matrix elements is only
available for the production of $\gamma^*/\Z^0/\W^{\pm}$
\cite{MiuZ} and $\g\g \to \H^0$ (in the infinitely-heavy-top-mass
limit). It turns out that the shower actually does a reasonable
job of describing radiation also harder than the mass scale of the
electroweak production process, i.e. the matrix-element reweighting
factors are everywhere of the order of unity. Unless there are
reasons to the contrary, for non-QCD processes it therefore makes
sense to start the shower from a $\pTmax = \sqrt{s}/2$. For a normal
QCD process this would lead to doublecounting, since the shower emissions
could be harder than the original hard process, but this risk does not
exist for particles like the $\Z^0$, which are not produced in the
shower anyway.
 
\subsubsection{Mass corrections}
 
Quark mass effects are seldom crucial for ISR: nothing heavier than
charm and bottom need be considered as beam constituents, unlike the
multitude of new massive particles one could imagine for FSR. Here the
mass effects are less trivial to handle, however, since we may get stuck
in impossible corners of phase space.
 
To illustrate this, consider $\g \to \Q\Qbar$, where we let $\Q$ denote
a generic heavy quark, charm or bottom. Then requiring the lightcone
$\pTslc = (1-z) Q^2 - z \mQ^2 > 0$, eq.~(\ref{eq:pTlightcone}) with
$m_a = 0$, $m_c = \mQ$ and $Q^2 = - m_b^2$, implies
$z < Q^2/(Q^2 + \mQ^2)$. Since $x_a = x_b/z < 1$ it follows that the
$\Q$ parton density must vanish for $x > Q^2/(Q^2 + \mQ^2)$. Many parton
density parameterizations assume vanishing $\Q$ density below
$Q^2 = \mQ^2$ and massless evolution above it, and so do not obey the
above constraint.
 
Actually, with our energy-sharing $z$ definition, now slightly modified
but still preserving $z = m_{br}^2/m_{ar}^2$, eq.~(\ref{eq:pTsbcspa})
is generalized to
\begin{equation}
\pTsbc = (1-z)Q^2 - \frac{Q^4}{m_{ar}^2} -
\mQ^2 \left( z + \frac{Q^2}{m_{ar}^2} \right) =
Q^2 - z \, \frac{(Q^2 + \mQ^2) (m_{br}^2 + Q^2)}{m_{br}^2} ~.
\end{equation}
which implies the somewhat tighter constraint
\begin{equation}
x_b < z <  \frac{Q^2}{Q^2 + \mQ^2} \, \frac{m_{br}^2}{m_{br}^2 + Q^2} ~.
\label{eq:zmaxmQ}
\end{equation}

For the backwards evolution of $\g \to \Q\Qbar$, the evolution variable
is chosen to be
\begin{equation}
\pTse = (1-z) (Q^2 + \mQ^2) = \mQ^2 + \pTslc ~,
\label{eq:pTisrmass}
\end{equation}
such that a threshold set at $\pTse = \mQ^2$ corresponds to
$\pTslc\to0$. Thereby, the evolution scale $\pTse$ may be used as argument
for $\alphas$ and for parton densities, while the physical $\pT$ will still
populate the full phase space. 
 
Writing the upper limit in terms of the evolution variable $\pTse$ rather
than $Q^2$, one obtains the analogue of eq.~(\ref{eq:zmax_massless}),
\begin{equation}
z_{\mrm{max}} = 1-\frac{\pTe}{m_{br}}\frac{1}{1-
\frac{\mQ^2}{m_{br}^2}\left(1+\frac{m_{br}^2}{\pTse}\right)}\left(
\sqrt{1+\frac{\pTse-\mQ^2}{4m_{br}^2}}-\frac{\pTe}{2m_{br}}
\left(1+\frac{\mQ^2}{\pTse}\right)  
\right)~.
\end{equation}
This expression would be rather cumbersome to deal with in practice,
but is bounded from above,
\begin{equation}
z_{\mrm{max}} < \frac{m_{br}(m_{br}-\mQ)}{m_{br}^2+\mQ m_{br}-\mQ^2}~,
\end{equation}
which we make use of in the evolution.

Should a hard-scattering configuration be inconsistent with these
constraints, it is rejected as unphysical. Should the shower end up in
such a region during the backwards evolution, a new shower is generated.
Even when no such disasters occur, the fact that the physically allowed
$z$ range is smaller than what has been assumed in
standard parton density parameterizations implies that more heavy quarks
can survive to the near-threshold region than ought to be the case.
This could be amended by an ad hoc compensating weight
factor in the splitting kernel, but currently we have not studied
this further.
 
Another technical problem is that, when performing the backwards
evolution, eq.~(\ref{eq:evbackwfin}), one needs to estimate from
above the ratio of parton densities, in order for the veto algorithm
to be applicable \cite{Pythia}. Normally, densities fall off with $x$
(the exception being valence quarks, for which some extra consideration
is required) and have a modest scale dependence, so that
\begin{equation}
\frac{x' f_a(x',\pTse)}{x f_b(x,\pTse)} <
\frac{x f_a(x,\pTse)}{x f_b(x,\pTse)} \simeq
\frac{x f_a(x,\pTsmax)}{x f_b(x,\pTsmax)}   ~.
\end{equation}
Now, however, the denominator $f_b = f_{\Q}$ vanishes for
$\pTse \to \mQ^2$, and so does not obey the above relation.
Given that $f_{\Q}(x,Q^2)$ increases roughly like $\ln(Q^2/\mQ^2)$,
a reasonable alternative approximation is
\begin{equation}
\frac{x' f_{\g}(x',\pTse)}{x f_{\Q}(x,\pTse)} <
\frac{x f_{\g}(x,\pTse)}{x f_{\Q}(x,\pTse)} \simeq
\frac{\ln(\pTsmax/\mQ^2)}{\ln(\pTse/\mQ^2)} \,
\frac{x f_{\g}(x,\pTsmax)}{x f_{\Q}(x,\pTsmax)}   ~.
\end{equation}
The $1/\ln(\pTse/\mQ^2)$ prefactor can be incorporated into the
choice of the next trial emission, so that steps taken in $\pTse$ get
shorter and shorter as the threshold is approached, until a valid
branching is found.
 
Finally, the $\g\to\Q\Qbar$ splitting function should be modified. 
The appropriate expressions may be identified by 
considering the collinear limit of relevant matrix elements.
Neglecting overall factors, $\g\to\Q\Qbar$ is equivalent to 
$\gamma\to\mu^+\mu^-$ with massive muons. Considering the $t\to 0$ 
limit of processes such as $\gamma\nu_\mu\to\mu^-\W^+$ and
$\gamma\mu^-\to\mu^-\H^0$, and letting
$m_\mu^2/m_{\W,\H}^2\to 0$, we thus obtain:
\begin{equation}
P_{\g \to \Q\Qbar}(z) = \frac{1}{2} \left( z^2 + (1-z)^2 +
2 z (1-z) \frac{\mQ^2}{\pTse} \right)  ~, 
\end{equation} 
which approaches a flat $1/2$ for $\pTse \to \mQ^2$. 
  
Since $\g \to \Q \Qbar$ and $\Q \to \Q \g$ compete in the backwards
evolution of a heavy quark, the $\pTse = (1-z) (Q^2 + \mQ^2)$ of
eq.~(\ref{eq:pTisrmass}) is used also here. The kinematics
interpretation is now slightly different, however. The branching $\Q$
is forced to be massless, so the kinematics is in this case identical 
to that of a light-quark $\q \to \q\g$ branching. However, since the 
massive $\pTse$ is different from the massless one, the $z$ limit 
expressed in terms of $\pTse$ also becomes different from 
eq.~(\ref{eq:zmax_massless}):
\begin{equation}
z_{\mrm{max}} = 1-\frac{\pTe}{m_{br}}\frac{1}{1-\frac{\mQ^2}{m_{br}^2}}
\left( \sqrt{1+\frac{\pTse}{4m_{br}^2}\left(1-\frac{\mQ^2}{\pTse}\right)^2} 
-\frac{\pTe}{2m_{br}}\left(1+\frac{\mQ^2}{\pTse}\right)\right)~.
\end{equation}

As before, also the splitting kernel receives a mass
correction. For $\Q \to \Q \g$, this may be obtained by considering the
equivalent processes $\mu^- \bar{\nu}_\mu \to \gamma \W^-$ and
$\mu^+\mu^-\to\gamma \H^0$ in the same limits as above, yielding:
\begin{equation}
 P_{\Q \to \Q\g}(z) = \frac{4}{3} \, \left( \frac{1+z^2}{1-z} - 2z(1-z)
\frac{\mQ^2}{\pTse}\right)   ~,
\end{equation}
i.e.\ the mass correction here has the same form but the opposite sign as
for $\g \to \Q\Qbar$. 

Finally, in the branching $\Q \to \g \Q$, a gluon is emitted by a
heavy quark, which in its turn must come from a $\g \to \Q \Qbar$
branching. Thus both the $Q$ and $\Qbar$ must be put on the mass
shell, which implies significant kinematical constraints. The
process is rare, however, and currently we have not considered it
further. 
 
\subsubsection{Algorithm tests}
 
While a FSR algorithm can be tested in $\e^+ \e^-$ annihilation
events, where only hadronization need be considered in addition,
the busier environment in hadron colliders makes ISR algorithms
more complicated to test. One of the few clean measurements is
provided by the $\pT$ spectrum of $\Z^0$ bosons. This quantity
has been studied for the new algorithm (without the inclusion of
incoming heavy flavours) \cite{Erik}, with the conclusion that it
there does at least as well as the old \textsc{Pythia} algorithm.
This is not surprising since the two are not so very different,
apart from the $Q^2$ vs.\ $\pTs$ ordering issue.
 
Actually, below and around the $\d\sigma/\d p_{\perp\Z}$ peak,
at $p_{\perp\Z}\approx 4$~GeV at the Tevatron, a difference would
have been welcome, since the old algorithm requires an uncomfortably
large primordial $\kT$ of around 2~GeV to provide a decent fit.
Unfortunately the new requires about the same. The number can be
reduced by using a larger $\Lambda$ in the algorithms than  that of
the parton densities. Such a procedure can be motivated by noting
that the actual evolution in a generator contains various kinematical
and dynamical suppressions not found in the leading-log parton
evolution equations \cite{Erik}. A fit to the whole
$\d\sigma/\d p_{\perp\Z}$ spectrum in the peak region does not
favour significant reductions of the primordial $\kT$, however.
This might be viewed as indications for the need of
physics beyond standard DGLAP \cite{smallx}.
 
\subsection{Combining spacelike and timelike showers}
 
The separation of ISR and FSR is not unambiguous: it is possible to
shuffle contributions between the two, i.e.\ take fewer but longer
steps in rapidity for the ISR and compensate that by more extensive
FSR radiation off those ISR partons that are emitted \cite{LDCMC}.
In part, compensation mechanisms of this kind automatically occur:
if ISR partons are more widely spaced then the colour dipoles spanned
between them become larger and thereby the FSR is increased, at least
to some extent.
 
We defer further studies of the optimal balance between the two, and
for now pick a simple strategy:
\begin{Itemize}
\item The initial-state shower is first handled in full. This provides
a set of final-state partons, from the hard interactions and from the
$c$ partons of all $a \to b c$ branchings in the ISR chains.
\item Each final-state parton is associated with a $\pT$ scale at which
it was formed, either the hard-scattering scale or the $\pTe$ of the
ISR evolution.
\item Each coloured final-state parton is also connected to other
final-state partons to form colour dipoles. Normally these dipole
partners would also act as recoilers. Top decay has been mentioned as
one example where this would not be the case, but such decays can be
considered separately from the production processes studied here,
and before the tops decay they can act both as radiators and recoilers.
When a colour-singlet particle like the $\Z^0$ is produced, there is
a freedom to admit this as a recoiler, to the hardest parton emitted
on either side of it, or to let those two partons act as each other's
recoilers, just like they are colour-connected. For now we choose the
latter strategy.
\item The lowest-$\pTe$ parton emitted on either side of the event is
colour-connected to the beam remnant. A remnant does not
radiate, but can act as recoiler; since the momentum transfer will
predominantly be in the longitudinal direction, it will not give rise
to any unphysical $\pT$ kicks. The internal structure of the remnant
then has to be resolved beforehand, since a small radiator+recoiler
invariant mass implies a restricted phase space for emissions. Such
a dependence of perturbative physics on nonperturbative assumptions
may be a bit uncomfortable. As an option, we have studied a
scenario without any emissions at all off this radiator+recoiler set.
Since the affected parton normally is a low-$\pTe$ one, and the
potential additional activity should occur below this already low scale,
one would not expect large differences, and indeed this is confirmed
by our studies.
\item The issue of what to do with loose colour ends is more important
if one intends to stop and restart the showers (both ISR and FSR) at
large $\pT$ scales, as in a (L)CKKW-style matching to higher-order
matrix-element programs \cite{CKKWL}.  We therefore consider two
alternatives for the FSR activity off the dipoles defined by the ISR
branchings. In one, each parton of a dipole radiates with a maximum 
$\pT$ scale set 
by its production $\pT$, phase space constraints permitting. In the
other, the maximum radiation scale in a dipole is set by the smaller of
the two endpoint parton production $\pT$ values, i.e.\ a dipole does not
radiate above the scale at which it is `formed'. Technically, the latter
option offers the possibility to combine ISR and FSR emissions in one
common sequence of decreasing $\pT$ values, certainly a boon for
matching procedures. The choice of maximum emission scale is not unique, 
since the shower language offers little guidance in the regions where 
several $\pT$ values are of comparable magnitude. In this case, that 
would be the emission or not of a hard FSR parton off the harder of the
ISR ones. Practical experience could tell which is preferable.
\item For now, however, all ISR activity is finished before the
system is evolved with the FSR algorithm, downwards in
$\pTe$. Initially only the hardest partons can therefore radiate, but
as $\pTe$ is reduced also more of the partons from the ISR cascades can
radiate, below the respective scale at which they themselves or their
dipole were produced, depending on the option used.
\end{Itemize}
 
\section{Interleaved Multiple Interactions}
\label{sec:imi}
 
\subsection{Multiple interactions}
 
Our basic framework for multiple interactions is the one presented in
ref.~\cite{multint}, which in turn builds on the work in ref.~\cite{Zijl}. 
We refer the reader to these for details, and here only provide a very 
brief summary.
 
\subsubsection{The basic formalism}
 
The cross section for $2 \to 2$ QCD scatterings is dominated by
$t$-channel gluon exchange and hence diverges roughly like
$\d \pT^2/ \pT^4$. Therefore the integrated interaction cross section
above some $\pTmin$ scale, $\sigma_{\mrm{int}}(\pTmin)$, exceeds the
total inelastic nondiffractive cross section $\sigma_{\mrm{nd}}$ when
$\pTmin \to 0$. The resolution of this apparently paradoxical situation 
probably comes in two steps.
 
Firstly, the interaction cross section is an inclusive number.
Thus, if an event contains two interactions it counts twice in
$\sigma_{\mrm{int}}$ but only once in $\sigma_{\mrm{nd}}$,
and so on for higher multiplicities. Thereby we may identify
$\langle n \rangle(\pTmin) = \sigma_{\mrm{int}}(\pTmin) /
\sigma_{\mrm{nd}}$ with the average number of interactions
above $\pTmin$ per inelastic nondiffractive event, and that number
may well be above unity.
 
As a starting point we will assume that all hadronic collisions are
equivalent, i.e.\ that there is no dependence on impact parameter, and that
the different parton--parton interactions take place independently of
each other, i.e.\ we disregard energy--momentum conservation effects.
The number of interactions above $\pTmin$ per event is then distributed
according to a Poisson distribution with mean $\langle n \rangle$,
$\mathcal{P}_n = \langle n \rangle^n \exp( - \langle n \rangle) / n!$.
 
Secondly, the incoming hadrons are colour singlet objects. Therefore,
when the $\pT$ of an exchanged gluon is made small and the transverse
wavelength correspondingly large, the gluon can no longer resolve the
individual colour charges, and the effective coupling is decreased.
Note that perturbative QCD calculations are always performed assuming
free incoming and outgoing quark and gluon states, rather than partons
inside hadrons, and thus do not address this kind of nonperturbative
screening effects.
 
The simplest solution to the second issue is to introduce a step
function $\theta(\pT - \pTmin)$, such that the perturbative cross
section is assumed to completely vanish below some $\pTmin$ scale.
Given the complexity of the nonperturbative physics involved,
$\pTmin$ cannot be calculated but has to be tuned to data.
A more realistic alternative is to note that the jet cross section
is divergent like $\alphas^2(\pT^2)/\pT^4$, and that therefore a
factor
\begin{equation}
\frac{\alphas^2(\pTo^2 + \pT^2)}{\alphas^2(\pT^2)} \,
\frac{\pT^4}{(\pTo^2 + \pT^2)^2}
\label{pTosmooth}
\end{equation}
would smoothly regularize the divergences, now with $\pTo$ as the
free parameter to be tuned to data. Later we will return to the issue 
of whether to do a similar replacement for the scale argument of parton
densities.
 
In an event with several interactions, it is convenient to order them
in $\pT$, as already discussed in the introduction.
The generation of a sequence $\sqrt{s}/2 > p_{\perp 1} >
p_{\perp 2} > \ldots > p_{\perp n} > \pTmin$ now becomes one of
determining $\pT = p_{\perp i}$ from a known $p_{\perp i-1}$,
according to the probability distribution
\begin{equation}
\frac{\d\mathcal{P}}{\d\pT} = \frac{1}{\sigma_{\mrm{nd}}}
\frac{\d\sigma}{\d\pT} \exp\left[ - \int_{\pT}^{p_{\perp i-1}}
\frac{1}{\sigma_{\mrm{nd}}} \frac{\d\sigma}{\d\pT'} \d\pT' \right] ~.
\label{eq:miformfactor}
\end{equation}
The exponential expression is the `form factor' from the requirement
that no interactions occur between $p_{\perp i-1}$ and $p_{\perp i}$,
cf.\ the Sudakov form factor of parton showers.
 
More realistically, one should include the possibility that each
collision also could be characterized by a varying impact parameter
$b$. Within the classical framework we use here, $b$ is to
be thought of as a distance of closest approach, not as the Fourier
transform of the momentum transfer. A small $b$ value corresponds to
a large overlap between the two colliding hadrons, and hence an
enhanced probability for multiple interactions. A large $b$, on the
other hand, corresponds to a grazing collision, with a large
probability that no parton--parton interactions at all take place.
 
Let $\mathcal{O}(b)$ denote the time-integrated matter overlap between
the two incoming hadrons at impact parameter $b$. The combined selection
of $b$ and a set of scattering $\p_{\perp i}$ values can be reduced to
a combined choice of $b$ and $p_{\perp 1}$, according to a generalization
of eq.~(\ref{eq:miformfactor})
\begin{equation}
\frac{\d\mathcal{P}}{\d p_{\perp 1} \, \d^2 b} =
\frac{\mathcal{O}(b)}{\langle \mathcal{O} \rangle} \,
\frac{1}{\sigma_{\mrm{nd}}} \frac{\d\sigma}{\d\pT}
\exp\left[ - \frac{\mathcal{O}(b)}{\langle \mathcal{O} \rangle} \,
\int_{\pT}^{\sqrt{s}/2}
\frac{1}{\sigma_{\mrm{nd}}} \frac{\d\sigma}{\d\pT'} \d\pT' \right] ~.
\label{eq:bandptsel}
\end{equation}
The subsequent interactions can be generated sequentially in falling
$\pT$ as before, with the only difference that $\d\sigma/\d\pT^2$ now
is multiplied by $\mathcal{O}(b)/\langle \mathcal{O} \rangle$,
where $b$ is fixed at the value chosen above.
 
\subsubsection{Correlated parton densities}
 
Consider a hadron undergoing multiple interactions in a collision.
Such an object should be described by multi-parton densities,
giving the joint probability of simultaneously finding $n$ partons with
flavours $f_1,\ldots,f_n$, carrying momentum fractions $x_1,\ldots,x_n$
inside the hadron, when probed by interactions at scales
$Q_1^2,\ldots,Q_n^2$, in our case with the association
$Q_i^2 = p_{\perp i}^2$. Having nowhere near sufficient experimental
information to pin down such distributions, and wishing to make maximal
use of the information that we \emph{do} have, namely the standard
one-parton-inclusive parton densities, we propose the following strategy.
 
The first and most trivial observation is that each interaction $i$
removes a momentum fraction $x_i$ from the hadron remnant. This momentum
loss can be taken into account by assuming a simple scaling ansatz for
the parton distributions, $f(x) \to f(x/X)/X$, where
$X = 1 - \sum_{i=1}^n x_i$ is the momentum remaining in the beam hadron
after the $n$ first interactions. Effectively, the PDF's are simply
`squeezed' into the range $x\in[0,X]$.
 
Next, for a given hadron, the valence distribution of flavour $f$ after $n$
interactions, $q_{f\mathrm{v} n}(x,Q^2)$, should integrate to the number
$N_{f\mathrm{v} n}$ of valence quarks of flavour $f$ remaining in the hadron
remnant. This rule may be enforced by scaling the original distribution down,
by the ratio of remaining to original valence quarks
$N_{f\mathrm{v} n}/N_{f\mathrm{v} 0}$, in addition to the $x$ scaling
mentioned above.
 
Also, when a sea quark is knocked out of a hadron, it must leave behind a
corresponding antisea parton in the beam remnant. We call this a companion
quark. In the perturbative approximation the sea quark $\qsea$ and its
companion $\qcmp$ come from a gluon branching $\g \to \qsea + \qcmp$
(it is implicit that if $\qsea$ is a quark, $\qcmp$ is its antiquark).
Starting from this perturbative ansatz, and neglecting other interactions
and any subsequent perturbative evolution of the $\qcmp$, we obtain the
$\qcmp$ distribution from the probability that a sea quark $\qsea$,
carrying a momentum fraction $x_{\mrm{s}}$, is produced by the branching
of a gluon with momentum fraction $y$, so that the
companion has a momentum fraction $x=y-x_{\mrm{s}}$,
\begin{equation}
q_{\mrm{c}}(x;x_{\mrm{s}}) \propto \int_0^1 g(y) \,
P_{\g \to \qsea \qcmp}(z) \, \delta(x_{\mrm{s}}-zy)~\mrm{d} z =
\frac{g(x_{\mrm{s}}+x)}{x_{\mrm{s}}+x} \, P_{\g \to \qsea \qcmp}
\left(\frac{x_{\mrm{s}}}{x_{\mrm{s}}+x}\right),
\end{equation}
with $P_{\g \to \qsea \qcmp}$ the usual DGLAP gluon splitting kernel.
A simple ansatz $g(x) \propto (1-x)^n/x$ is here used for the gluon.
Normalizations are fixed so that a sea quark has exactly one companion.
 
Without any further change, the reduction of the valence
distributions and the introduction of companion distributions, in the
manner described above, would result in a violation of the total
momentum sum rule, that the $x$-weighted parton densities should
integrate to $X$: by removing a valence quark from the parton
distributions we also remove a total amount of momentum corresponding
to $\langle x_{f\mathrm{v}} \rangle$, the average momentum fraction
carried by a valence quark of flavour $f$, and by adding a companion
distribution we add an analogously defined momentum fraction.
To ensure that the momentum sum rule is still respected, we assume that
the sea and gluon normalizations fluctuate up when a valence distribution
is reduced and down when a companion distribution is added, by a
multiplicative factor. The requirement of a physical $x$ range is of
course still maintained by `squeezing' all distributions into the
interval $x\in[0,X]$.
 
After the perturbative interactions have taken each their fraction of
longitudinal momentum, the remaining momentum is to be shared between
the beam remnant partons. Here, valence quarks receive an $x$ picked at
random according to a small-$Q^2$ valence-like parton density, while sea
quarks must be companions of one of the initiator quarks, and hence should
have an $x$ picked according to the $q_{\mrm{c}}(x ; x_{\mrm{s}})$
distribution introduced above. In the rare case that no valence quarks
remain and no sea quarks need be added for flavour conservation, the beam
remnant is represented by a gluon, carrying all of the beam remnant
longitudinal momentum. 
 
Further aspects of the model include the possible formation of composite
objects in the beam remnants (e.g.\ diquarks) and the addition
of non-zero primordial $k_{\perp}$ values to the parton shower
initiators. Especially the latter introduces some complications, to
obtain consistent kinematics. More complete descriptions may be found in
\cite{multint,briefmult}.
 
\subsubsection{Colour correlations}
 
The initial state of a baryon may be represented by three valence quarks,
connected antisymmetrically in colour via a central junction, which acts
as a switchyard for the colour flow and carries the net baryon number.
 
The colour-space evolution of this state into the initiator and remnant
partons actually found in a given event is not predicted by perturbation
theory, but is crucial in determining how the system hadronizes; in the
Lund string model \cite{string}, two colour-connected final-state
partons together define a string piece, which hadronizes by
successive non-perturbative breakups along the string. Thus, the colour
flow of an event determines the topology of the hadronizing strings,
and consequently where and how~many hadrons will be produced.
The question can essentially be reduced to one of choosing a fictitious
sequence of gluon emissions off the initial valence topology, since sea
quarks together with their companion partners are associated with parent
gluons, by construction.
 
The simplest solution is to assume that gluons are attached to the
initial quark lines in a random order. If so, the junction of an
incoming baryon would rarely be colour-connected directly to two
valence quarks in the beam remnant, and the initial-state baryon number
would be able to migrate to large $p_{\perp}$ and small $x_F$ values.
While such a mechanism should be present, there are reasons to believe
that a purely random attachment exaggerates the migration effects.
Hence a free parameter is introduced to suppress gluon attachments onto
colour lines that lie entirely within the remnant.
 
This still does not determine the order in which gluons are attached to
the colour line between a valence quark and the junction. We consider a
few different possibilities: 1) random, 2) gluons are ordered according
to the rapidity of the hard scattering subsystem they are associated with,
and 3) gluons are ordered so as to give rise to the smallest possible
total string lengths in the final state. The two latter possibilities
correspond to a tendency of nature to minimize the total potential
energy of the system, i.e.\ the string length. Empirically such a
tendency among the strings formed by multiple interactions is supported
e.g.\ by the observed rapid increase of $\langle p_{\perp} \rangle$ with
$n_{\mathrm{charged}}$ \cite{meanptfornch}
 
It appears, however, that a string minimization in the initial state is
not enough, and that also the colours inside the initial-state cascades
and hard interactions may be nontrivially correlated. Currently this is
handled by a reassignment among a fraction of the colours in the
final state, chosen so as to reduce the total string length.
 
\subsection{Multiple interactions and initial-state radiation}
 
Each multiple interaction is associated with its set of initial- and
final-state radiation. We have already argued that, to a good
approximation, the addition of FSR can be deferred until after ISR and
MI have been considered in full. Specifically, FSR does not modify the
total amount of energy carried by perturbatively defined partons, it
only redistributes that energy among more partons. By contrast, both
the addition of a further ISR branching and the addition of a further
interaction implies more perturbative energy, taken from the limited
beam-remnants reservoir. These two mechanisms therefore are in direct
competition with each other.
 
\begin{figure}[t]
\mbox{\epsfig{file=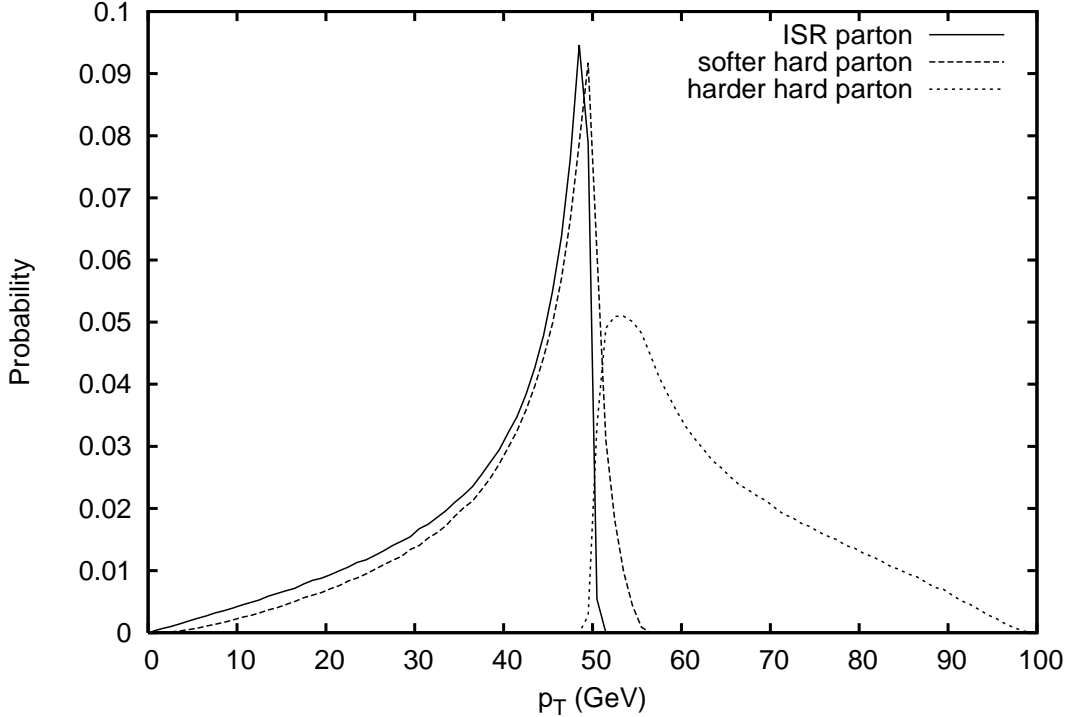,width=14.5cm}}
\caption{Parton $\pT$ spectra when 2-parton events of a fixed
$\pT = 50$~GeV, for an 1800~GeV $\p\pbar$ collider, are modified
by a single ISR branching with $\pTe = 50$~GeV, using  CTEQ5L parton
distributions and the standard DGLAP splitting kernels. Owing to
$\pTe \neq \pT$, the parton emitted at the ISR branching has a tail
to $\pT$ values well below 50 GeV. However, this spectrum is comparable
with the lower-$\pT$ of the two hard-scattering partons, after the recoil
from the ISR has been taken into account, so there is a certain symmetry
if it all is viewed as a $2 \to 3$ process.
\label{fig:pTjetforced}}
\end{figure}
 
We have advocated for $\pT$ as a convenient ordering variable, with smaller
$\pT$ values corresponding to `later times'. The $\pT$ measure used for
MI fills a similar function as the $\pTe$ variable used for ISR, such that
the two can be viewed as measuring the same kind of `time ordering'. To wit,
kinematically $\pTe$ agrees well with the standard $\pT$, except in the
corner of high virtualities, where there is little multiple activity anyway.
An example of this mapping is shown in Fig.~\ref{fig:pTjetforced}. Further,
the generation of a new interaction, eq.~(\ref{eq:miformfactor}) (or its
extension to varying impact parameters), can be viewed as an evolution
downwards in a $\pTe = \pT$, in a similar form-factor formalism as for the
backwards evolution of ISR.
 
Starting from a hard interaction, a common sequence of subsequent
evolution steps --- interactions and branchings mixed --- can therefore
be found. Assuming that the latest step occurred at some
$p_{\perp i-1}$ scale, this sets the maximum $\pTmax = p_{\perp i-1}$ for
the continued evolution. What can happen next is then 
either a new interaction
or a new ISR branching on one of the two incoming sides in one of the
existing interactions. The probability distribution for
$\pT = p_{\perp i}$ is given by
\begin{equation}
\frac{\d \mathcal{P}}{\d \pT} =
\left( \frac{\d \mathcal{P}_{\mrm{MI}}}{\d \pT} + \sum
\frac{\d \mathcal{P}_{\mrm{ISR}}}{\d \pT} \right) \;
\exp \left( - \int_{\pT}^{p_{\perp i-1}} \left(
\frac{\d \mathcal{P}_{\mrm{MI}}}{\d \pT'} + \sum
\frac{\d \mathcal{P}_{\mrm{ISR}}}{\d \pT'} \right)
\d \pT' \right)
\label{intermiiisr}
\end{equation}
in simplified notation. Technically, the $p_{\perp i}$ can be found by
selecting a new trial interaction according to
$\d \mathcal{P}_{\mrm{MI}} \, \exp ( - \int \d \mathcal{P}_{\mrm{MI}})$,
and a trial ISR branching in each of the possible places according to
$\d \mathcal{P}_{\mrm{ISR}} \, \exp ( - \int \d \mathcal{P}_{\mrm{ISR}})$.
The one of all of these possibilities that occurs at the largest $\pT$
preempts the others, and is allowed to be realized. The whole process is
iterated, until a lower cutoff is reached, below which no further
interactions or branchings are allowed.
 
If there were no momentum constraints linking the different subsystems, it
is easy to see that such an interleaved evolution actually is equivalent
to considering the ISR of each interaction in full before moving on to
the next interaction. Competition is introduced via the correlated parton
densities already discussed. Thus distributions are squeezed to be
nonvanishing in a range $x\in[0,X]$, where $X < 1$ represents the fraction
of the original beam remnant momentum still available for an interaction
or branching. When a trial $n$'th interaction is considered,
$X = 1 - \sum_{i=1}^{n-1} x_i$, where the sum runs over all the already
existing interactions. The $x_i$ are the respective momentum fractions
of the ISR shower initiators at the current resolution scale, i.e.,
an $x_i$ is increased each time an ISR branching is backwards-constructed
on an incoming parton leg. Similarly, the flavour content is modified to
take into account the partons already extracted by the $n-1$ previous
interactions, including the effects of ISR branchings. When instead a
trial shower branching is considered, the $X$ sum excludes the interaction
under consideration, since this energy \textit{is} at the disposal of the
interaction, and similarly for the flavour content.
 
We have already discussed the choice of $\pTmax$ scale for ISR showers,
and that now generalizes. Thus, for minimum-bias QCD events the full phase
space is allowed, while the $\pT$ scale of a QCD hard process sets the
maximum for the continued evolution, in order not to doublecount. When the
hard process represents a possibility not present in the MI/ISR machinery
--- production of $\Z^0$, top, or supersymmetry, say --- there is no risk of
doublecounting, and again the full (remaining) phase space is available.
 
There is also the matter of a lower $\pTmin$ scale. Customarily such scales
are chosen separately for ISR and MI, and typically lower for the former than
the latter. Both cutoffs are related to the resolution of the incoming
hadronic wave function, however, and in the current formalism ISR and MI
are interleaved, so it makes sense to use the same regularization procedure.
Therefore also the branching probability is smoothly turned off at a $\pTo$
scale, like for MI, by a factor the square root of eq.~(\ref{pTosmooth}),
since only one vertex is involved in a shower branching relative to the two
of a hard process. Thus the $\alphas(\pTse) \, \d\pTse / \pTse$ divergence is
tamed to $\alphas(\pTo^2 + \pTse) \, \d\pTse / (\pTo^2 + \pTse)$. 
The scale of parton densities in ISR and MI alike is maintained at
$\pTse$, however, the argument being that the actual evolution of the
partonic content is given by standard DGLAP evolution, and that it is 
only when this content is to be resolved that a dampening is to be 
imposed. This also has the boon that flavour thresholds appear where
they are expected. 

The cutoff for FSR still kept separate and lower, since that scale deals 
with the matching between perturbative physics and the nonperturbative 
hadronization at long time scales, and so has a somewhat different function.
 
\section{Some First Results}
\label{sec:results}
 
\subsection{Simple tunes}

In this section, some first tests of the new framework are presented. We 
compare Tune A \cite{Field} of the old multiple interactions scenario 
\cite{Zijl} and the ``Rap'' tune of \cite{multint} with three rough 
`tunes' of the new framework. These preliminary new tunes all take the 
parameters of the ``Rap'' model as a starting point:
\begin{Itemize}
\item A matter overlap profile proportional to $\exp(-b^{1.8})$, where $b$ is
  the impact parameter.
\item Rapidity-ordered initial-state colour connections.
\item Shower initiator attachments between two partons both in the beam
  remnant are suppressed by a factor 0.01 relative to others.
\item Only valence quarks are allowed to participate in the formation of
  diquarks in the beam remnants, and these diquarks are then assumed to
  acquire total $x$ values twice as large as the naive sum of $x$ values of
  their constituents. 
\item As for Tune A, the regularization scale $\pTo$ is given at a
  reference cm energy of 1800~GeV, with an energy rescaling proportional to
  $E_{\mrm{cm}}^{1/4}$.
\end{Itemize}
These choices have been made for convenience, to keep down the number of 
free parameters to be tuned. Very likely, an improved agreement with data 
can be obtained by relaxing this, e.g.\ by varying the matter overlap
profile. We also have indications that the energy dependence of $\pTo$
may be smaller than in Tune A but, since we only show comparisons at
1.8--1.96 TeV, this will be of no importance here. 

In addition, the three new tunes differ in the parameters listed in 
Table~\ref{tab:tunes}, which also show the resulting average numbers of
interactions, and ISR and FSR branchings for each model in a 
`minimum-bias' sample of inelastic nondiffractive events.  
One may view ``High FSR'' as our preferred
new scenario, with ``Sharp ISR'' and ``Low FSR'' representing two
variations, as a check of the sensitivity to some key assumptions.

\begin{table}[t]\vskip2mm
\begin{center}
\begin{tabular}{|clcccc|ccc|}
\hline
 & Model  &  ISR    & FSR    & $\pTo$  & & & &\\
Kind & name \hspace*{5mm} & cutoff & scale & [GeV]  & $F'$
& \avg{\nint} & \avg{\nisr} & \avg{\nfsr} \\ 
\hline
old & Tune A   & sharp & --      & 2.00    & --    & 5.8 & 2.0 & 3.6 \\
''   & Rap      & sharp & --      & 2.40    &($F=0.55$)&3.6&4.4& 5.5 \\
\hline
new & Sharp ISR   & sharp & radiator& 2.70   & 1.9 & 1.8 & 3.9 & 15.9\\
''    & Low FSR  & smooth& lowest  & 2.30    & 0.8 & 2.9 & 2.2 & 9.2\\
''    & High FSR & smooth& radiator& 2.50    & 1.3 & 2.4 & 1.7 & 14.0\\
\hline
\end{tabular}
\caption{The parameters distinguishing the new tunes, compared to Tune A and
the ``Rap'' model where meaningful. Also shown are the mean numbers of
interactions, \avg{\nint} (including the hardest), ISR branchings, 
\avg{\nisr}, and FSR branchings, \avg{\nfsr}, for each model.
\label{tab:tunes}}
\end{center}
\end{table}

The ``Sharp ISR'' model uses a threshold regularization of the ISR evolution
(at $\pTe = 1$~GeV), similarly to the old models, rather than the smooth 
dampening, (the square root of) eq.~(\ref{pTosmooth}), used in the other 
new tunes. (The multiple interactions
cross sections are regularized by eq.~(\ref{pTosmooth}) in all cases.) In
addition, both the ``High FSR'' and the ``Sharp ISR'' tunes let the 
maximum scale for final-state
emissions off a given parton be determined by the $\pTe$ of that
parton, while for the ``Low FSR'' tune the scale is given by whichever
has the lowest $\pTe$ of the two partons spanning the radiating dipole.

The parameter $F'$ controls the strength of colour
reconnections in the final state. Essentially, this is a fudge parameter,
required in the new framework in order to approximately reproduce the effect
of the rather extreme parameter settings controlling the final-state colour
correlations between different scatterings in Tune A. We still have not
penetrated to the details of the underlying mechanism here, 
i.e.\ why data seem to prefer such an extreme behaviour, hence the
appearance of effective parameters controlling these correlations in both
types of models. 
$F'$  has a slightly different meaning than $F$ of the ``Rap'' model, 
as follows. In \cite{multint}, the colour
reconnections were performed after all the perturbative activity had been
generated, including final-state radiation. 
In the new framework, the colour reconnections are performed
\emph{before} the final-state showers, since \emph{a priori} we believe it is
mostly a lack of correlation in the initial-state colour flows that we are
trying to make up for by this procedure. 

\begin{figure}[tp]
\begin{center}\vspace*{-12mm}%
\includegraphics*[scale=0.92]{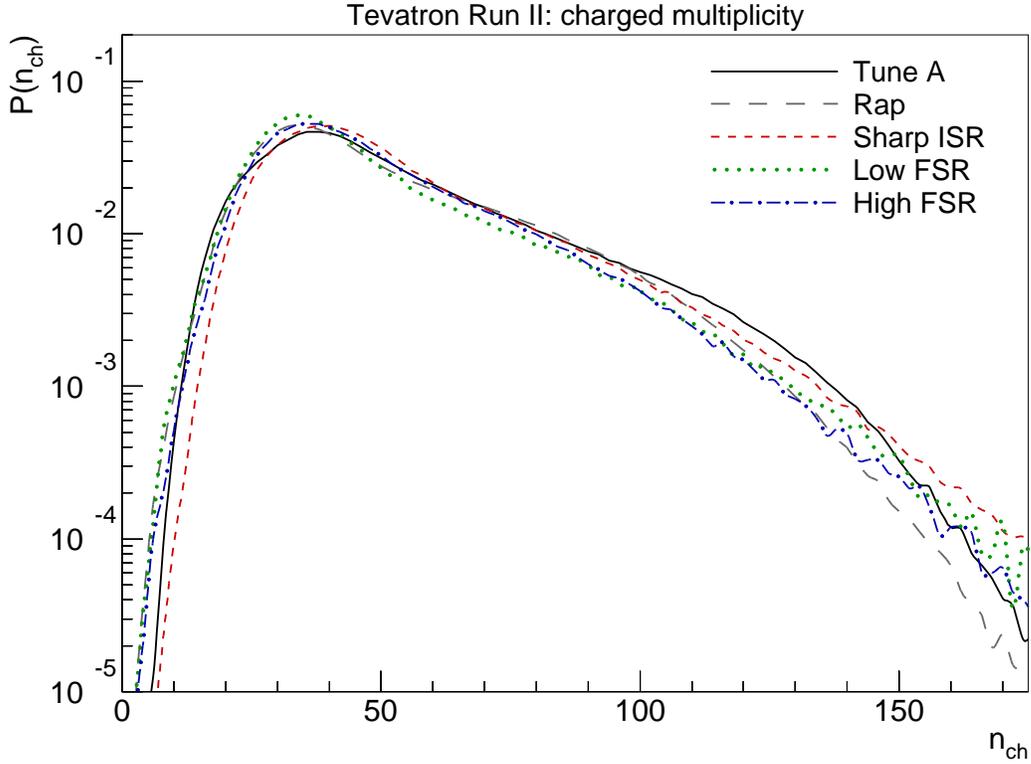}\\[-12mm]
\caption{Charged multiplicity distributions, for
  1.96~TeV $\p\pbar$ minimum-bias events.\label{fig:tmult}}
\end{center}\vspace*{-3mm}
\end{figure}

\begin{figure}[tp]
\begin{center}\vspace*{-12mm}%
\includegraphics*[scale=0.92]{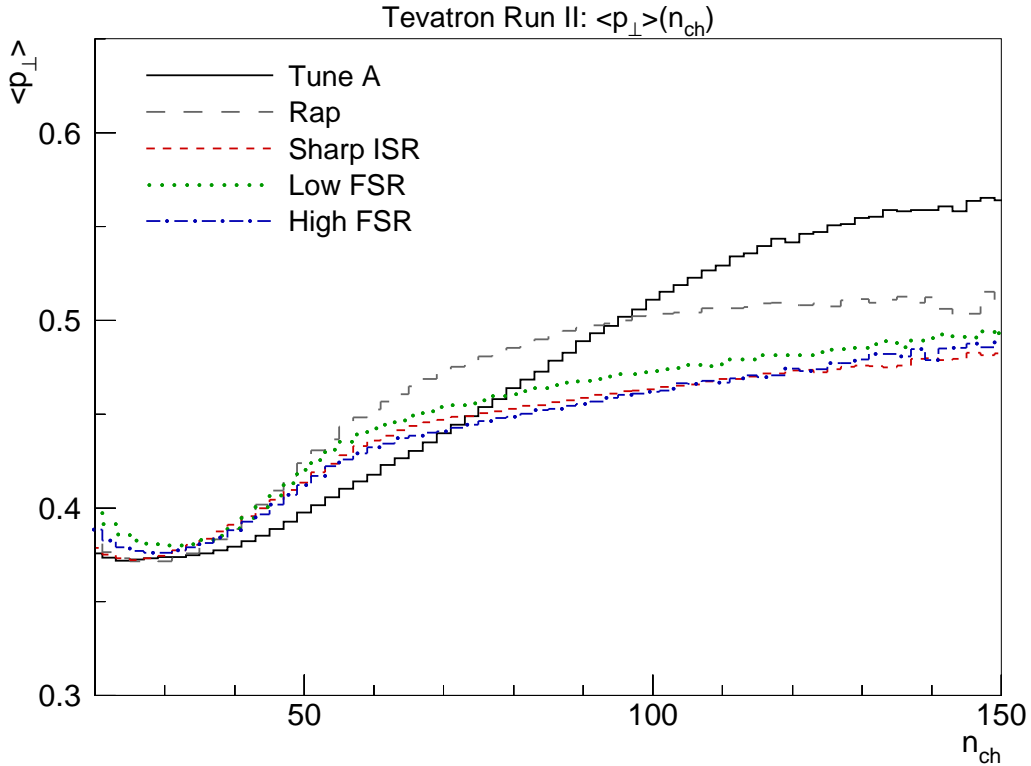}\\[-12mm]
\caption{Average $\pT$ as a function of charged multiplicity,
  $\avg{\pT}(n_{\mrm{ch}})$, for
  1.96~TeV $\p\pbar$ minimum-bias events. Note that the origo of the plot
  is \emph{not} at (0,0).\label{fig:tavgpt}}
\end{center}
\end{figure}

The tunes have been produced by adjusting $\pTo$ and $F'$ so as to 
simultaneously describe the Tune A charged
multiplicity and $\avg{\pT}(n_{\mrm{ch}})$
distributions as well as possible, since these in turn give good fits to
Tevatron data. Results are shown in Figs.~\ref{fig:tmult} and \ref{fig:tavgpt}. 

While the multiplicity distributions have been brought into 
fair agreement with each other, the Tune A $\avg{\pT}(n_{\mrm{ch}})$
is very difficult to duplicate in the new framework. This problem was also
present for the models presented in \cite{multint}. 
Our interpretation is that this particular distribution is
highly sensitive to the colour correlations, and we have so far been
unsuccessful in identifying a physics mechanism that could explain the rather
extreme correlations that are present in Tune A. Since data seems to be in
fair agreement with Tune A here, the bottom line is that \emph{some}
kind of more or less soft colour correlations working \emph{between} the
scattering chains is likely to be present, beyond what our primitive fudge
parameters $F$ and $F'$ are capable of describing at this point. 

\subsection{Event activity}

\begin{figure}[tp]
\begin{center}\vspace*{-12mm}%
\begin{tabular}{cc}\hskip-0.8cm
\includegraphics*[scale=0.72]{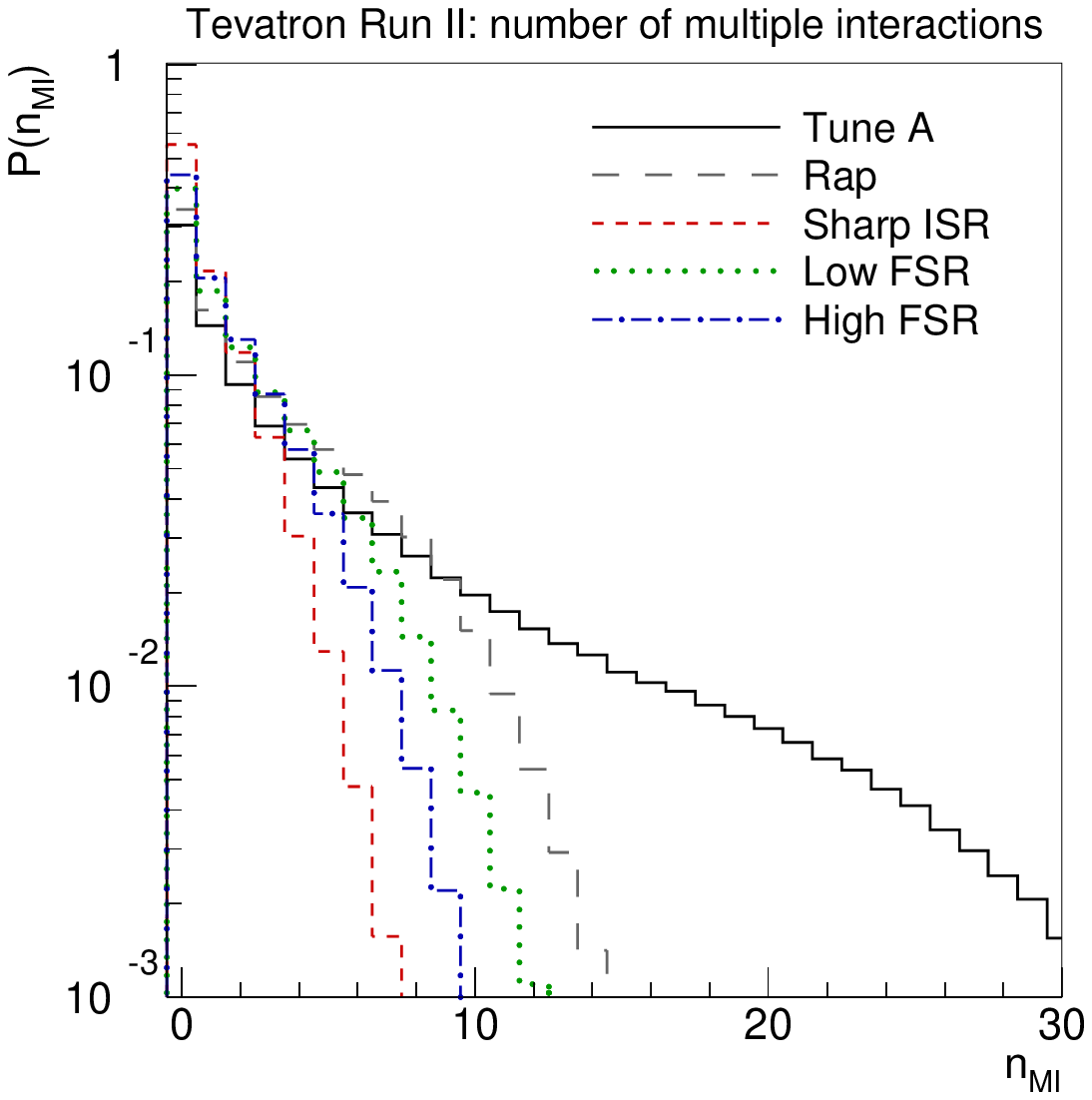}\hskip-1.9cm&\hskip-1.9cm
\includegraphics*[scale=0.72]{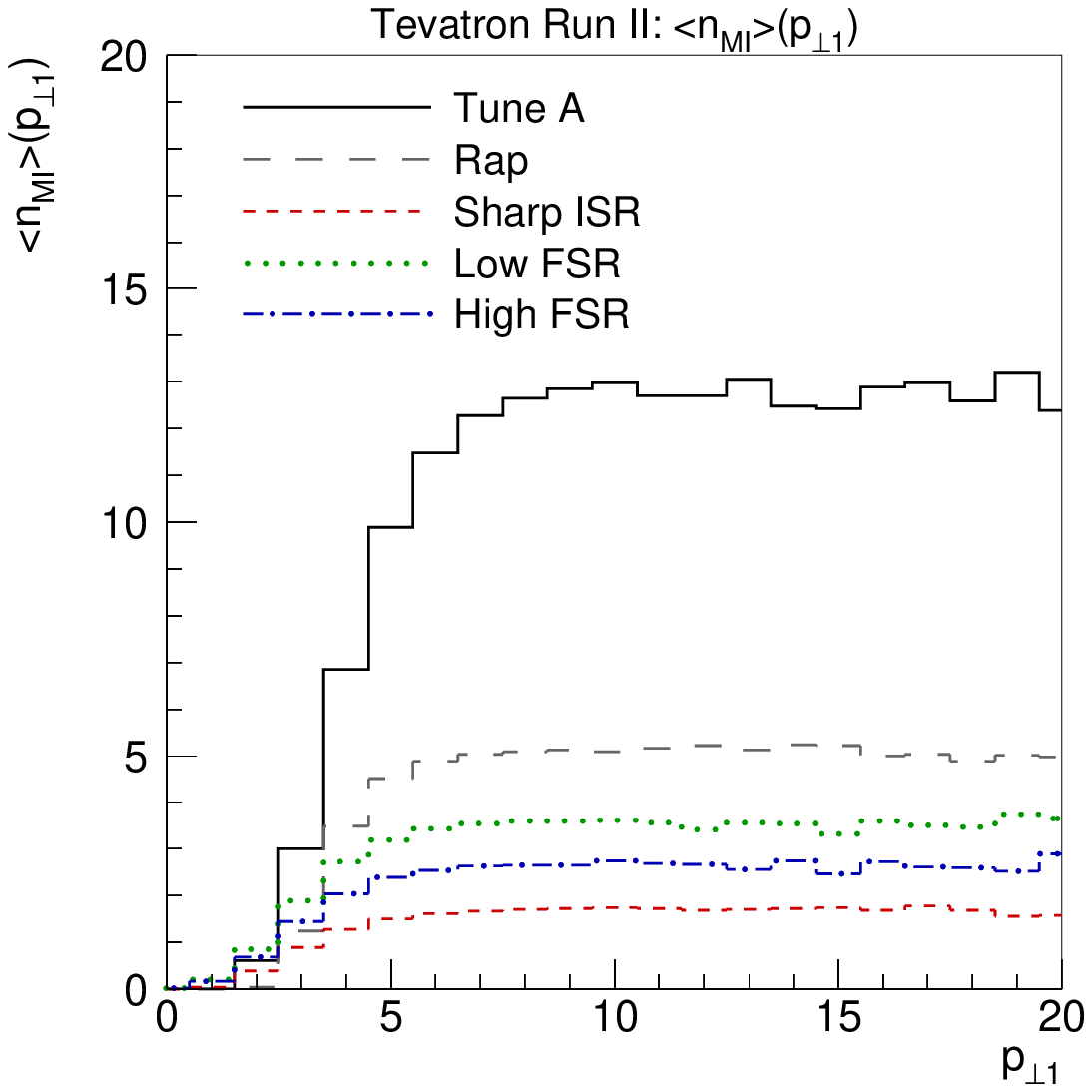}\hskip-1cm\\[-10mm]
\hskip-0.8cm{\it a)}\hskip-1.9cm & \hskip-1.9cm{\it b)}\hskip-1cm
\end{tabular}
\caption{{\it a)} Number of multiple interactions (in addition to the hardest
  one) and {\it b)} the average number of additional interactions as a
  function of the $\pT$ of the hardest interaction, both for
1.96~TeV $\p\pbar$ minimum-bias events.  
\label{fig:nint}}
\end{center}\vspace*{-3mm}
\end{figure}
\begin{figure}[tp]
\begin{center}\vspace*{-12mm}
\includegraphics*[scale=0.92]{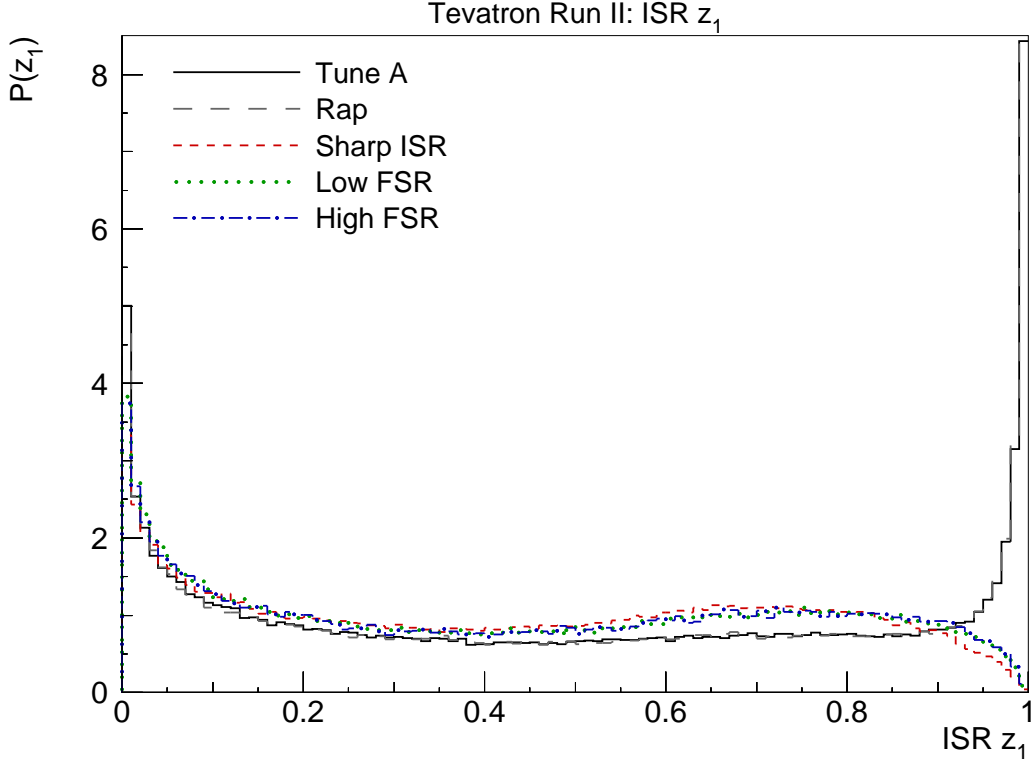}\\[-12mm]
\caption{$z$ distributions for the first ISR branching, $z_1$, in 
  1.96~TeV $\p\pbar$ minimum-bias events.
\label{fig:zdist}}
\end{center}\vspace*{-3mm}
\end{figure}

We now take a closer look at the relative proportions of the MI, ISR, and FSR
make-up of minimum-bias events, for the models in Table \ref{tab:tunes}. 
Firstly, the number of multiple interactions (excluding the hardest) is shown
in Fig.~\ref{fig:nint}a, and the dependence of the average number of extra
interactions on the $\pT$ of the hardest interaction in Fig.~\ref{fig:nint}b. 
The relatively low $\pTo$ and slightly more peaked matter distribution of
Tune A gives a tail towards very large multiplicities which is substantially
reduced both in the new models and in the Rap tune. Surprisingly, 
the Low FSR scenario lies somewhat below the Rap model, even though the
latter has a higher $\pTo$ scale. A sanity check is to switch off
ISR and then compare the four models with the same matter overlap.
Without the ISR evolution competing for phase space, the $\nmi$ distribution
then looks as would be expected, with the lower $\pTo$ scenario exhibiting the
broadest distribution. Thus, the ISR branchings `eat up' phase space more
quickly in the new framework than before, leaving less room for 
multiple interactions. This conclusion is verified in Fig.~\ref{fig:zdist},
which compares the distribution of $z$ values for the first, i.e.\ hardest,
ISR branching in an event. The soft-gluon enhancement of ISR near $z=1$ in 
the old models is absent in the new ones! This comes from the use of an 
evolution variable $\pTse = (1-z) Q^2$ in the latter ones, which favours 
larger $1-z$ in a branching than an evolution in $Q^2$, cf.\ the 
$z_{\mrm{max}}$ expression in eq.~(\ref{eq:zmax_massless}). 

\begin{figure}[tp]
\begin{center}\vspace*{-12mm}%
\begin{tabular}{cc}\hskip-0.8cm
\includegraphics*[scale=0.72]{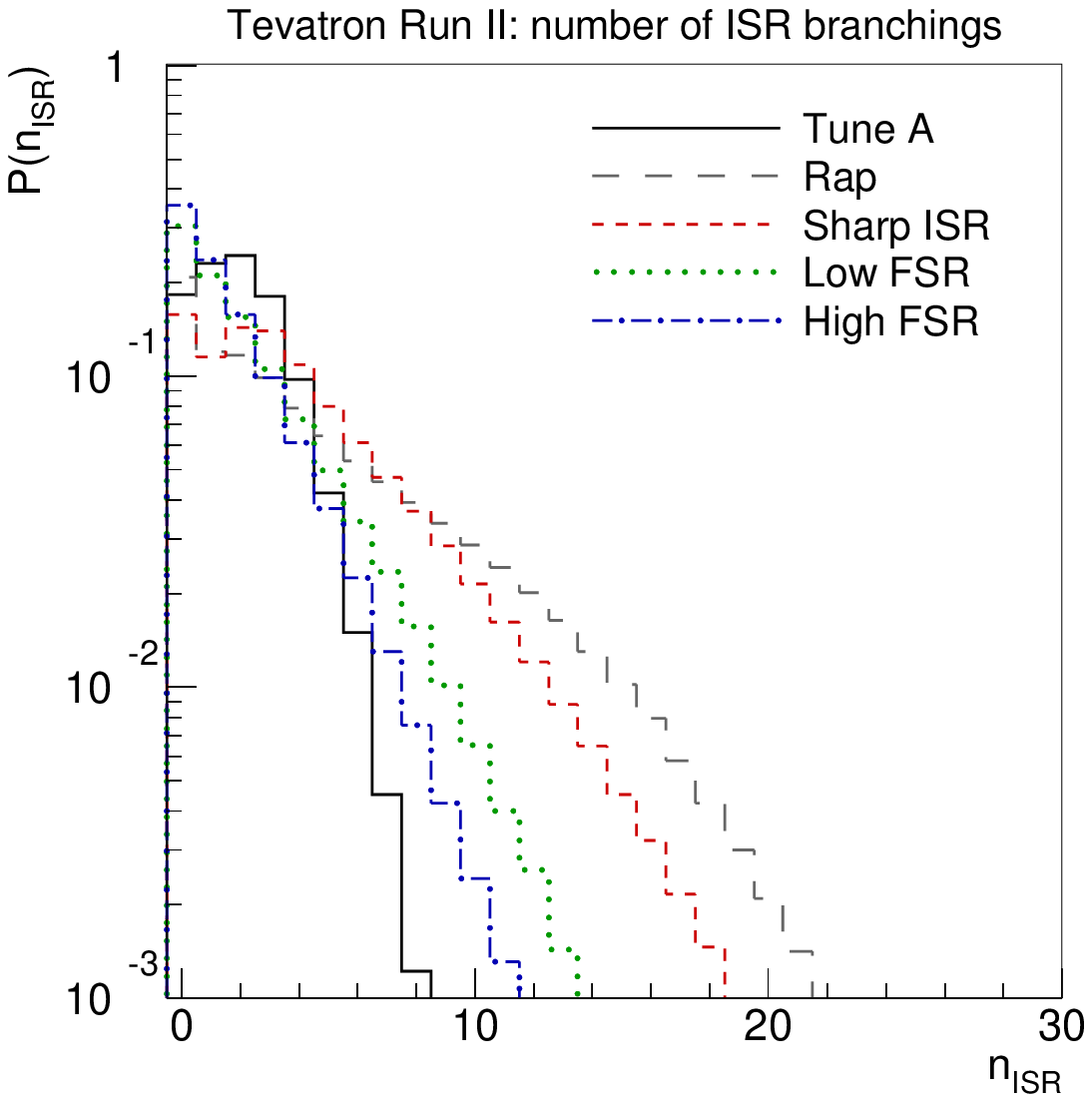}\hskip-1.9cm&\hskip-1.9cm
\includegraphics*[scale=0.72]{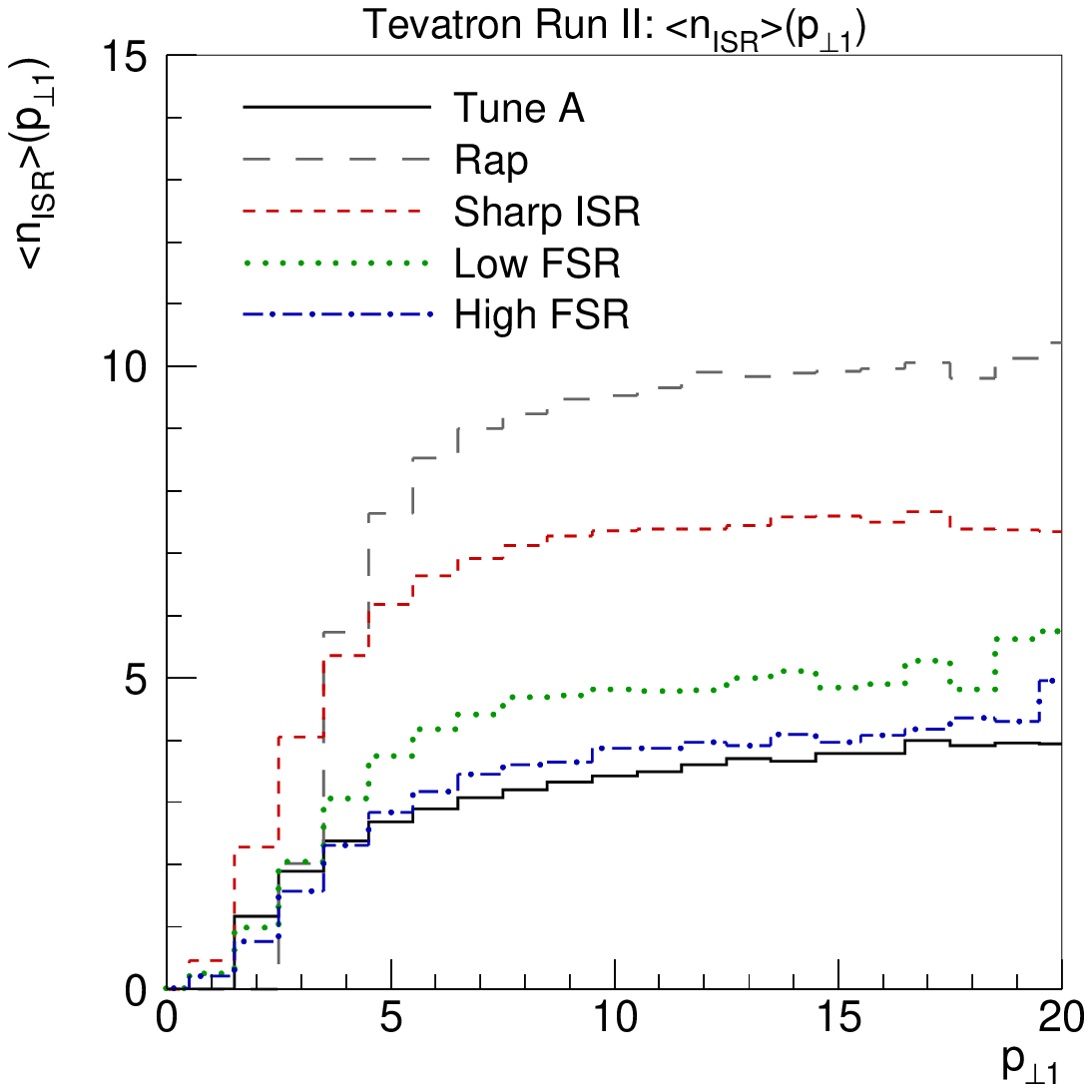}\hskip-1cm\\[-10mm]
\hskip-0.8cm{\it a)}\hskip-1.9cm & \hskip-1.9cm{\it b)}\hskip-1cm
\end{tabular}
\caption{{\it a)} Number of ISR branchings and {\it b)} 
the average number of ISR branchings as a
  function of the $\pT$ of the hardest interaction, both for
1.96~TeV $\p\pbar$ minimum-bias events.  \label{fig:nISR}}
\end{center}\vspace*{-3mm}
\end{figure}

\begin{figure}[tp]
\begin{center}\vspace*{-12mm}%
\begin{tabular}{cc}\hskip-0.8cm
\includegraphics*[scale=0.72]{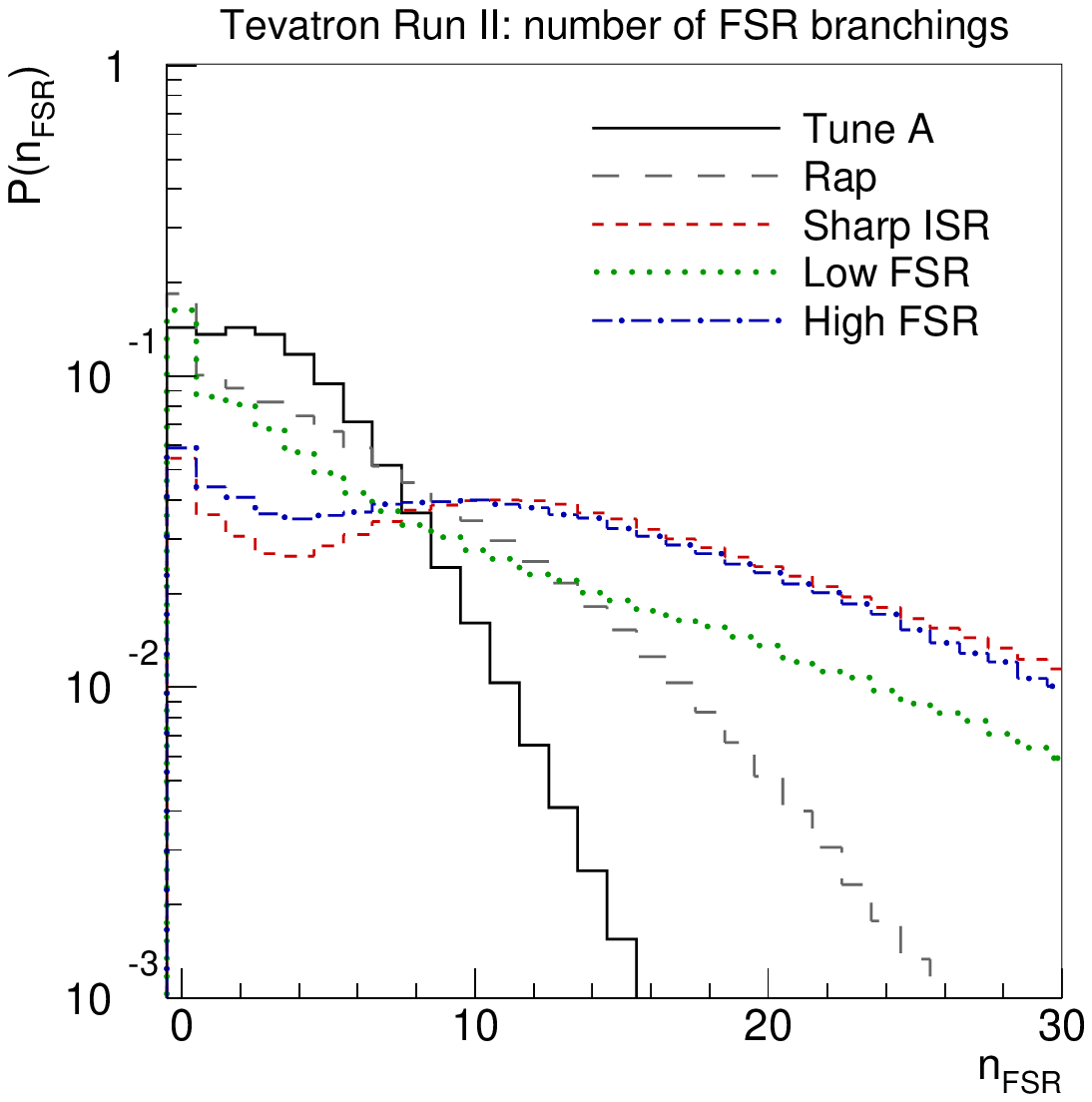}\hskip-1.9cm&\hskip-1.9cm
\includegraphics*[scale=0.72]{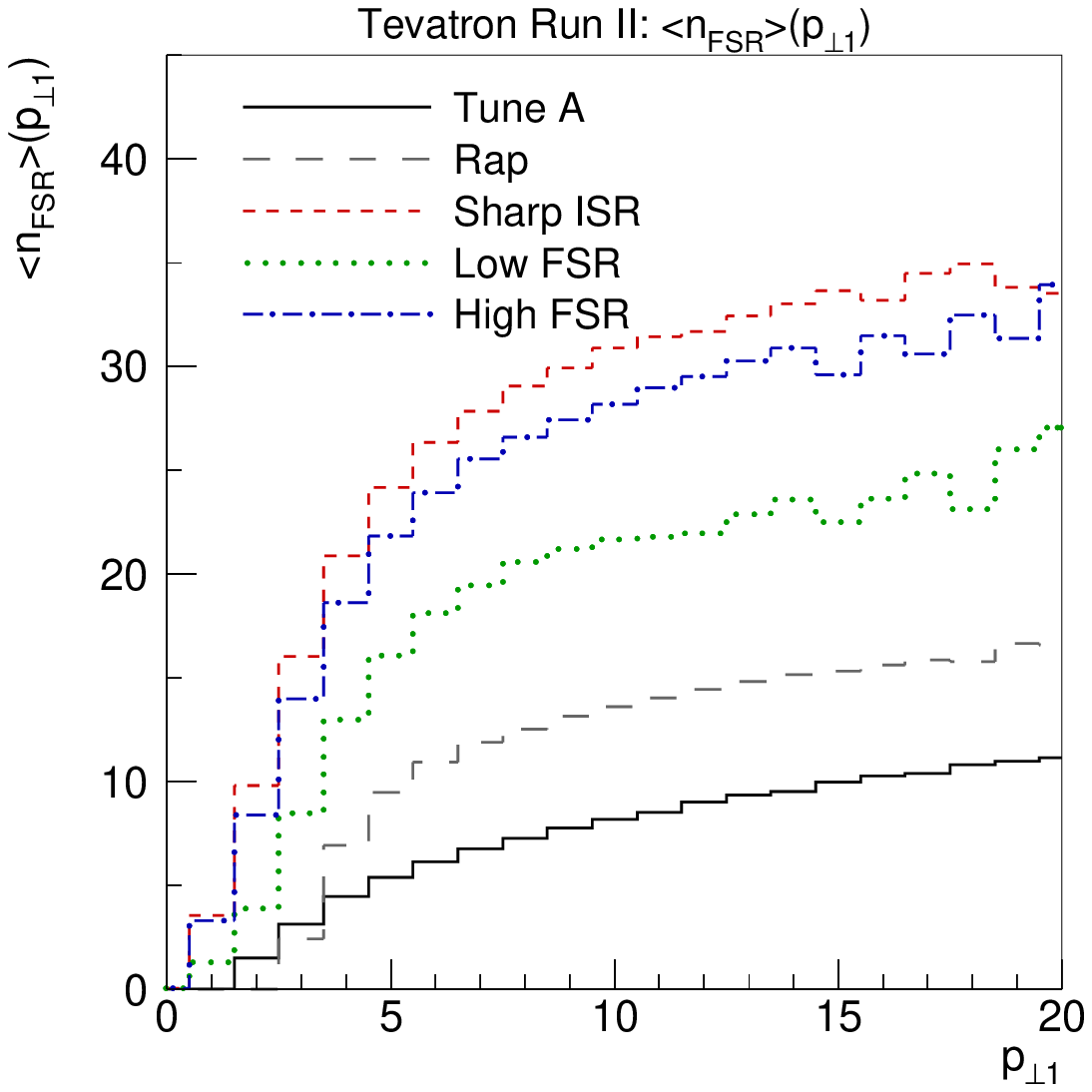}\hskip-1cm\\[-10mm]
\hskip-0.8cm{\it a)}\hskip-1.9cm & \hskip-1.9cm{\it b)}\hskip-1cm
\end{tabular}
{\it b)}
\caption{Number of FSR branchings and {\it b)} 
the average number of FSR branchings as a
  function of the $\pT$ of the hardest interaction, both for
  1.96~TeV $\p\pbar$ minimum-bias events.\label{fig:nFSR}}
\end{center}\vspace*{-3mm}
\end{figure}

In analogy with Fig.~\ref{fig:nint}, the multiplicities of ISR and FSR
branchings are depicted in Figs.~\ref{fig:nISR} and  \ref{fig:nFSR},
respectively. For ISR as well as for FSR, 
Tune A has by far the narrowest distributions, since only
the hardest interactions are associated with parton showers. Concentrating on
the ISR distribution, Fig.~\ref{fig:nISR}, again the Rap model exhibits a
very broad distribution, together with the Sharp ISR model. This behaviour is
characteristic of the threshold regularization of the ISR cascade employed in
these models, which gives a larger number of fairly soft emissions than 
the smoothly regularized models, Low and High FSR. Also note that the smaller 
number of branchings in these models partly is compensated by the larger
$\langle 1-z \rangle$ for the branchings that do occur.


The large number of FSR branchings, Fig.~\ref{fig:nFSR}, is related to 
the use of a very small cutoff here, of the order of $\pTmin = 0.5$~GeV, 
and so it cannot be compared directly with the MI and ISR multiplicities. 
The new models clearly have much broader FSR distributions than both 
Tune A and the Rap model. As one would expect from the choice of maximum 
scale of emission, the Low FSR model is the narrowest of the new models. 
We also recall that there is a built-in compensation mechanism: if the 
number of ISR  branchings is reduced then, other things being the same,
this results in fewer but larger dipoles that therefore can radiate more.
Although the old and new shower algorithms do not allow a straightforward
comparison, the difference between Rap and the new models is at least
consistent with such a partial compensation. 

\begin{figure}[tp]
\begin{center}\vspace*{-12mm}%
\includegraphics*[scale=0.92]{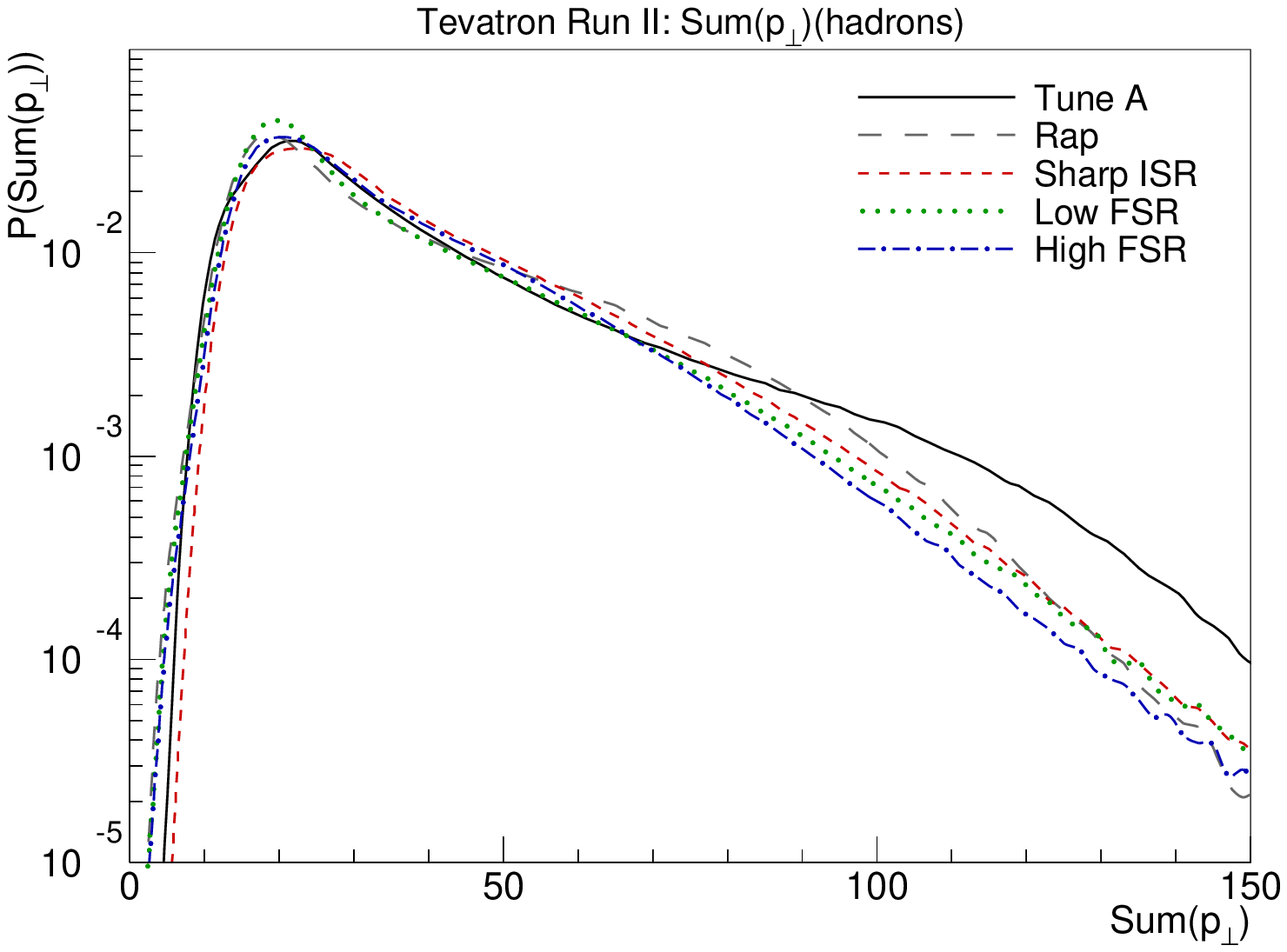}\\[-12mm]
\caption{Total $\pT$ sum for hadrons in
  1.96~TeV $\p\pbar$ minimum-bias events.\label{fig:t120}}
\end{center}\vspace*{-3mm}
\end{figure}

Returning now to observable distributions, the fact that less $\pT$ is kicked
into events with large multiplicities in the new frameworks, cf.\
Fig.~\ref{fig:tavgpt},  while the multiplicity distributions are similar, also
implies that there should be fewer events with large total $E_\perp$ than in
Tune A. This is corroborated by Fig.~\ref{fig:t120}, which shows the scalar
sum of hadron $\pT$ values in 1.9 ~TeV $\p\pbar$ minimum-bias events. Both
the new models and the Rap model have noticeably fewer events in the region
above $\sim 100$~GeV than does Tune A.

\begin{figure}[tp]
\begin{center}\vspace*{-12mm}%
\includegraphics*[scale=0.92]{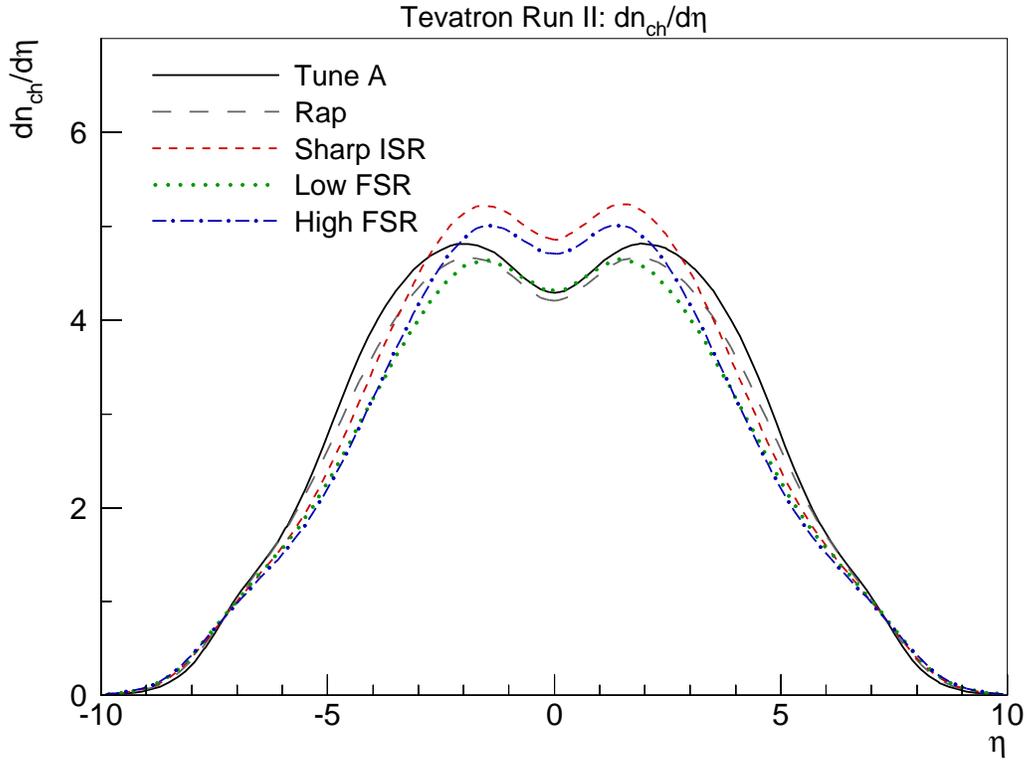}\\[-12mm]
\caption{Charged multiplicity as a function of pseudorapidity $\eta$ in
  1.96~TeV $\p\pbar$ minimum-bias events.\label{fig:t30}}
\end{center}\vspace*{-3mm}
\end{figure}

In addition, Fig.~\ref{fig:t30} shows that the pseudorapidity
distribution has become narrower, i.e.\ the particle
production has become more central. The normalization differences are in this
context not very interesting, arising from small differences in the
average charged multiplicity of the tunes.

\begin{figure}[tp]
\begin{center}\vspace*{-12mm}%
\includegraphics*[scale=0.92]{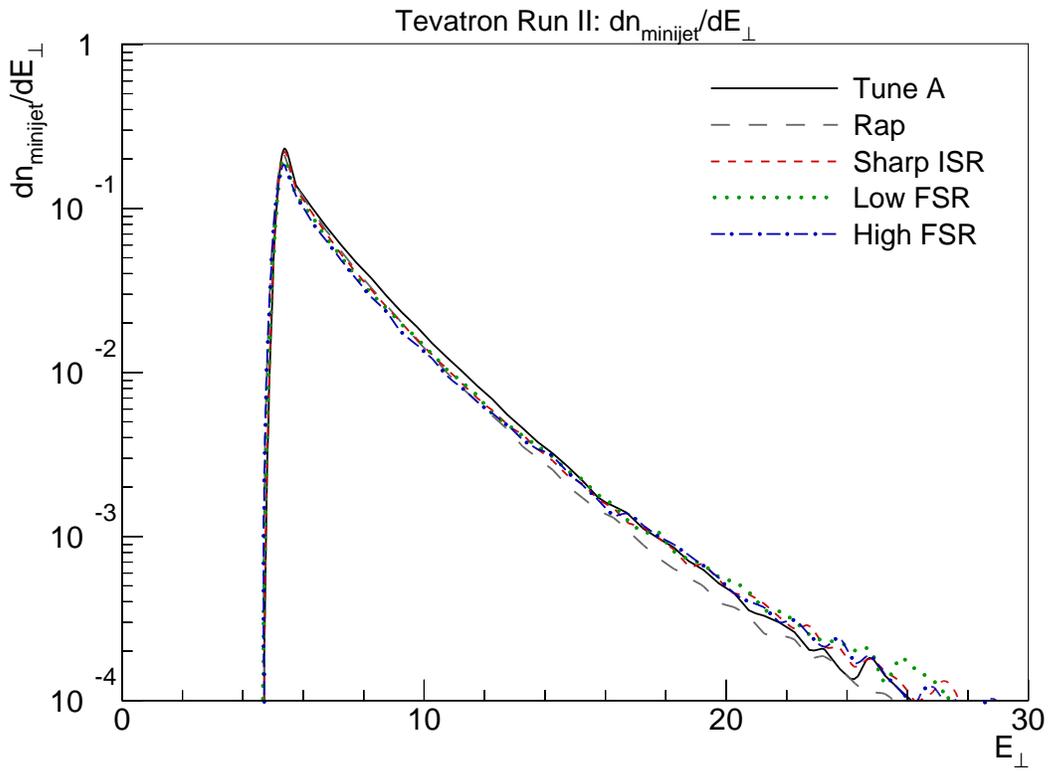}\\[-12mm]
\caption{Minijet rates, for
  1.96~TeV $\p\pbar$ minimum-bias events.\label{fig:t100}}
\end{center}\vspace*{-3mm}
\end{figure}

\begin{figure}[tp]
\begin{center}\vspace*{-12mm}%
\includegraphics*[scale=0.92]{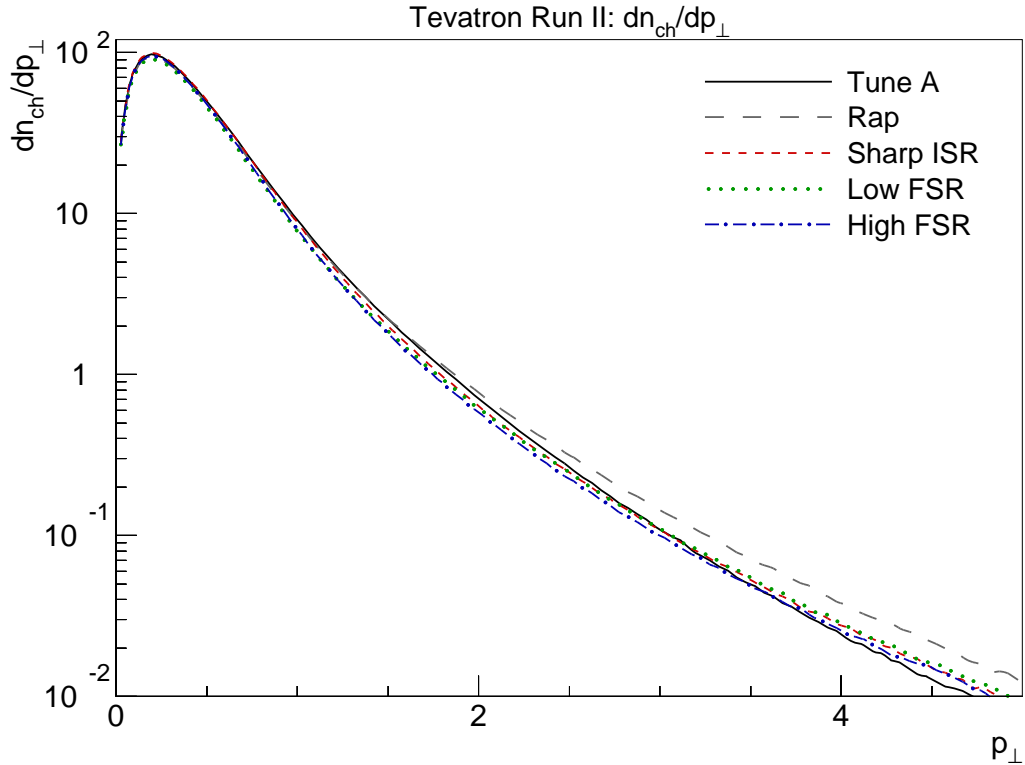}\\[-12mm]
\caption{Charged hadron $\pT$ spectra in 1.96~TeV $\p\pbar$ minimum-bias
  events. The normalization corresponds the total average
  charged multiplicity.\label{fig:t40}}  
\end{center}
\end{figure}

However, these difference do not have a large impact on most other
observables. Thus e.g.\ the minijet rates and charged hadron $\pT$
distributions in Figs.~\ref{fig:t100} and \ref{fig:t40} are hardly
distinguishable between Tune A and the new models. 
The minijet $E_\perp$ spectrum, defined by a simple cone algorithm 
with a cone radius of $R = \sqrt{(\Delta \eta)^2 + (\Delta \phi)^2} = 0.7$
and an $E_{\perp\mrm{min}} = 5$~GeV, 
which was slightly softer in the Rap model
than in Tune A, has become slightly harder. On the other hand, the charged
hadron $\pT$ spectrum, which was slightly harder in the Rap model than in
Tune A, has dropped back down fairly close to the Tune A level. 

\subsection{Jet events and profiles}
 
\begin{figure}[p]
\mbox{\epsfig{file=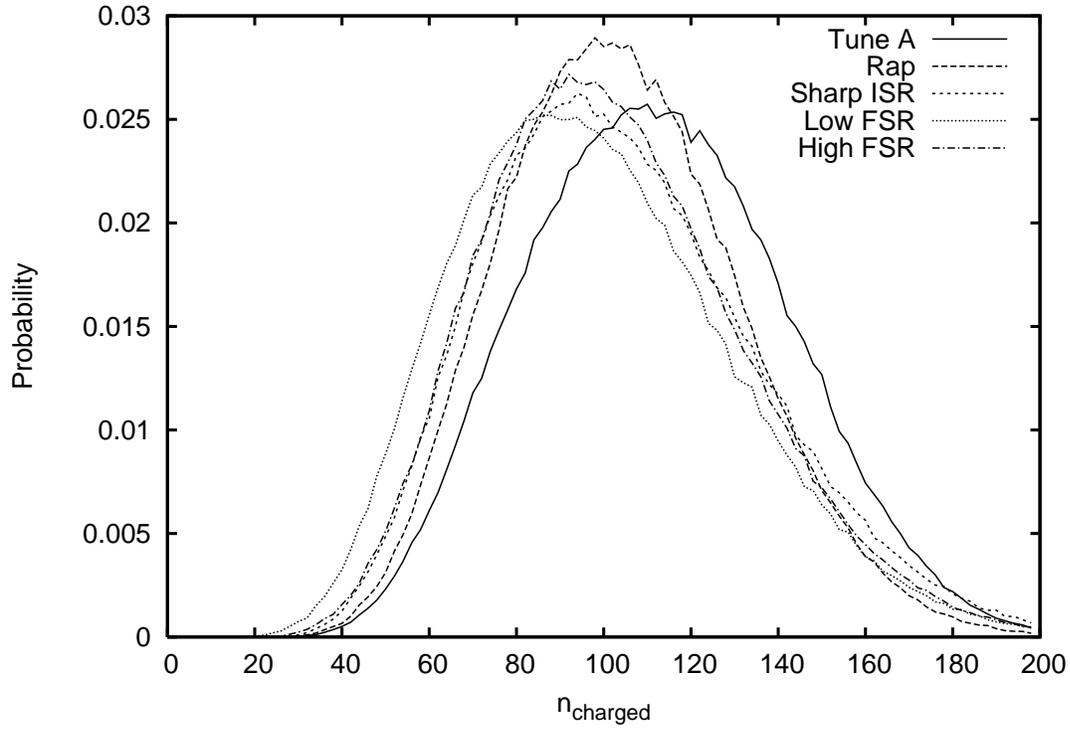,width=14.5cm}}
\vspace{-3mm}
\caption{Charged multiplicity distribution for 1.96~TeV 
$\p\pbar$ events with $\pThard > 100$~GeV.
\label{fig:nchhundred}}
\end{figure}
\begin{figure}[p]
\mbox{\epsfig{file=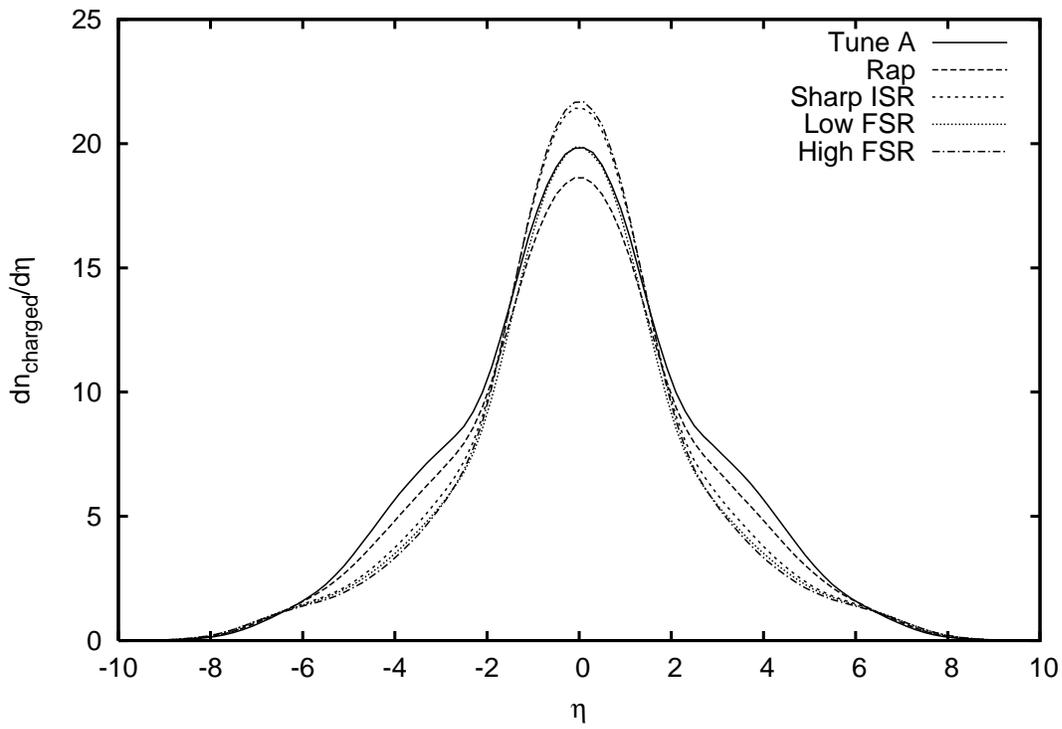,width=14.5cm}}
\vspace{-3mm}
\caption{Pseudorapidity distribution for 1.96~TeV 
$\p\pbar$ events with $\pThard > 100$~GeV.
\label{fig:etahundred}}
\end{figure}
 
\begin{figure}[p]
\mbox{\epsfig{file=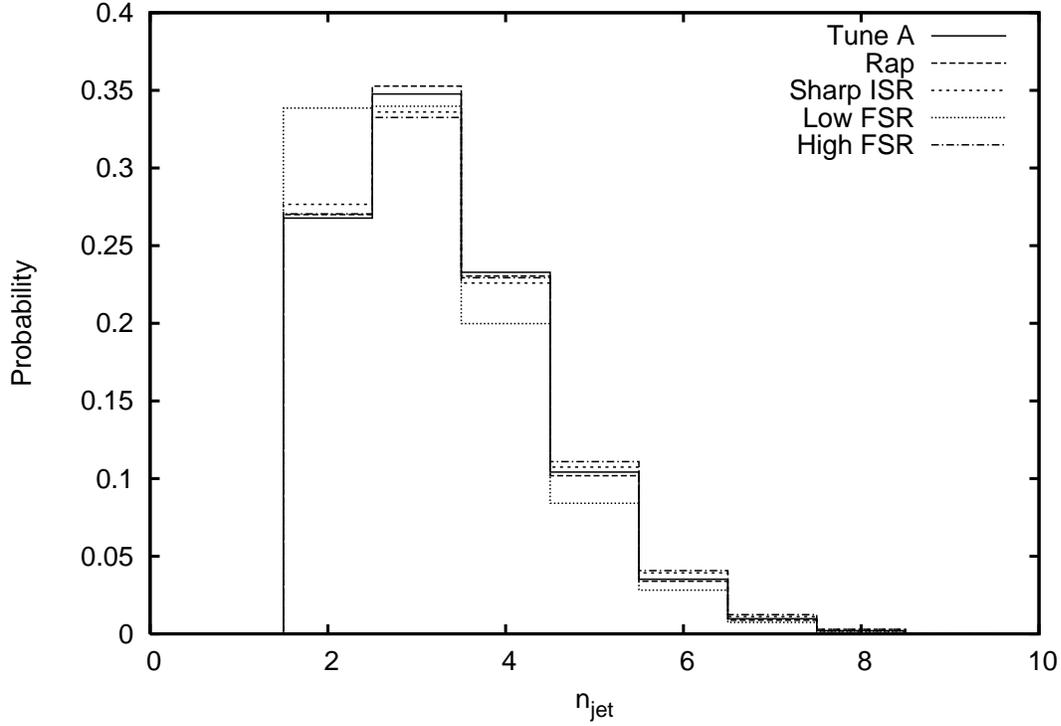,width=14.5cm}}
\vspace{-3mm}
\caption{Jet multiplicity distribution for 1.96~TeV 
$\p\pbar$ events with $\pThard > 100$~GeV,
using a cone clustering algorithm with $R = 0.7$ and 
$E_{\perp\mrm{min}} = 10$~GeV.
\label{fig:njethundred}}
\end{figure}
\begin{figure}[p]
\mbox{\epsfig{file=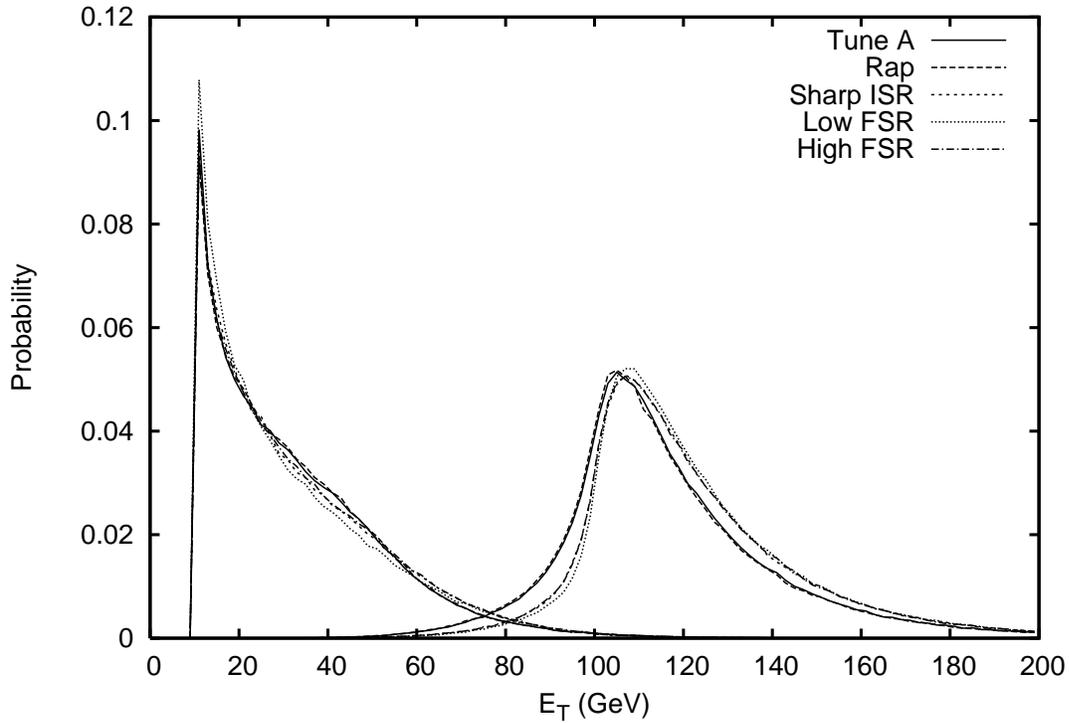,width=14.5cm}}
\vspace{-3mm}
\caption{$E_{\perp}$ spectra for the hardest and third hardest
jet in 1.96~TeV $\p\pbar$ events with $\pThard > 100$~GeV.
\label{fig:etonethree}}
\end{figure}
 
\begin{figure}[tp]
\mbox{\epsfig{file=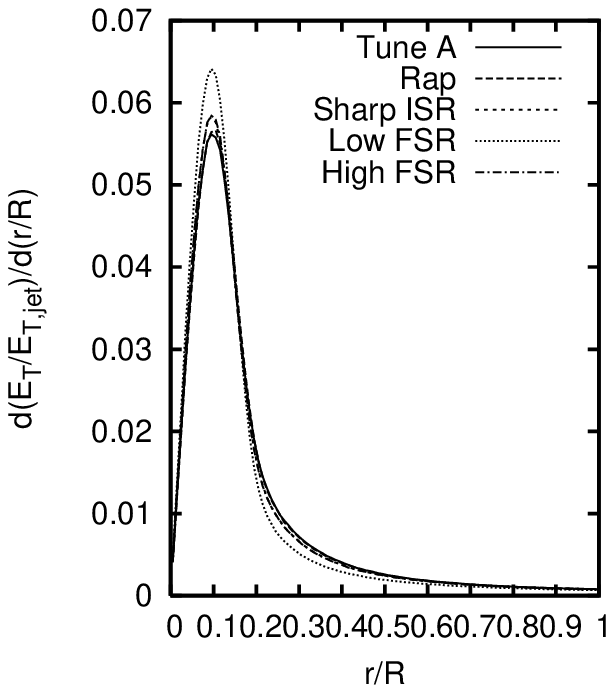,width=8cm}}\hfill%
\mbox{\epsfig{file=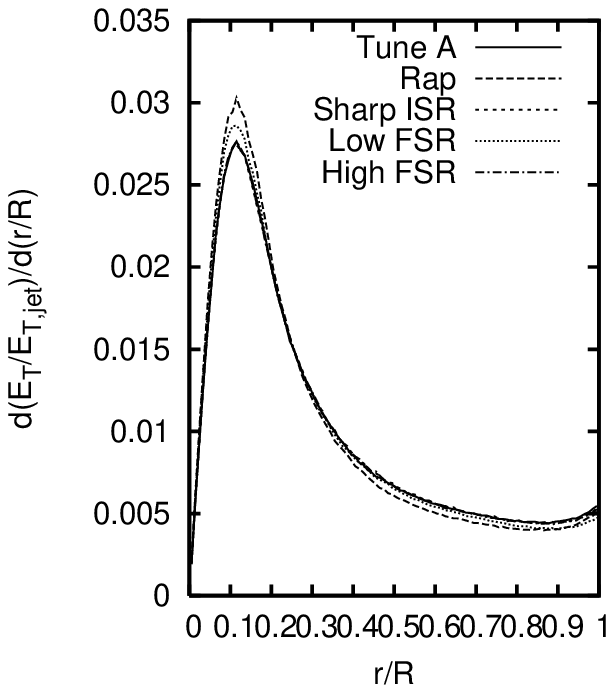,width=8cm}}
\vspace{-3mm}
\hspace*{4cm}\textit{a)}\hspace{8cm}\textit{b)}
\caption{The \textit{a)} hardest and \textit{b)} third hardest jet 
$E_{\perp}$ profile for 1.96~TeV $\p\pbar$ events with 
$\pThard > 100$~GeV.
\label{fig:etprof}}
\end{figure}

\begin{figure}[tp]
\mbox{\epsfig{file=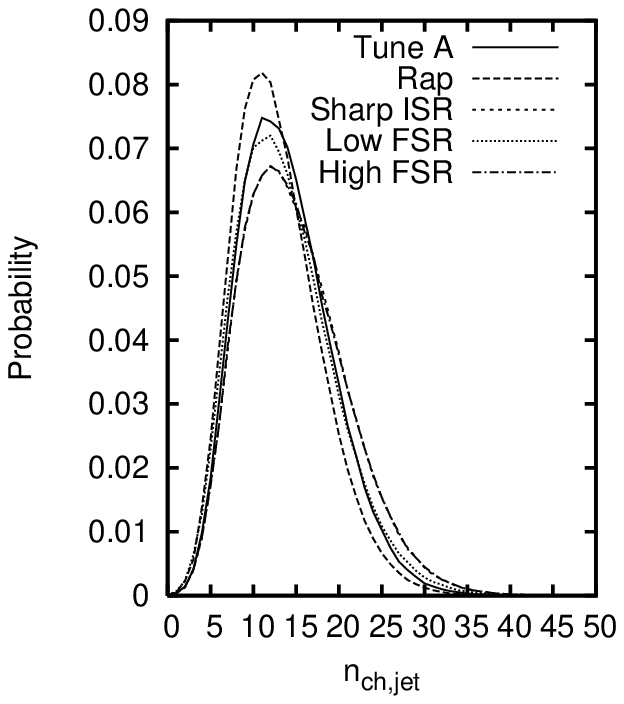,width=8cm}}\hfill%
\mbox{\epsfig{file=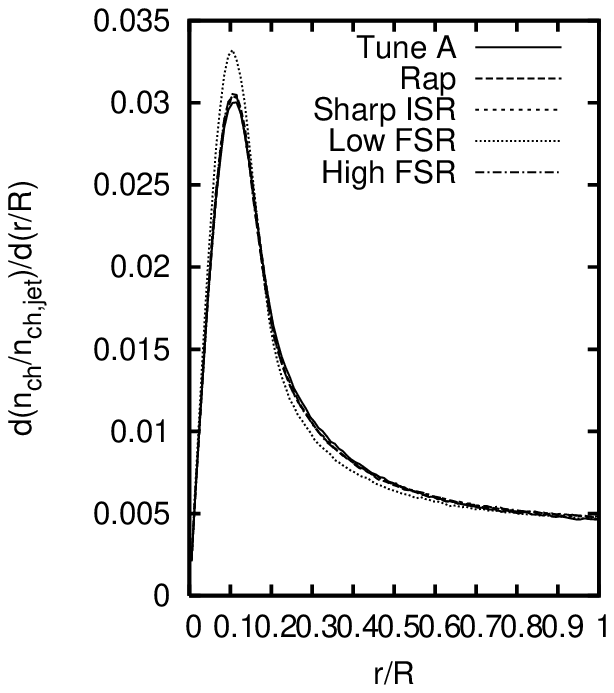,width=8cm}}
\vspace{-3mm}
\hspace*{4cm}\textit{a)}\hspace{8cm}\textit{b)}
\caption{The \textit{a)} charged multiplicity distribution 
and \textit{b)} charged particle profile of the hardest jet in 
1.96~TeV $\p\pbar$ events with $\pThard > 100$~GeV.
\label{fig:nchone}}
\end{figure}

Complementary to the above are studies of events with hard
jets and their properties. As an example of this, we have considered 
1.96~TeV $\p\pbar$ events where the hardest $2 \to 2$ interaction has a 
$\pThard > 100$~GeV, without any further requirements.
The charged multiplicity distribution of such events is shown in 
Fig.~\ref{fig:nchhundred}, and their pseudorapidity distribution in
Fig.~\ref{fig:etahundred}. Given that the models have been tuned to 
each other exclusively for a minimum-bias event sample, the differences
are less than could have been expected. We note a clear difference 
at mid-rapidities, however, where Tune~A shows more activity than any 
of the newer scenarios, cf.\ also Fig.~\ref{fig:t30}. 
This is likely to be related to the way 
strings are connected from the central interactions to the beam remnants.

The jet multiplicity in these events, obtained by a combination of
MI, ISR and FSR activity, is shown in Fig.~\ref{fig:njethundred}.
The Low FSR scenario stands out by having significantly less jet 
activity than any of the other ones, clearly indicating the impact of 
the reduced FSR in these events. The other rates come surprisingly 
close, given that both the ISR and the FSR algorithms are quite different
between the old and the new scenarios. At high jet multiplicities the 
new ones are somewhat above the older ones.

Next we study the properties of the jets produced. Since the two hardest
jets both arise already as a consequence of the hard interaction, they have
similar properties, while further jets are related to the additional
activity and thus internally similar. Therefore only results for the hardest 
and (when present) third hardest jet are shown here. The respective jet 
$E_{\perp}$ spectra are shown in Fig.~\ref{fig:etonethree}. The hardest jet 
is harder in all the three new scenarios than in the two old ones, while the 
third and subsequent ones are more similar. Again, given the changed ISR and 
FSR algorithms, the similarities for the third jet are more surprising than 
the differences for the first. Notably, the lower jet activity in the 
Low FSR scenario is not reflected in a reduced tail out to 
high-$E_{\perp}$ third jets.

The energy flow inside a jet can be plotted as a function of the 
distance $r$ away from the center of the jet, or better as a function of
$r/R$. Such profiles are shown for the hardest and third jet in 
Fig.~\ref{fig:etprof}.
For the hardest jet, again Low FSR stand out by producing narrower 
jets, while for the third Rap is even more narrow. Generally, the
differences are small, however. 

Turning to charged multiplicity distributions inside jets, the Rap 
scenario tends to have the least, and the High FSR and Sharp ISR
the most. This is illustrated in Fig.~\ref{fig:nchone}\textit{a} for the hardest 
jet, but the same pattern repeats also for the softer one. Comparing with
the total charged multiplicity of these events, Fig.~\ref{fig:nchhundred}
above, which does not show the same pattern,we conclude that the balance
between activity inside and outside the identified jets differs, possibly
reflecting the amount of softer jet activity. 

By contrast, the charged particle number jet profile follows the same
pattern as observed above for the $E_{\perp}$ profile. That is, Low FSR
gives the most narrow hardest jet, Fig.~\ref{fig:nchone}\textit{b}, while
Rap gives the most narrow third jet, not shown.

In summary, differences are smaller than might have been guessed,
considering the changes especially in the ISR and FSR algorithms.
Specifically, with the new algorithms the upper scale $\pTmax$ for
ISR and FSR evolution is unambiguously set by the $\pThard$ 
of the hard interaction, while the older ones did involve an ambiguous 
choice of a $Q^2_{\mrm{max}} = 4 \pThard^2$, intended roughly 
to give $\pT$ ordering, but not in the guaranteed sense of the new 
algorithms.   

\subsection{$\Z^0$ production}
 
\begin{figure}[p]
\mbox{\epsfig{file=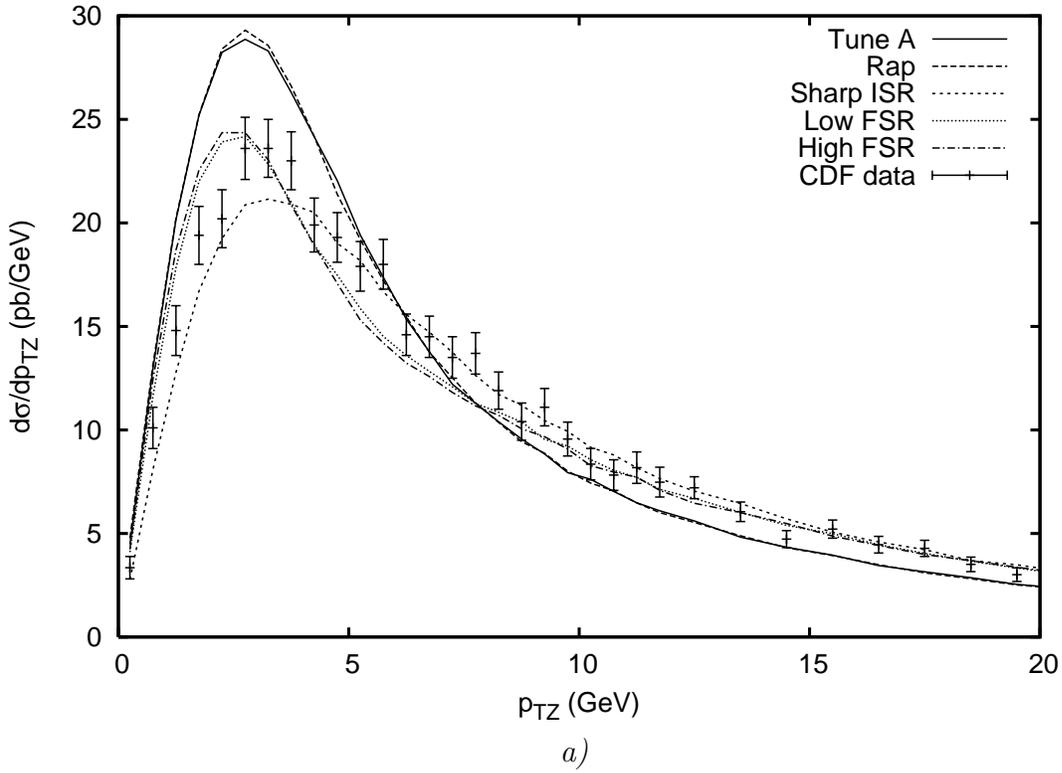,width=14.5cm}}\\[-3mm]
\hspace*{7.5cm}\textit{a)}\\[6mm]
\mbox{\epsfig{file=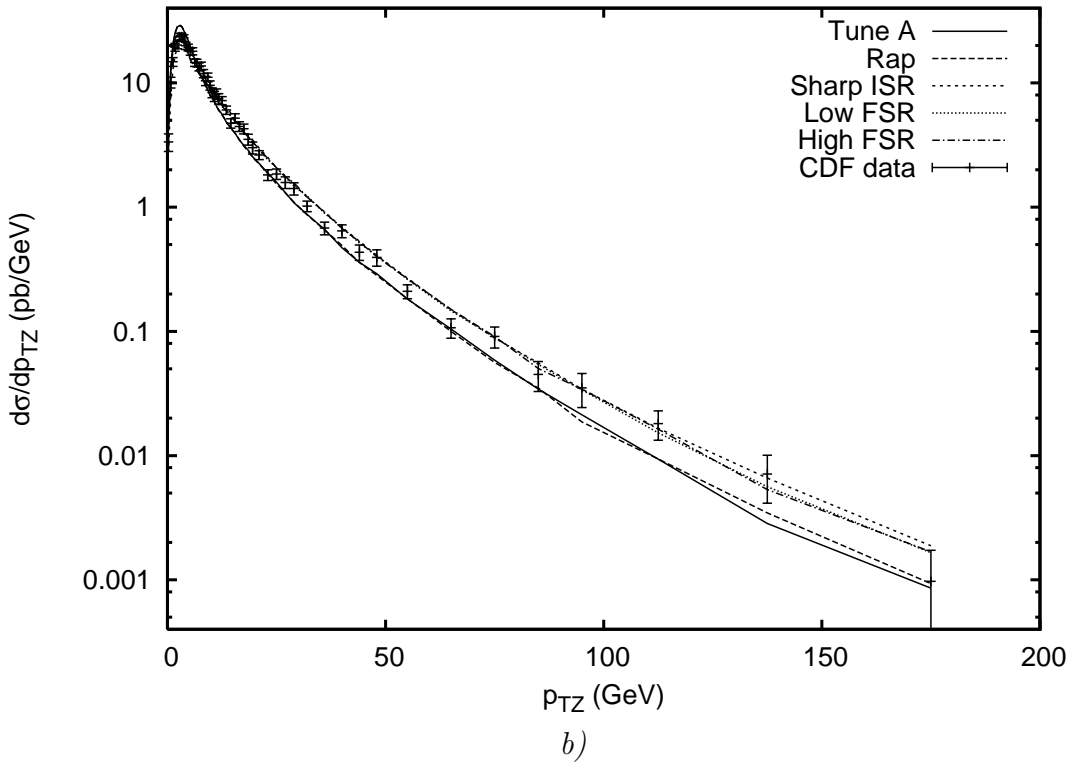,width=14.5cm}}\\[-3mm]
\hspace*{7.5cm}\textit{b)}
\caption{The $\pT$ spectrum at \textit{a)} low and \textit{b)} all $\pT$
for dilepton pairs in the 66--116~GeV mass range, $\gamma^*/\Z^0$ 
simulation compared with CDF corrected data at 1.8~TeV \cite{CDFZ}.
\label{fig:ZpT}}
\end{figure}

A slightly different test is to study the $\pT$ spectrum of high-mass
dileptons coming from the decay of a $\gamma^*/\Z^0$. We can here compare
with the CDF $\pTZ$ spectrum at 1.8~TeV \cite{CDFZ}, normalizing the curves 
to the experimental integrated cross section, Fig.~\ref{fig:ZpT}. 

Since the high-$\pTZ$ behaviour is constrained by our use of first-order 
matrix-element corrections \cite{MiuZ}, it is not surprising that 
differences here are small. That the three new scenarios are above 
the two older ones presumably is a consequence of the different 
treatment of FSR, which does not at all influence $\pTZ$ in the new
models, while the $\pT$ of an ISR branching is reduced by FSR in the 
older ones. This is a degree of freedom that could be studied further
when FSR is interleaved with MI and ISR.

More interesting is the improvement in the low-$\pT$ region,
similarly to what has been found earlier \cite{Erik}, in a study of 
the new ISR algorithm without any MI. However, note that we in all cases 
make use of a Gaussian primordial $\kT$ with a 2~GeV width (thus deviating 
from the pure Tune A, where it is kept at the \textsc{Pythia} default of 
1~GeV). The implementation of this $\kT$ is more complex with the new
beam-remnant implementation of \cite{multint}, and e.g.\ could depend on 
the number of multiple interactions, but actually the distributions
turn out to be quite similar. The problem therefore remains that this
primordial $\kT$ is larger than can physically be well motivated based 
on purely nonperturbative physics. We observe that, among the new models, 
the Sharp ISR could have been combined with a smaller primordial $\kT$  
since its peak is shifted towards too large $\pTZ$, while the High FSR 
and Low FSR (which here only differ by their $\pTo$ values) could have 
used an even larger primordial $\kT$. In part, this makes sense: with 
ISR being turned off at larger $\pT$ values in the latter models, it is 
then also easier to motivate a larger primordial $\kT$. 

The complete comparison of algorithms is rather complicated, however.
The primordial $\kT$ that reaches the hard interaction is diluted by 
the ISR activity, and so scales down like the ratio of the $x$ value of 
the incoming parton at the hard interaction to that of the initiator,
$z_{\mrm{tot}} = x_{\mrm{in}}/x_{\mrm{init}}$. This ratio is 
approximately the same for the old ISR shower 
($\langle z_{\mrm{tot}} \rangle \approx 0.59$) and 
Sharp ISR ($\langle z_{\mrm{tot}} \rangle \approx 0.62$), indicating 
that the fewer ISR branchings and smaller $z$ per branching in the
new algorithm rather well cancel. The smooth turnoff of High and Low FSR 
gives less branchings ($\langle z_{\mrm{tot}} \rangle \approx 0.75$)
and thus more primordial $\kT$ survives in these scenarios.

In summary, the new MI+ISR scheme gives an improved description of $\Z^0$
production, but does not remove the need for an uncomfortably large
primordial $\kT$.
 
\section{Outlook}
\label{sec:outlook}

In this article we have considered the consequences of interleaved
multiple interactions and initial-state radiation, and paved the way 
for interleaving also final-state radiation in this framework, but 
that does not exhaust the list of perturbative processes in `normal' 
hadronic events. One further possibility is that a parton from one of 
the incoming hadrons scatters twice, against two different partons
from the other hadron, rescattering or `$3 \to 3$'. Another possibility 
is that two partons participating in two separate hard scatterings 
may turn out to have a common ancestor when the backwards evolution 
traces the prehistory to the hard interactions, 
\textit{joined interactions} (JI). 

The $3 \to 3$ processes have been considered in the literature 
\cite{threeproc}, with the conclusion that they should be less important 
than multiple $2 \to 2$ processes, except possibly at large $\pT$ values,
where QCD radiation anyway is expected to be the dominant source
of multijet events. The reason is that one $3 \to 3$ scattering and 
two $2 \to 2$ ones have similar parton-level cross sections, but the 
latter wins by involving one parton density more. Nevertheless, at 
some point, there ought to be a more detailed modelling, in order 
better to quantify effects. 

The joined interactions are well-known in the 
context of the evolution of multiparton densities \cite{intertwpdf}, 
but have not been applied to a multiple interactions framework. We 
will therefore here carry out a first study, to quantify roughly how 
common JI are and how much activity they contribute with. A full 
implementation of the complete kinematics, intertwining MI, ISR and JI 
all possible ways, is a major undertaking, worth the effort only if 
the expected effects are non-negligible. Given the many uncertainties 
in all the other processes at play, one would otherwise expect that 
the general tuning of MI/ISR/FSR/\ldots to data would hide the effects 
of JI, as well as of $3 \to 3$ processes. 
  
\subsection{Joined interactions: theory}
 
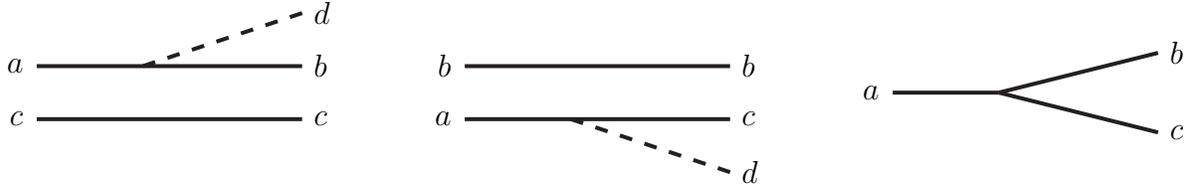
\begin{figure}[t]
\begin{picture}(140,80)(0,-40)
\SetWidth{1.5}
\Line(20,10)(120,10)\Text(15,10)[r]{$a$}\Text(125,10)[l]{$b$}
\Line(20,-10)(120,-10)\Text(15,-10)[r]{$c$}\Text(125,-10)[l]{$c$}
\DashLine(60,10)(120,30){5}\Text(125,30)[l]{$d$}
\end{picture}\hfill
\begin{picture}(140,80)(0,-40)
\SetWidth{1.5}
\Line(20,10)(120,10)\Text(15,10)[r]{$b$}\Text(125,10)[l]{$b$}
\Line(20,-10)(120,-10)\Text(15,-10)[r]{$a$}\Text(125,-10)[l]{$c$}
\DashLine(60,-10)(120,-30){5}\Text(125,-30)[l]{$d$}
\end{picture}\hfill
\begin{picture}(140,80)(0,-40)
\SetWidth{1.5}
\Line(20,0)(60,0)\Text(15,0)[r]{$a$}
\Line(60,0)(120,15)\Text(125,15)[l]{$b$}
\Line(60,0)(120,-15)\Text(125,-15)[l]{$c$}
\end{picture}
\caption{Illustration of the three terms in the two-parton density
evolution, eq.~(\protect\ref{twoevoleq}).
\label{fig:twoevoleq}}
\end{figure}

Just like the starting point for a discussion of ISR is the DGLAP
evolution equations for the single-parton densities, the starting 
point for JI is the evolution equations for the two-parton 
densities. Relevant forwards-evolution formulae are available in 
the literature in integrated form \cite{intertwpdf}. 
Here, however, we will choose a differential form,
that can then be applied to our backwards evolution framework. To this
end, define the two-parton density $f^{(2)}_{bc}(x_b, x_c, Q^2)$ as
the probability to have a parton $b$ at energy fraction $x_b$ and
a parton $c$ at energy fraction $x_c$ when the proton is probed at a
scale $Q^2$. The evolution equation for this distribution is
\begin{eqnarray}
\d f^{(2)}_{bc}(x_b, x_c, Q^2)
 & = & \frac{\d Q^2}{Q^2} \, \frac{\alphas}{2\pi}
       \int \!\! \! \int \d x_a \, \d z \, \left\{ \,
       f^{(2)}_{ac}(x_a, x_c, Q^2) \, P_{a \to b d}(z) \,
       \delta(x_b - z x_a) \right. \nonumber \\
 &   & + \left. f^{(2)}_{ba}(x_b, x_a, Q^2) \, P_{a \to c d}(z) \,
       \delta(x_c - z x_a)  \right. \nonumber \\
 &   & + \left. f_a(x_a, Q^2) \, P_{a \to bc}(z) \,
       \delta(x_b - z x_a) \, \delta(x_c - (1-z) x_a) \right\}~.
\label{twoevoleq}
\end{eqnarray}
As usual, we assume implicit summation over the allowed flavour
combinations; thus the last term is absent when there is no suitable
mother $a$ for a given set of $b$ and $c$. An illustration of the three
terms is given in Fig.~\ref{fig:twoevoleq}. The first two are the
standard ones, where $b$ and $c$ evolve independently, up to flavour and
momentum conservation constraints, and are already taken into account in
the ISR framework. It is the last term that describes the new possibility
of two evolution chains having a common ancestry.
 
Carrying out the $\delta$ integrations, which imply that
$x_a = x_b + x_c$ and $z = x_b/(x_b + x_c)$, the probability for
the unresolution of $b$ and $c$ into $a$ when $Q^2$ is decreased
(cf.\ the step from eq.~(\ref{dglap}) to eq.~(\ref{dglapback}))
can be rewritten as
\begin{eqnarray}
\d \mathcal{P}_{bc}(x_b, x_c, Q^2)
& = & \left|
\frac{\d f^{(2)}_{bc}(x_b, x_c, Q^2)}{f^{(2)}_{bc}(x_b, x_c, Q^2)}\right|
\,  \nonumber \\
& = & \left| \frac{\d Q^2}{Q^2} \right| \, \frac{\alphas}{2 \pi} \,
\frac{f_a(x_a, Q^2)}{f^{(2)}_{bc}(x_b, x_c, Q^2)} \,
\frac{1}{x_b + x_c} \, P_{a \to bc}(z) \nonumber \\
& = & \left| \frac{\d Q^2}{Q^2} \right| \, \frac{\alphas}{2 \pi} \,
\frac{x_a f_a(x_a, Q^2)}{x_b x_c f^{(2)}_{bc}(x_b, x_c, Q^2)} \,
z (1-z) P_{a \to bc}(z) \nonumber \\
& \simeq & \left| \frac{\d Q^2}{Q^2} \right| \, \frac{\alphas}{2 \pi} \
\frac{x_a f_a(x_a, Q^2)}{x_b f_b(x_b, Q^2) \, x_c f_c(x_c, Q^2)} \,
z (1-z) P_{a \to bc}(z) ~.
\label{fcorrevol}
\end{eqnarray}
In the last step we have introduced the approximation
$f^{(2)}_{bc}(x_b, x_c, Q^2) \simeq f_b(x_b, Q^2) \, f_c(x_c, Q^2)$
to put the equation in terms of more familiar quantities. Just like
for the other processes considered, a form factor is given by
integration over the relevant $Q^2$ range and exponentiation.
 
The strategy now is clear. Previously we have introduced a scheme
wherein events are evolved downwards in $\pT$. At each step a new
trial multiple interaction competes against trial ISR branchings
on the existing interactions, and the one with largest $\pT$ `wins'.
Now a third option is added, competing with the first two in the
same way, i.e.\ eq.~(\ref{intermiiisr}) is extended to
\begin{eqnarray}
\frac{\d \mathcal{P}}{\d \pT} & = &
\left( \frac{\d \mathcal{P}_{\mrm{MI}}}{\d \pT} + \sum
\frac{\d \mathcal{P}_{\mrm{ISR}}}{\d \pT} + \sum
\frac{\d \mathcal{P}_{\mrm{JI}}}{\d \pT} \right) \; \times
\nonumber \\
& & \times \;
\exp \left( - \int_{\pT}^{p_{\perp i-1}} \left(
\frac{\d \mathcal{P}_{\mrm{MI}}}{\d \pT'} + \sum
\frac{\d \mathcal{P}_{\mrm{ISR}}}{\d \pT'} + \sum
\frac{\d \mathcal{P}_{\mrm{JI}}}{\d \pT'} \right)
\d \pT' \right) ~.
\label{intermiiisrji}
\end{eqnarray}
The JI sum runs over all pairs of initiator
partons with allowable flavour combinations, separately for the
two incoming hadrons. A gluon line can always be joined with a quark
or another gluon one, and a sea quark and its companion can be joined
into a gluon. For each of these possibilities,
$\d \mathcal{P}_{\mrm{JI}} \, \exp ( - \int \d \mathcal{P}_{\mrm{JI}})$
can be used to do a backwards evolution from the
$\pTmax = p_{\perp i-1}$ scale given by the previous step. If such a
trial joining occurs at a larger $\pT$ scale than any of the other
trial possibilities, then it is allowed to occur. Also the
regularization procedure at small $\pT$ values is the same as for
MI and ISR.
 
The parton densities we will use are defined in the same spirit as
previously discussed, e.g.\ $f_b(x_b,\pTs)$ and $f_c(x_c,\pTs)$ are
squeezed into ranges $x\in[0,X]$, where $X$ is reduced from unity
by the momentum carried away by all but the own interaction, and
for $f_a(x_a,\pTs)$ by all but the $b$ and $c$ interactions. Note
that companion distributions are normalized to unity. Therefore,
for heavy quarks, the branching probability $\g \to \Q\Qbar$ goes
like $1/\ln(\pTs/\mQ^2)$ for $\pTs \to \mQ^2$, as it should, rather
than like $1/\ln^2(\pTs/\mQ^2)$, which would have been obtained if
$f_{\Q}$ and $f_{\Qbar}$ independently were assumed to vanish in
this limit.
 
\begin{figure}[t]
\begin{picture}(150,80)(-60,-50)
\SetWidth{1.5}
\LongArrow(-50,15)(20,15)\Text(-15,25)[]{$\mbf{p}_b$}
\LongArrow(-50,-15)(0,-15)\Text(-25,-25)[]{$\mbf{p}_c$}
\LongArrow(60,15)(20,15)\Text(40,25)[]{$\mbf{p}_d$}
\LongArrow(80,-15)(0,-15)\Text(40,-25)[]{$\mbf{p}_e$}
\Text(15,-45)[]{(a)}
\end{picture}\hfill
\begin{picture}(270,80)(-180,-50)
\SetWidth{1.5}
\LongArrow(-170,0)(-50,0)\Text(-110,10)[]{$\mbf{p}_a$}
\LongArrow(-50,0)(20,15)\Text(-15,20)[]{$\mbf{p}'_b$}
\LongArrow(-50,0)(0,-15)\Text(-25,-20)[]{$\mbf{p}'_c$}
\LongArrow(60,15)(20,15)\Text(40,25)[]{$\mbf{p}_d$}
\LongArrow(80,-15)(0,-15)\Text(40,-25)[]{$\mbf{p}_e$}
\Text(-45,-45)[]{(b)}
\end{picture}
\caption{Kinematics of the $b+d$ and $c+e$ colliding systems
(a) before and (b) after the $a \to b + c$ branching is reconstructed.
\label{fig:kinemtwoevol}}
\end{figure}
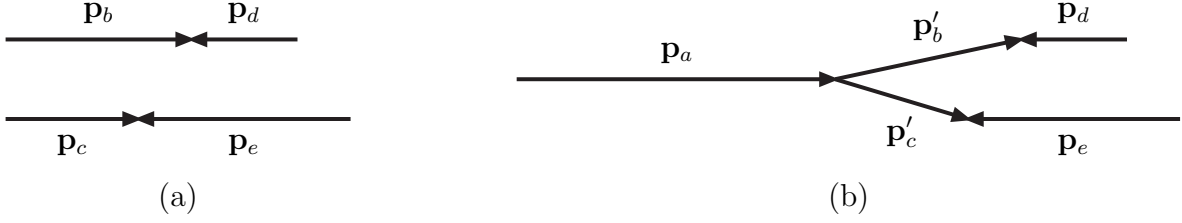
 
Unfortunately the kinematics reconstruction offers a complication.
Consider a system with recoilers $d$ and $e$ to $b$ and $c$,
respectively, as depicted in Fig.~\ref{fig:kinemtwoevol}. Use a prime
to denote the modified $b$ and $c$ four-momenta after the
$a \to b + c$ branching has been constructed, while $d$ and $e$ should
be unchanged. From $x_a = x_b + x_c$ it then follows that
$p_a = p_b + p_c = p'_b + p'_c$, and from the $z$ definition that
\begin{eqnarray}
(p'_b + p_d)^2 = (p_b + p_d)^2 & = & z (p_a + p_d)^2 ~, \\
(p'_c + p_e)^2 = (p_c + p_e)^2 & = & (1-z) (p_a + p_e)^2 ~.
\end{eqnarray}
Further, $p'_b$ and $p'_c$ should have opposite and compensating
transverse momenta given by the $\pT$ scale above, and spacelike
virtualities to be determined. Now, it turns out that these requirements
overconstrain the system. The basic problem is illustrated by
eq.~(\ref{pbspacebranch}): the spacelike parton needs to pick up a
larger $p_{\parallel}$ component than its $z$ share, in order to retain
the invariant mass with the recoiler when the $\pT$ is introduced. So,
if both daughters should be spacelike, not both of them can pick up more
$p_{\parallel}$ than $E$, given that $a$ is massless. (A solution where
one of $p'_b$ and $p'_c$ is timelike sometimes exists, but not always,
and is anyway rather contrived.)
 
We see two alternative ways out of this dilemma.
\begin{Itemize}
\item Retain the $x_a = x_b + x_c$ expression, at the expense of not
giving any $\pT$ or virtualities in the branching, i.e.\ $p'_b = p_b$
and $p'_c = p_c$. Then $\pT$ only plays the role of a formal evolution
parameter, denoting the scale above which $b$ and $c$ may radiate and
interact separately.
\item Insist on having a $\pT$ kick in the branching. Then a sensible
(but not unique) choice is to put ${p'_b}^2 = {p'_c}^2 = - \pT^2$,
such that both have $m_{\perp} = 0$ and thus $p_{\parallel} = E$.
These energies must now be scaled up somewhat, to
$E'_b = (1+\pTs/m_{bd}^2) E_b$ and $E'_c = (1+\pTs/m_{ce}^2) E_c$, for
the invariant masses with the recoiler to be preserved, and therefore
\begin{equation}
x_a = \left( 1 + \frac{\pTs}{m_{bd}^2} \right) x_b
    + \left( 1 + \frac{\pTs}{m_{ce}^2} \right) x_c ~.
\end{equation}
It is then this $x_a$ that should be used in parton densities, to ensure
that the probability of a joining is suppressed near the kinematical limit.
\end{Itemize}
Given that no joinings are possible until after (at least) two
interactions have been generated, and that the rate increases roughly
quadratically with the number of interactions, this physics mechanism
becomes more important at smaller $\pT$ values. Therefore we do not
expect the above two extremes to differ that significantly for practical
applications.
  
\subsection{Joined interactions: results}
 
Although an algorithm implementing the full kinematics for joined
interactions has not yet been constructed, it is still possible to gauge the
order of magnitude of the effects such joinings could have. We do this 
by formally performing the backwards evolution according to
eq.~(\ref{intermiiisrji}), i.e.\ including the joining term
eq.~(\ref{fcorrevol}) in competition with the ordinary ISR and MI terms, 
without actually letting the generated joinings occur physically. Thereby we 
still obtain an estimate for how often and at which $\pT$ values joinings would
occur. 

Since we do not perform the joinings physically, 
the backwards evolution could in principle attempt joinings involving the
\emph{same} initial state shower chain more than once. Such joinings are
of course rejected; only the first joining involving a particular chain is
kept track of.  

\begin{figure}[tp]
\begin{center}\vspace*{-12mm}%
\begin{tabular}{cc}\hskip-0.8cm
\includegraphics*[scale=0.72]{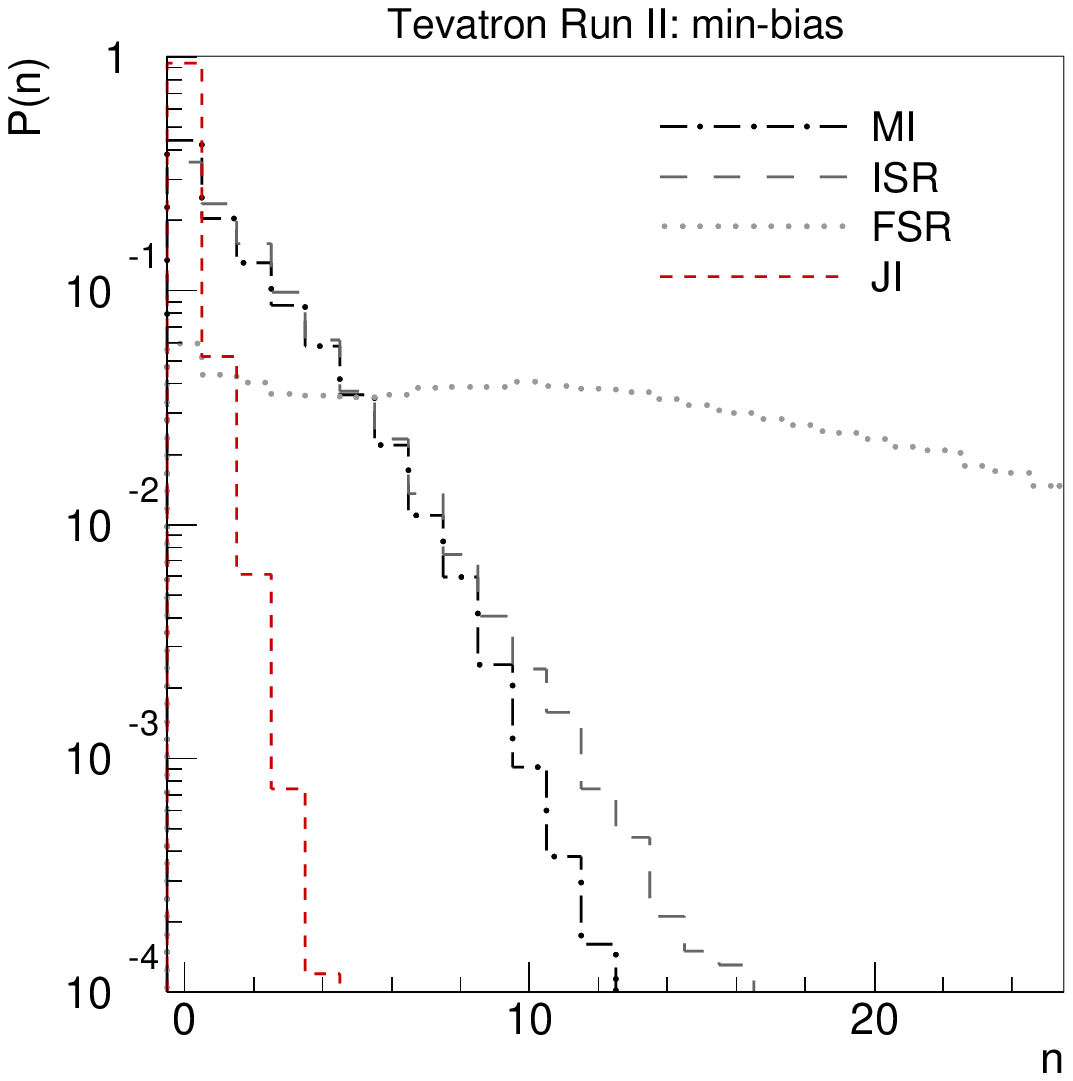}\hskip-1.9cm&\hskip-1.9cm
\includegraphics*[scale=0.72]{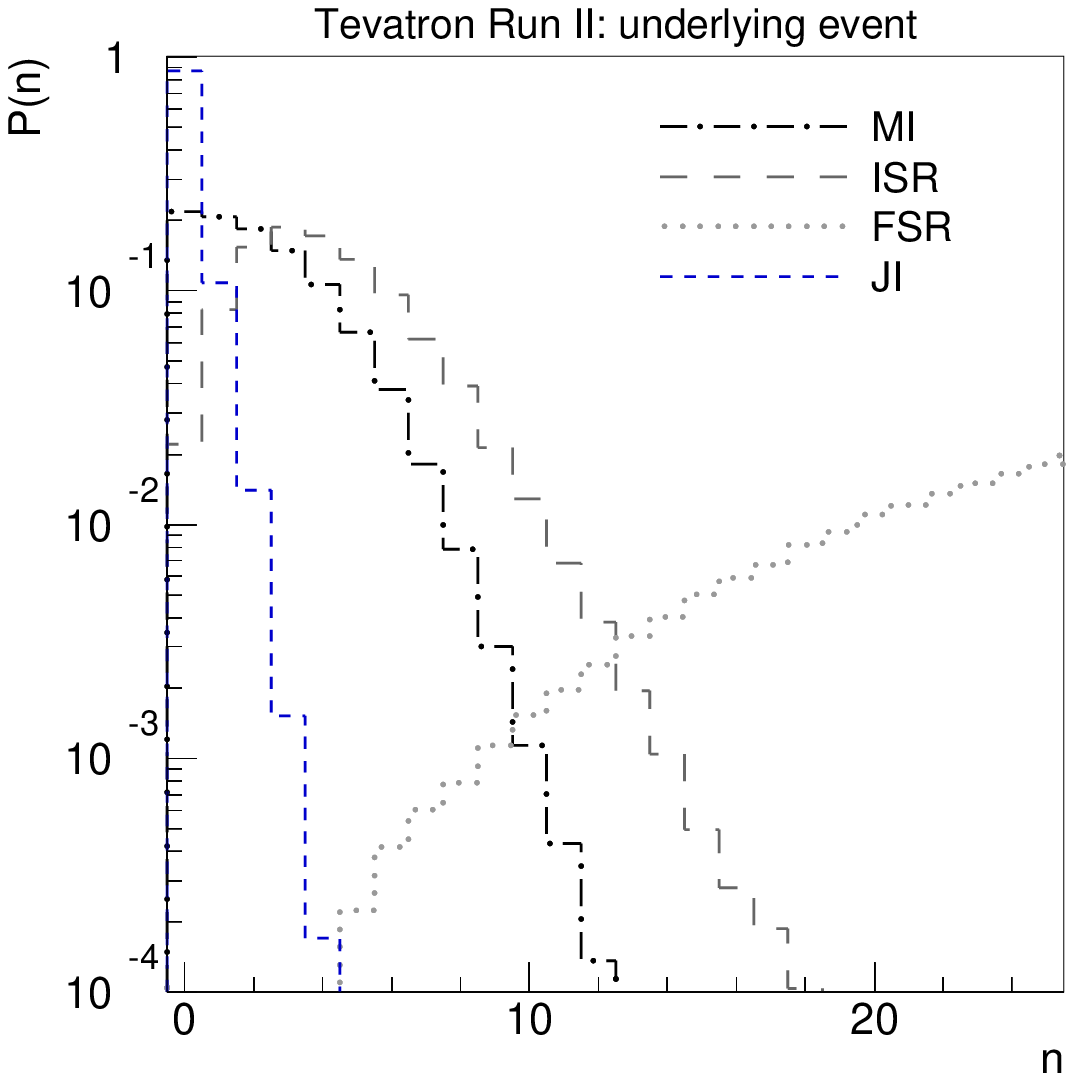}\hskip-1cm\\[-10mm]
\hskip-0.8cm{\it a)}\hskip-1.9cm &\hskip-1.9cm {\it b)}\hskip-1cm
\end{tabular}
\caption{Probability distributions of MI, ISR, FSR, and JI in \emph{a)}
 min-bias events and \emph{b)} events with $\pThard>100$~GeV, for 1.96~TeV
 $\p\pbar$ events. Note that the MI distribution does \emph{not} include the
 hardest scattering.\label{fig:c6mb0}} 
\end{center}\vspace*{-3mm}
\end{figure}

Taking the High FSR model in Table \ref{tab:tunes} as a fair 
representative of the evolution in the new framework, we show the number
distributions of 
multiple interactions excluding the first (MI), ISR branchings, FSR
branchings, and trial joinings (JI) in 1.96~TeV $\p\pbar$
minimum-bias events, Fig.~\ref{fig:c6mb0}a, and for events
where the $\pT$ of the hard interaction is above 100 GeV,
Fig.~\ref{fig:c6mb0}b. Below, we refer to the former as the ``min-bias''
sample and to the latter as the ``UE'' (underlying event) sample.
FSR is shown mainly for reference here, the important graphs being the ones
illustrating the evolution in the initial state: MI, ISR, and JI. 
One clearly observes that joinings are much less frequent than the other
types of evolution steps, averaging at roughly one joining per
15 events for the min-bias sample and one per 7 events for the UE
sample. Thus, even when relatively hard physics is involved, shower joinings
do not appear to take a very prominent role in the evolution.

\begin{figure}[tp]
\begin{center}\vspace*{-10mm}%
\begin{tabular}{cc}\hskip-0.8cm
\includegraphics*[scale=0.72]{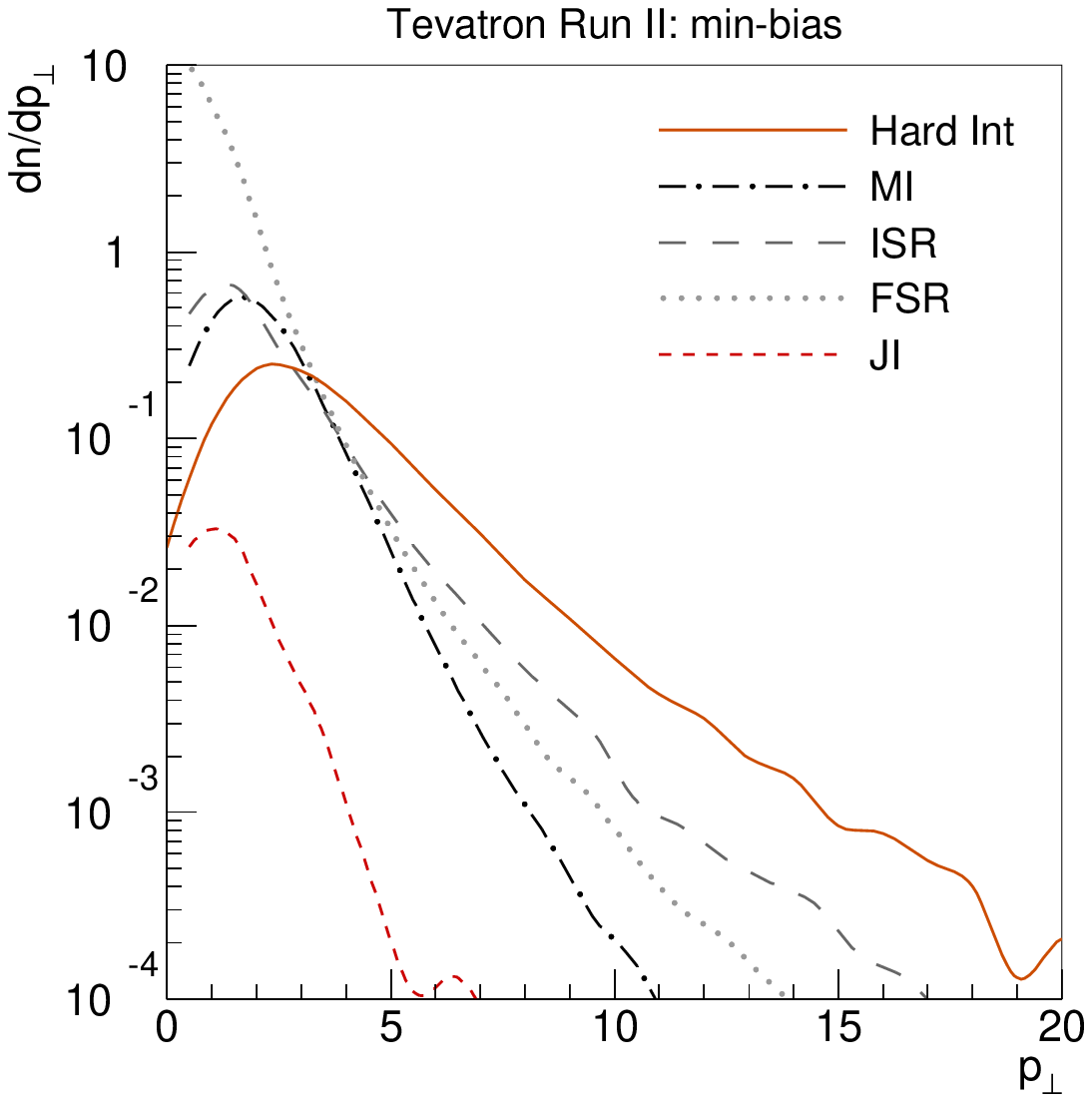}\hskip-1.9cm&\hskip-1.9cm
\includegraphics*[scale=0.72]{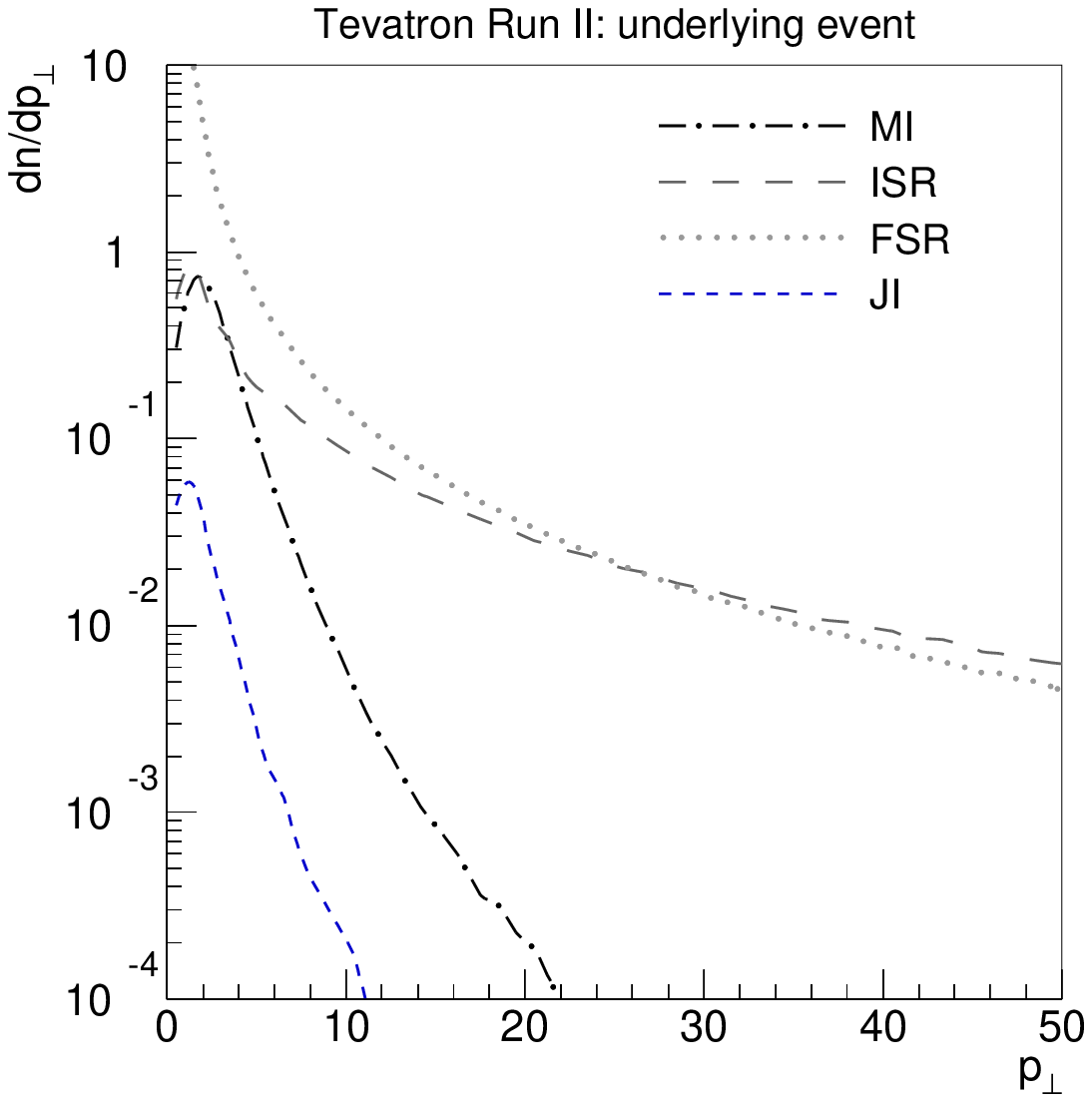}\hskip-1cm\\[-10mm]
\hskip-0.8cm{\it a)}\hskip-1.9cm & \hskip-1.9cm{\it b)}\hskip-1cm
\end{tabular}
\caption{$\pTe$ distributions showing the scale at which multiple
  interactions (MI), ISR branchings, FSR branchings, and joined interactions
  (JI) occur, in 1.96~TeV $\p\pbar$ min-bias events. \emph{a)} Minimum-bias
  events, with the $\pT$
  scale of the hardest interaction shown for reference (solid
  line). \emph{b)} events with $\pThard>100$~GeV. Note that the $\pT$ axis
  goes out to 20 GeV in \emph{a)} and to 50 GeV in \emph{b)}.
\label{fig:c6mb2}} 
\end{center}\vspace*{-3mm}
\end{figure}

To complement the number distributions, Fig.~\ref{fig:c6mb2} shows
\emph{where} the evolution steps occur in $\pTe$. 
As expected, the joinings occur at comparatively low values of $\pTe$. Also
notice that both the ISR, MI, and JI distributions exhibit a turnover around
$\pTo$, characteristic of the smooth regularization used in the High FSR
model.  

\begin{figure}[tp]
\begin{center}\vspace*{-10mm}%
\begin{tabular}{cc}\hskip-0.8cm
\includegraphics*[scale=0.72]{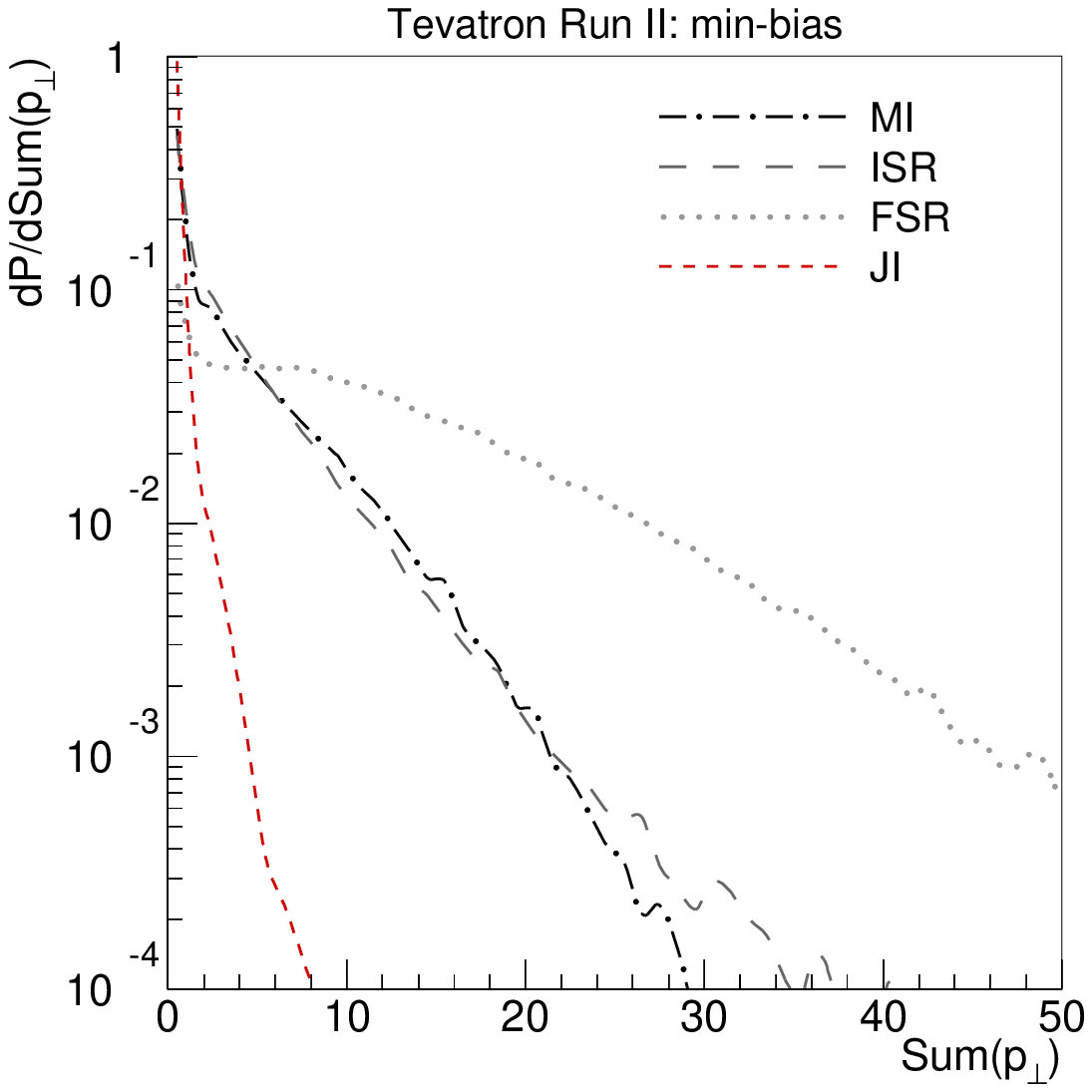}\hskip-1.9cm&\hskip-1.9cm
\includegraphics*[scale=0.72]{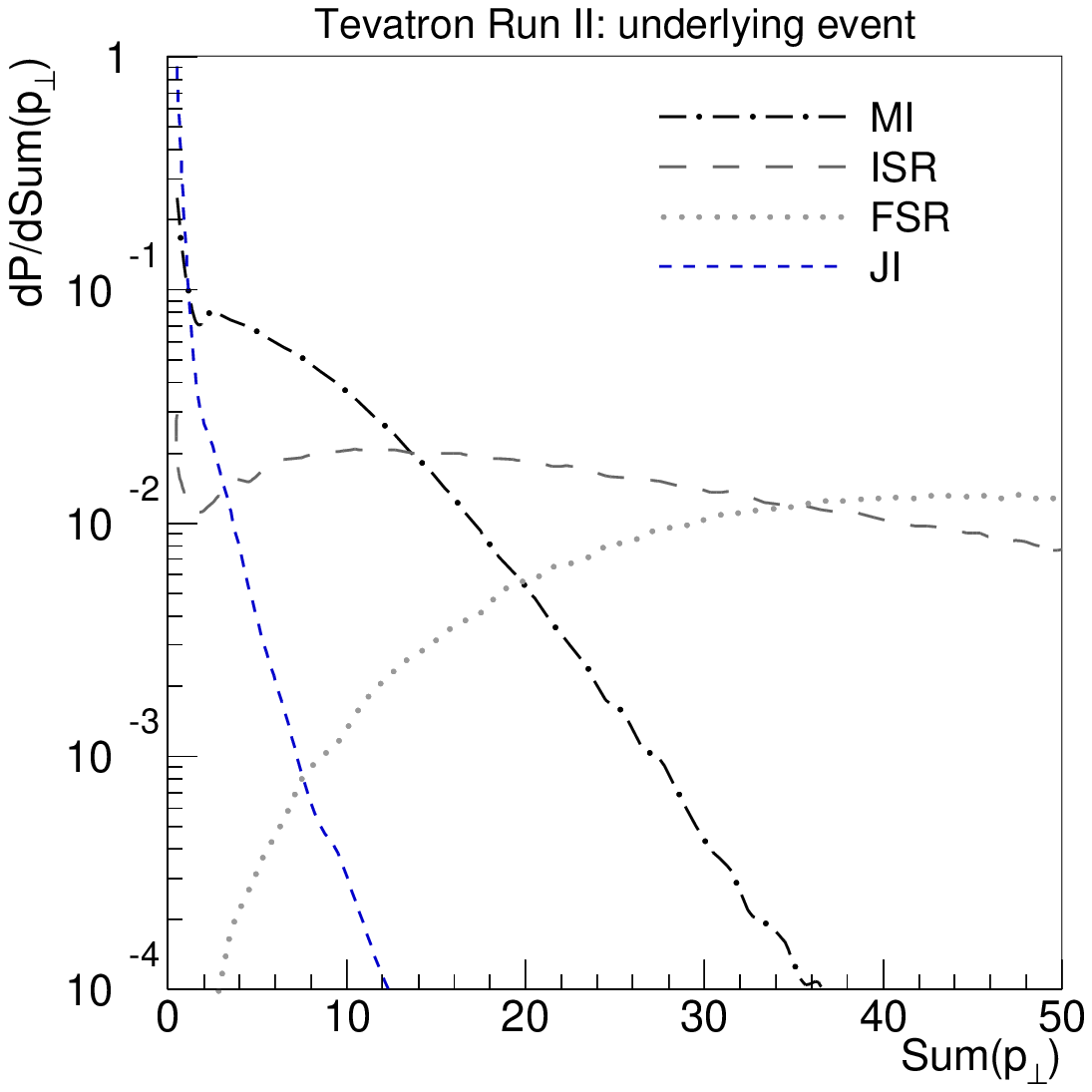}\hskip-1cm\\[-10mm]
\hskip-0.8cm{\it a)}\hskip-1.9cm & \hskip-1.9cm{\it b)}\hskip-1cm
\end{tabular}
\caption{$\sum\pT$ distributions for MI, ISR, FSR and JI, in 1.96~TeV 
  $\p\pbar$  \emph{a)} minimum-bias  events and \emph{b)} events with 
  $\pThard>100$~GeV. Note that the MI distribution does not include the 
  hardest scattering.
\label{fig:c6mb6}} 
\end{center}\vspace*{-3mm}
\end{figure}

Finally, Fig.~\ref{fig:c6mb6} shows the total $\pT$ sum of MI, ISR, FSR, and JI
activity, respectively. That is, for each interaction, branching, or joining,
a scalar $\pT$ is defined, which is added to a cumulative sum. For MI and JI 
this $\pT$ is defined with respect to the beam axis, while the ISR and FSR
$\pT$ is defined with respect to the branching parton, which for ISR is 
roughly along the beam direction, but for FSR normally not. Therefore
FSR mainly broadens jets, i.e.\ redistributes the existing $\ET$, whereas the 
other mechanisms increase the total $\ET$ of the event.
Again, the cumulative effect of joinings is small, with only about 1\% of
the min-bias sample and 5\% of the UE one exhibiting more than 2 GeV of
total $\pT$ from joinings.

\section{Conclusions}
\label{sec:end}

It would seem natural to consider the evolution of a high-energy event
in the normal time order. In such a framework, the incoming hadrons are 
evolved from a simple partonic configuration at a low $Q_0$ scale, up 
through a number of short-lived fluctuations at different virtuality scales.
At the moment of collision, the two sets of partons may undergo several
independent interactions. The scattered partons can radiate in the final 
state, while fluctuations inside which no scatterings occurred may 
recombine. Finally the set of outgoing low-virtuality partons hadronize.

Such an approach has the advantage that it inherently provides multiparton
distributions, and thereby automatically contains correlations between
interactions, including what is here called joined interactions. It does 
not remove the need to consider possible scatterings in some order of 
hardness, however: the partons of a high-virtuality fluctuation may either 
interact individually or collectively, the latter as the unresolved mother 
parton at a lower resolution scale, and the former should preempt the 
latter. 

There is also the well-known problem that it is difficult to generate 
rare processes, since there is no straightforward way to preselect the 
forwards evolution to give the desired configuration. The nightmare example 
is the production of a narrow Higgs state, where the incoming partons must 
match very precisely in invariant mass for a reaction to be possible. More 
generally, efficiency suffers from the need to consider a wealth of virtual 
fluctuations that in the end lead to nothing. The assignment of individual 
virtualities and transverse momenta to partons in such fluctuations is also 
not unique, and does affect the kinematics reconstruction. And, of 
course, the whole plethora of coherence effects need to be considered.
 
The alternative is to start with the hardest interaction, and then 
`work outwards' to longer timescales both in the past and future, 
i.e.\ to (re)construct less hard steps in the evolution of the event. 
This makes the preselection of desired events straightforward, and in 
general implies that the most striking aspects of the event are considered 
`up front'. The price is a tougher task of reconstructing the soft
associated activity in the initial state, while final-state radiation and
hadronization offer about the same challenges in the two scenarios.

In a set of articles we have begun the task of providing an improved 
description of events along this latter philosophy. The first step 
\cite{BNV} was to develop a model for the hadronization of junction 
topologies, thereby allowing complicated beam remnants. 
The second step \cite{multint} was to develop a framework for correlated
parton densities, in flavour, colour and transverse and longitudinal 
momentum, thereby allowing initial-state radiation to be considered in 
full. In this, the third step, we have interleaved multiple interactions
and initial-state radiation in a common transverse-momentum-ordered 
sequence, with a common dampening procedure at low transverse momenta
to handle destructive interference in this region. Our lack of explicitly 
implemented joined interactions has been shown not to be a major shortcoming, 
since such joinings are reasonably rare. That is, taken together, we now have 
a framework that should provide a complete description of all aspects that 
could have been covered by a forwards-in-time evolution scenario, along 
with the traditional advantages of the backwards-evolution approach.
In addition, the new framework makes use of new algorithms for $\pT$-ordered 
evolution in initial- and final-state radiation, which should further 
improve the quality of the description.
 
It may then be somewhat disappointing that we here have used \textsc{Pythia} 
Tune A \cite{Field} as a reference, well knowing that Tune A is able to 
describe a host of jet and minimum-bias data at the Tevatron, in spite of
it being based on a much more primitive approach \cite{Zijl}. The hope, 
of course, is that our new approach will be able to explain --- and predict 
--- much more data than Tune A can. For sure we know of many aspects of the
old framework that are unreasonable, but that either have not been probed
or that may have been fixed up by a contrived choice of tuned parameters.
Ultimately this is for experimentalists to tell, as tests become increasingly
more sophisticated. Certainly, one should not expect the advantages of the 
new model to become apparent unless a similar effort is mounted as went into 
producing Tune A in the first place.

There are also a few issues still hanging over us, awaiting a `fourth step'.
One is the implementation of joined interactions and $3 \to 3$ rescattering
processes, to see what their real impact is, whether negligible or not. But 
the main one we believe to be the interleaving of final-state radiation with 
multiple interactions and initial-state radiation. On the one hand, such an 
interleaving may not be required, since the competition between 
FSR and MI+ISR is less direct than that between MI and ISR: an FSR emission
at a high $\pT$ scale does not affect the probability for MI or ISR activity
at lower $\pT$ values. On the other hand, there would then also not seem to be 
any disadvantage to having a commonly ordered $\pT$ sequence of MI+ISR+FSR, 
and such an ordering would come in handy for a consistent interfacing to 
higher-order matrix elements. Furthermore, a $\pT$-ordered FSR algorithm is 
available, well matched to the $\pT$-ordering of MI and ISR. 

There is, however, one major open question related to FSR interleaving: 
which parton takes the momentum recoil when a FSR branching pushes a parton 
off the mass shell? The problem is not so much the momentum transfer itself, 
but that the size of the radiating dipole sets the maximum scale for allowed 
emissions. We have in this article illustrated how such a choice can affect 
e.g.\ the jet multiplicity and jet profiles. The crucial distribution is the 
$\langle \pT \rangle (n_{\mrm{ch}})$ one, however. In order to provide a 
reasonable description of the experimental data, we are forced to arrange 
colours in the final state to have a smaller string length than colour 
correlations in the initial state alone would suggest. This problem has 
`always' been there \cite{Zijl}, and is accentuated in Tune A, where as much 
as 90\% of the partons added by multiple interactions are connected so as to 
minimize the string length. The hope that an improved treatment of other 
aspects would remove the need for a special string-length minimization 
mechanism has so far failed to materialize. We therefore need to understand
better how the colour flow is set, and how this influences the evolution
of an event, especially the FSR activity.

The related fields of minimum bias physics and underlying events thus are
further explored but not 
solved with this article, and likely not with the next one either. This 
should come as no surprise: in the world of hadronic physics, there are 
few simple answers. Everything that is not explicitly forbidden is bound 
to happen, and often at a significant rate. To reflect reality, 
the theoretical 
picture therefore has to become more and more complex, as one 
consideration after the next is pulled into the game. However, if the 
journey is interesting and educational, why despair that the end station 
is not yet reached?
 
\subsection*{Acknowledgements}
 
We are grateful to G.~Rudolph for having tuned our new 
$\pT$-ordered final-state shower algorithm to LEP data. 
Communications with Tevatron and LHC physicists, especially 
R.D.~Field, J.~Huston and A.~Moraes, have spurred us on.


\begin{thebibliography}{99}
 
\bibitem{multint}
T. Sj\"ostrand and P.Z. Skands, JHEP {\bf 03} (2004) 053
 
\bibitem{Zijl}
T. Sj\"ostrand and M. van Zijl, Phys. Rev. {\bf D36} (1987) 2019
 
\bibitem{BNV}
T. Sj\"ostrand and P.Z. Skands, Nucl. Phys. {\bf B659} (2003) 243
 
\bibitem{backwards}
T. Sj\"ostrand, Phys. Lett. {\bf 157B} (1985) 321; \\
M. Bengtsson, T. Sj\"ostrand and M. van Zijl, Z. Phys. {\bf C32}
(1986) 67
 
\bibitem{pTshower}
T. Sj\"ostrand, LU TP 04--05 [hep-ph/0401061],
in the QCD/SM Working Group Summary Report of the Physics at TeV
Colliders Workshop, Les Houches, France, 26 May -- 6 Jun 2003
[hep-ph/0403100]
 
\bibitem{Pythia}
T. Sj\"ostrand, P. Ed\'en, C. Friberg, L. L\"onnblad, G. Miu, S. Mrenna
and E. Norrbin, Computer Phys. Commun. {\bf 135} (2001) 238;\\
T. Sj\"ostrand, L. L\"onnblad, S. Mrenna and P. Skands, LU TP 03-38
[hep-ph/0308153].
 
\bibitem{DGLAP}
V.N. Gribov and L.N. Lipatov, Sov. J. Nucl. Phys. {\bf 15} (1972) 438,
{\it ibid.} 75; \\
G. Altarelli and G. Parisi, Nucl. Phys. {\bf B126} (1977) 298;\\
Yu.L. Dokshitzer, Sov. J. Phys. JETP {\bf 46} (1977) 641
 
\bibitem{string}
B. Andersson, G. Gustafson, G. Ingelman and T. Sj\"ostrand,
Phys. Rep. {\bf 97} (1983) 31;\\
T. Sj\"ostrand, Nucl. Phys. {\bf B248} (1984) 469;\\
B. Andersson, `The Lund Model' (Cambridge University Press, 1998)
 
\bibitem{Sudakov}
V.V. Sudakov, Zh.E.T.F. {\bf 30} (1956) 87 (Sov. Phys. J.E.T.P.
{\bf 30} (1956) 65)
 
\bibitem{PythiaFSR}
M. Bengtsson and T. Sj\"ostrand,
Nucl. Phys. {\bf B289} (1987) 810
 
\bibitem{PythiaMEmatch}
E. Norrbin and T. Sj\"ostrand, Nucl. Phys. {\bf B603} (2001) 297
 
\bibitem{Herwig}
G. Marchesini and B.R. Webber, Nucl. Phys. {\bf B238} (1984) 1;\\
G. Corcella, I.G. Knowles, G. Marchesini, S. Moretti, K. Odagiri,
P. Richardson, M.H. Seymour and B.R. Webber,
JHEP {\bf 01} (2001) 010, hep-ph/0210213
 
\bibitem{pTcoherence}
G. Gustafson, Phys. Lett. {\bf B175} (1986) 453
 
\bibitem{GGUP}
G. Gustafson and U. Pettersson, Nucl. Phys. {\bf B306} (1988) 746
 
\bibitem{Ariadne}
L. L\"onnblad, Computer Physics Commun. {\bf 71} (1992) 15
 
\bibitem{coherence}
A.H. Mueller, Phys. Lett. {\bf 104B} (1981) 161; \\
B.I. Ermolaev, V.S. Fadin, JETP Lett. {\bf 33} (1981) 269
 
\bibitem{HerwigMEmatch}
M.H. Seymour, Computer Physics Commun. {\bf 90} (1995) 95
 
\bibitem{GGonPythia}
G. Gustafson, private communication
 
\bibitem{CKKWL}
S. Catani, F. Krauss, R. Kuhn and B.R. Webber,
JHEP {\bf 11} (2001) 063;\\
L. L\"onnblad, JHEP {\bf 05} (2002) 046
 
\bibitem{briefmult}
P.~Skands and T.~Sj\"ostrand, LU TP 03-45 [hep-ph/0310315], in the
proceedings of International Europhysics 
Conference on High-Energy Physics (HEP 2003), Aachen, Germany, 17-23 Jul
2003, European Physical Journal {\bf C} Direct, DOI:
10.1140/epjcd/s2003-03-520-7;\\
T.~Sj\"ostrand and P.~Skands, LU TP 04-04 [hep-ph/0401060], 
in The QCD/SM working group: Summary report, 
3rd Les Houches Workshop: Physics at TeV Colliders, Les Houches, France, 26
May - 6 Jun 2003, M.~Dobbs \emph{et al.} [hep-ph/0403100]

\bibitem{LEPstudy}
OPAL Collaboration, M.Z. Akrawy et al.,
Z. Phys. {\bf C47} (1990) 505;\\
L3 Collaboration, B. Adeva et al.,
Z. Phys. {\bf C55} (1992) 39;\\
I.G. Knowles et al., in `Physics at LEP2', eds. G. Altarelli,
T. Sj\"ostrand and F. Zwirner, CERN 96--01 (Geneva, 1996),
Vol. 2, p. 103;\\
DELPHI Collaboration, P. Abreu et al.,
Z. Phys. {\bf C73} (1996) 11;\\
ALEPH Collaboration, R. Barate et al.,
Phys. Rep. {\bf 294} (1998) 1
 
\bibitem{smallx}
B. Andersson et al., Eur. Phys. J. {\bf C25} (2002) 77;\\
J. Andersen et al., Eur. Phys. J. {\bf C35} (2004) 67
 
\bibitem{AriadneIS}
B. Andersson, G. Gustafson, L. L\"onnblad and U. Pettersson,
C. Phys. {\bf C43} (1989) 625
 
\bibitem{LDCMC}
B. Andersson, G. Gustafson and J. Samuelsson,
Nucl. Phys. {\bf B467} (1996) 443;\\
B. Andersson, G. Gustafson and H. Kharraziha,
Phys. Rev. {\bf D57} (1998) 5543;\\
H. Kharraziha and L. L\"onnblad, JHEP {\bf 03} (1998) 006
 
\bibitem{CCFM}
M. Ciafaloni, Nucl. Phys. {\bf B296} (1987) 249;   \\
S. Catani, F. Fiorani and G. Marchesini, Nucl. Phys. {\bf B336}
(1990) 18
 
\bibitem{Luclus}
 T. Sj\"ostrand, Computer Physics Commun. {\bf 28} (1983) 229
 
\bibitem{Durham}
S. Catani, Yu. L. Dokshitzer, M. Olsson, G. Turnock and B.R. Webber,
Phys. Lett. {\bf B269} (1991) 432
 
\bibitem{asscale}
D. Amati, A. Bassetto, M. Ciafaloni, G. Marchesini and G. Veneziano,
Nucl. Phys. {\bf B173} (1980) 429; \\
G. Curci, W. Furmanski and R. Petronzio, Nucl. Phys. {\bf B175}
(1980) 27
 
\bibitem{Webberrev}
B.R. Webber, Ann. Rev. Nucl. Part. Sci. {\bf 36} (1986) 253
 
\bibitem{Rudolph}
G. Rudolph, private communication
 
\bibitem{MiuZ}
G. Miu and T. Sj\"ostrand, Phys. Lett. {\bf B449} (1999) 313
 
\bibitem{Erik}
E. Thom\'e, master's thesis, Lund University, LU TP 04--01
[hep-ph/0401121];\\
J. Huston, I. Puljak, T. Sj\"ostrand, E. Thom\'e, LU TP 04--07
[hep-ph/0401145], in the QCD/SM Working Group Summary Report of
the Physics at TeV Colliders Workshop, Les Houches, France,
26 May -- 6 Jun 2003 [hep-ph/0403100]
 
\bibitem{meanptfornch}
A. Breakstone et al., Phys. Lett. {\bf 132B} (1983) 463;\\
UA1 Collaboration, C. Albajar et al., Nucl. Phys. {\bf B335} (1990) 261;\\
E735 Collaboration, T. Alexopoulos et al.,
Phys. Lett. {\bf B336} (1994) 599;\\
CDF Collaboration, D. Acosta et al., Phys. Rev. {\bf D65} (2002) 072005

\bibitem{Field}
R.D. Field, presentations at the `Matrix Element and Monte Carlo Tuning
Workshop', Fermilab, 4 October 2002 and 29--30 April 2003, talks available
from webpage \texttt{http://cepa.fnal.gov/CPD/MCTuning/}, and further
recent talks available from 
\ttt{http://www.phys.ufl.edu/}$\sim$\ttt{rfield/cdf/}

\bibitem{CDFZ}
CDF Collaboration, T. Affolder et al., Phys. Rev. Lett. {\bf 84} (2000) 845

\bibitem{threeproc}
N. Paver and D. Treleani, Nuovo Cimento {\bf 70A} (1982) 215,
Phys. Lett. {\bf 146B} (1984) 252, Z. Phys. {\bf C28} (1985) 187

\bibitem{intertwpdf}
K. Konishi, A. Ukawa and G. Veneziano,
Nucl. Phys. {\bf B157} (1979) 45;\\
R. Kirschner, Phys. Lett. {\bf 84B} (1979) 266;\\
V.P. Shelest, A.M. Snigirev and G.M. Zinovjev,
Phys. Lett. {\bf 113B} (1982) 325;\\
A.M. Snigirev, Phys. Rev. {\bf D68} (2003) 114012;\\
V.L. Korotkikh and A.M. Snigirev, hep-ph/0404155
 
\end{thebibliography}
\end{document}